\documentclass[a4paper,10pt,reqno]{amsart}
\usepackage{latexsym}
\usepackage{amssymb}
\usepackage{amsmath}
\usepackage{amsmath,amssymb,amsthm}
\usepackage{mathrsfs}
\begin{document}  

\renewcommand{\baselinestretch}{1.2}
\newcommand{\x}{\times}
\newcommand{\<}{\langle}
\renewcommand{\>}{\rangle}
\renewcommand{\i}{\infty}
\newcommand{\p}{\partial}
\newcommand{\mat}[4]{\ensuremath{{#1 \, #2 \choose #3 \, #4}}}
\newcommand{\dmat}{\left( \begin{array}{ccc}}
\newcommand{\ea}{\end{array} \right)}
\newcommand{\mbf}{\mathbf}
\newcommand{\mbb}{\mathbb}
\newcommand{\dia}{\diamond}
\newcommand{\tbf}{\textbf}
\newcommand{\rta}{\rightarrow}
\newcommand{\Rta}{\Rightarrow}
\newcommand{\mcl}{\mathcal}
\newcommand{\mclh}{\mathcal{H}}
\newcommand{\mcloi}{\mathcal{O}(-\infty)}
\newcommand{\mcls}{\mathcal{S}}
\newcommand{\mcld}{\mathcal{D}}
\newcommand{\ox}{\otimes}
\newcommand{\dg}{\dagger}
\newcommand{\wdg}{\wedge}
\newcommand{\halb}{\frac{1}{2}}
\newcommand{\prl}{\parallel}
\newcommand{\vc}[2]{(^{#1}_{#2})}
\newcommand{\inv}[1]{\frac{1}{#1}}
\newcommand{\lo}{\leqno}
\newcommand{\ex}{\begin{equation}}
\newcommand{\ey}{\end{equation}}
\newcommand{\eu}{\begin{eqnarray}}
\newcommand{\ev}{\end{eqnarray}}
\newcommand{\XXX}{$\clubsuit \clubsuit \clubsuit$}
\newcommand{\sqre}{\fbox{\rule{0cm}{.15cm}   }}
\newcommand{\grad}{\mbox{ grad }}
\newcommand{\curl}{\mbox{ curl }}

\renewcommand{\a}{\alpha}
\renewcommand{\b}{\beta}
\renewcommand{\d}{\delta}
\newcommand{\D}{\Delta}
\newcommand{\e}{\varepsilon}
\newcommand{\g}{\gamma}
\newcommand{\G}{\Gamma}
\newcommand{\io}{\iota}
\renewcommand{\l}{\lambda}
\renewcommand{\L}{\Lambda}
\newcommand{\vi}{\varphi}
\newcommand{\s}{\sigma}
\newcommand{\Sig}{\Sigma}
\renewcommand{\t}{\tau}
\renewcommand{\th}{\theta}
\newcommand{\Th}{\Theta}
\renewcommand{\o}{\omega}
\renewcommand{\O}{\Omega}
\renewcommand{\u}{\upsilon}
\newcommand{\U}{\Upsilon}
\newcommand{\z}{\zeta}
\newcommand{\kp}{\kappa}


\newcommand{\lpd}{(2\pi)^{-1/2}}
\newcommand{\kpd}{(2\pi)^{-3/2}}
\newcommand{\kpdz}{\inv{8\pi^3}}
\newcommand{\Opsc}{Op\psi c}
\newcommand{\opsc}{Op\psi c}
\newcommand{\pdo}{$\psi$do}
\newcommand{\trace}{\mbox{trace }}

\newtheorem{theorem}{Theorem}[section]
\newtheorem{lemma}[theorem]{Lemma}
\newtheorem{corollary}[theorem]{Corollary}
\newtheorem{proposition}[theorem]{Proposition}
\newtheorem{definition}[theorem]{Definition}
\newtheorem{remark}[theorem]{Remark}
\newtheorem{thm}[theorem]{Theorem}
\newtheorem{prop}[theorem]{Proposition}
\newtheorem{cor}[theorem]{Corollary}
\newtheorem{rem}[theorem]{Remark}
\newtheorem{defn}[theorem]{Definition}
\newtheorem{note}[theorem]{Note}
\newtheorem{obs}[theorem]{Observation}

{Investigations in Mathematical Sciences}\quad\quad\quad\quad\quad\quad\quad\quad\quad\quad\quad\quad\quad\quad\quad\quad\quad{ISSN:2250-1436}

{ Vol.4(2), 2014, 1-44 } 

\bigskip
\bigskip
\title[MATHEMATICAL ANALYSIS OF DIRAC EQUATION PHYSICS]{A MATHEMATICAL ANALYSIS OF DIRAC EQUATION PHYSICS}
\author[H.O.CORDES]{H.O.CORDES\copyright 2014}

\maketitle
       \begin{center}

To the Memory of Lars H\"{o}rmander        

       \end{center}
       \LARGE

        \bigskip

         \scriptsize

\begin{abstract}
This paper analyzes time-propagation of Dirac observables --- using Heisenberg representation --- in the 
light of various pseudodifferential operator algebras. We have discussed such matters earlier (cf. [Co3],
[Co15,[Co16]), observing the elegant relation to classical physics coming into play, also giving insight
into a (sort of) magnetic moment, representing the spin.\

Presently we analyze this more carefully --- looking at the Physical aspects. Our theory gives (i) a mechanical
angular momentum (the spin) and (ii) another real 3-vector travelling with the particle with magnetic properties
(its motion guided by the magnetic field around it, but not in the proper relativistic way). This questions the
interpretation of the magnetic moment of the particle being generated by rotation of the charge, as
suggested by macroscopic arguments.

All the above was proven under assumptions on potentials, making them vanish  at infinity. But we now also
look at a Dirac particle under the influence of a plane polarized
X-ray-wave, trying to analyze the Compton effect. 
What we can derive there might also be surprising: Looking at the total energy $E$ and the orbital
momentum  coordinate $P_1$ in the direction of the radiation, we find that these two observables are
coupled. Their time propagation shows a number of discrete possibilities: Either there is no change in time,
of both $E$ and $P_1$ or there is a change by $nh\nu$ of $E$ and $nh\nu/c$ of $P_1$ with an integer
$n=1,2,\cdots$ --- with same $n$ for $E$ and $P_1$. This is valid for large frequencies --- i.e., large
values of the momentum coordinates.

We need not point out the possible interpretation: There may be a collision of the electron-positron-particle
with one -- or two --- or $n$ --- Photons of total energy $h\nu$ each, effecting a sudden change of 
energy and momentum. Observe, this does not require 
any use of QFT. 
\end{abstract}

\normalsize

\bigskip


\bigskip

\bigskip

\section{Introduction}

In this paper we try to apply rigorous mathematics to analyze two different physical problems,
attached to Dirac's first order symmetric hyperbolic $4\x 4$-system of partial differential
equations, using calculus of pseudodifferential operators, resp. Fourier integral operators.
In sections 3 through 6 we have a class of electro-magnetic potentials vanishing at $|x|=\i$,
including the Coulomb potential with its singularity smoothened out. In sections 7 to 12
we deal with a Dirac particle under the (time-dependent) potential of an electro-magnetic wave,
such as occurring at the Compton effect.

In the first case we mainly focus on the spin of the particle: We can establish a mechanical spin,
as a 3-vector, travelling with the particle, behaving just like a mechanical angular momentum
should, in this relativistic environment.

But, on the other hand, there is another 3-vector $\vec{\kp}$, also travelling with the particle, with its
motion along the particles orbit entirely determined by the two components $\mcl{B}$ and 
$\dot{x}\x \mcl{E}$, combined in a way not expected for the magnetic field, the moving
particle see's. Actually if either $\mcl{E}=0$ or $\mcl{B}=0$, then the movement of $\vec{\kp}$
fits that of a magnetic moment. But then there is a difference in strength of these two magnetic moments
by a factor $(1+\sqrt{1-\dot{x}^2})$.     
That factor\footnotemark will be $\approx 2$, for relativistically small $\dot{x}$. 
While we think that , perhaps a better mathematical construction might correct this,
so that the vector $\vec{\kp}$ might be regarded as the magnetic moment generated by the spinning
charge of the particle, we are left open, with this problem.

\footnotetext{Reading in R.Becker [Be1], p.85, we note that there also seems to be a factor 2-discrepancy 
in the theoretical interpretation of the Einstein-de Haas experiment, and electron spin.}
 
In the second case  --- an electron under an X-ray-wave --- we also end up with a contradiction to
general expectation: a possible mathematical rediscovery of simple (or multiple) collision between
the Dirac particle and photons of energy $h\nu$ and momentum $h\nu/c$, from Dirac's and Maxwells
equations only. Looking at old standard
text, such as Sommerfeld [So1] , ch.1, sec.7,  this was believed to be impossible  to explain from Dirac- or 
electro-magnetic wave theory. But we believe now, it probably can be explained --- and without using
second quantization, i.e., without quantizing the electro-magnetic field.

The organization of the paper seems clear, after these remarks. In sec.2 we give some basics of 
Dirac's equation; in sec.3 we try to give hints about 3 different algebras of pseudodifferential
operators, with the main effort on explaining various asymptotically convergent Leibniz formulas:
the asymptotic convergence to be regarded none other than that of the well known
Hankel-asymptotic expansions for Bessel-functions at infinity: totally divergent, but still extremely useful.

Unfortunately, as a retired mathematician, working alone, we feel quite helpless in examining the huge physical
literature on the subject. We are very grateful to have available the large reference section in the book of 
B. Thaller [Th1] of 1992 on Dirac's equation,  but apologize in advance to anyone who
might have worked in similar directions without our knowledge.

        \normalsize

        \bigskip

\section{Elementary Facts on Dirac Operators}

We depart from the non-relativistic Dirac equation $\dot{\psi}+iH\psi=0$ with
$\dot{\psi}=\p\psi/\p t$, and the `Dirac operator'
$$
H=\sum_{j=1}^3\a_j(D_j-\mbf{A}_j(t,x))+\b+\mbf{V}(t,x)\ ,\ D_j=\inv{i}\p/\p x_j\ ,\ 
\lo{(2.1)}
$$
with a set $\a_j,\b$ of self-adjoint $4\x 4$-(Dirac)-matrices satisfying
$$\a_j\a_l+\a_l\a_j=2\d_{jl}\ ,\  \b^2=1\ ,\ \a_j\b+\b\a_j=0\ ,\ j,l=1,2,3,
\lo{(2.2)}
$$
and with real-valued potentials $\mbf{V}(t,x),\mbf{A}_j(t,x)\ ,\ j=1,2,3.$.

The first order differential operator $H$ in the 3 variables $x_1,x_2,x_3$ has `symbol'
$$
h(t,x,\xi)= \sum_{j=1}^3\a_j(\xi_j-\mbf{A}_j(t,x))+\b+\mbf{V}(t,x)\ ,\ D_j=\inv{i}\p/\p x_j\ ,\ 
\lo{(2.3)}
$$
so that we may write $H=h(t,x,D)$. For the mathematics of the differential equation $\dot{\psi}+iH\psi=0$
the spectral behaviour of the $4\x 4$-matrix-valued function $h(t,x,\xi)$ is important. 
 Clearly $h(t,x,\xi)$, as a self-adjoint $4\x 4$-matrix, has real eigenvalues.
We get $(h-\mbf{V})^2=1+|\xi-\mbf{A}|^2=\<\xi-\mbf{A}\>^2$, a scalar multiple of $1$, as a consequence of relations (2.2).
Accordingly, $h$  can only have the eigenvalues $\l_\pm=\mbf{V}\pm\<\xi-\mbf{A}\>$, and the orthogonal projections on corresponding
eigenspaces are given by
$$
p_\pm(t,x,\xi)=\halb(1\pm\inv{\<\xi-\mbf{A}(t,x)\>}h(t,x,\xi))\ .
\lo{(2.4)}
$$
A calculation shows that both eigenspaces are two-dimensional, for every $t,x,\xi$.

There even will be a need for a unitary $4\x 4$-matrix $\Upsilon$ diagonalizing  the self-adjoint $h(t,x,\xi)$, then
also supplying a natural orthonormal set of eigenvectors. For this we introduce the $4\x 4$-matrix
$$
\Upsilon(t,x,\xi)=\inv{\sqrt{2(1+\u_0)}}(1+\u_0-\b\a\u)\ ,\
\u(x,\xi)=\frac{\xi-\mbf{A}(t,x)}{\<\xi-\mbf{A}(t,x)\>}
\ ,\  \u_0(x,\xi)=\inv{\<\xi-\mbf{A}(t,x)\>}\ .
\lo{(2.5)}
$$
Using (2.2) again, a calculation shows that we have
$$
\Upsilon^*\Upsilon=1\ ,\ \Upsilon^*h\Upsilon=\mbf{V}(t,x)+\<\xi-\mbf{A}(t,x)\>\b\ \mbox{ for all } t,x,\xi\ .
\lo{(2.6)}
$$
Accordingly, the matrix $\Upsilon$ will diagonalize $h(t,x,\xi)$ for every $t,x,\xi$, if we select a set of Dirac matrices
such that $\b$ equals the diagonal matrix with entries $1,1,-1,-1$.

Actually, we are going to use two kinds of Dirac matrices $\a_j,\b$. Introducing the $2\x 2$-Pauli matrices
$$
\s_1=\mat{\ 0}{\ i}{-i}{\ 0}\ ,\ \s_2=\mat{0}{1}{1}{0}\ ,\
\s_3=\mat{1}{\ \ 0}{0}{-1} \ ,
\lo{(2.7)}
$$
we  may define
$$
\a=\mat{0\ \ \ }{i\s}{-i\s}{\ 0}\ ,\ \b=\mat{1\ \ }{\ 0}{0\ }{-1}\ ,
\lo{(2.8)}
$$
writing the $4\x 4$-matrices as $2\x 2$-matrices of $2\x 2$-blocks.
This indeed checks with the conditions (2.2), while, indeed, $\b$ is the diagonal matrix  with entries as desired above.

Another set of Dirac matrices will be used in sections 7-11. There we set
$$
\a_1=\mat{-1\ \ }{\ \ 0\ }{\ \ 0\ \ }{\ \ 1\ }\ ,\ 
\a_2=i\mat{\ \ 0\ \ }{\ \s_3\ }{-\s_3\ \ }{\ \ 0\ \ }\ ,\
\a_3=i\mat{\ \ 0\ \ }{\ \s_2\ }{-\s_2\ \ }{\ \ 0\ \ }\ ,\
\b=\mat{\ 0\ }{\ 1\ }{\ 1\ }{\ 0\ }\ ,
\lo{(2.9)}
$$
again checking with (2.2). The set (2.9) will not have $\b$ diagonal but, instead, have $\a_1$ with that property,
this being helpful when we use the positive $x_1$-direction as the direction of an incoming X-ray.
The set (2.9) may be related to (2.8) by conjugating each matrix (2.8) with a certain constant real 
orthogonal $4\x 4$-matrix.
 
The lemma, below, is valid for any choice of Dirac matrices $\a,\b$ satisfying (2.2), and the corresponding projections
$p_\pm(\xi)$ of (2.4), setting $\mbf{V}=\mbf{A}_j=0\ ,\ j=1,2,3$. Its proof is a calculation.

\begin{lemma}
We have
$$
p_\pm(\xi)\a_j p_\pm(\xi)=\pm s_j(\xi)p_\pm(\xi)\ ,\ j=1,2,3\ ,\ \ \ \ \ 
p_\pm(\xi)\b p_\pm(\xi)=\pm s_0(\xi)p_\pm(\xi)\ ,\
\lo{(2.10)}
$$
where we have set $s_j(\xi)=\xi_j/\<\xi\>\ ,\ s_0(\xi)=1/\<\xi\>\ ,\ j=1,2,3.$ 

\end{lemma}

\bigskip
 
It is known that the Dirac equation $\dot{\psi}+H\psi=0$ has a solution $\psi(t,x)$ satisfying $\psi(0,x)=\psi_0(x)$
where $\psi_0(x)$ may be any complex 4-vector-valued function satisfying $\int |\psi_0(x)|^2dx<\i$. In fact,
we get 
$$
\int |\psi(t,x)|^2dx=\int |\psi_0(x)|^2dx \ ,\  \mbox{ for all } t \  .
\lo{(2.11)}
$$
Defining a linear operator $U(t)$ in the Hilbert space $\mclh$ of squared integrable 4-vector-functions by setting 
$U(t)\psi_0(x)=\psi(t,x)$ one finds that $U(t)$ is unitary. We call $U(t)$ the propagator of Diracs equation.

Coming to Quantum Mechanics, we first notice\footnotemark that one may introduce physical units for length, 
time, energy and electrical charge making  $\hbar=c=m_e=|e|=1$, denoting charge and mass of the electron
by $e$ and $m_e$. That will give the Dirac operator the form (2.1).

\footnotetext{The physical constants
usually found in the Dirac equation have been absorbed by choosing
proper units: The unit of length is the \emph{Compton wave length} of
the electron $\hbar/mc \approx 3.861 \x 10^{-13} m$. The unit of time is
$\hbar/mc^2 \approx 1.287 \x 10^{-21} sec $. The unit of energy is
$mc^2 \approx 0.5 MeV$. This will make $c=m=\hbar=|e|=1$. Furthermore, we must
choose units of electromagnetic field strength to absorb the factor $e$
- rather $|e|$ - the elementary charge (while $e$ (of course) counts as a
negative charge). Note that, with these units,
we get $\mcl{E}=-$grad $\mbf{V}-\mbf{A}_{|t}\ ,\ \mcl{B}=$curl $\mbf{A}$
as electrostatic and magnetic
field strength, resp. Also, for the Coulomb potential we get
$\mbf{V}(x)=-\frac{c_f}{|x|}$ with the fine structure constant
$c_f\approx \inv{137}$.}

A `state' (of the electron-positron system) then is described by a unit-vector in $\mclh$ --- a 4-vector-function
$\psi_0(x)$ with $\|\psi\|^2= \int |\psi_0(x)|^2dx=1$ . The observable quantities --- called `observables' ---
are given by (unbounded) self-adjoint operators (acting on a subspace of $\mclh$).
The theory predicts the statistical expectation value
$$
\breve{A}_{\psi_0}=\<\psi_0,A\psi_0\>
\lo{(2.12)}
$$
for the observable $A$ in the state $\psi_0$, where $\<.,.\>$  denotes the inner product in the Hilbert space $\mclh$.

One may predict such expectation-value  of the observable $A$ for a future time, starting with the state $\psi_0$ at time
$t=0$,  by using the state $\psi_t(x)=\psi(t,x)$, with above solution $\psi(t,x)$ of the Dirac equation, starting with $\psi_0$
at $t=0$. Or else, we get
$$
\breve{A}_{\psi_t}=\<\psi_t,A\psi_t\>=\<U(t)\psi_0,AU(t)\psi_0\>=\<\psi_0,A_t\psi_0\>=\breve{A_t}_{\psi_0}\ ,\ 
\lo{(2.13)}
$$
 with above `propagator' of Dirac's equation, setting $A_t=U^*(t)AU(t)$.

So, for future predictions of $A$ in the state $\psi_0$ at $t=0$ , we either must obtain the solution $\psi_t=\psi(t,x)$ of
Dirac's equations, or else, the observable $A_t=U^*(t)AU(t)$. Traditionally, getting $\psi_t$ is called the `Schr\"{o}dinger
representation', and,  getting $A_t$ the 'Heisenberg representation'.

\bigskip

While a general unbounded self-adjoint operator of $\mclh$ will qualify as observable, we should emphasize the two 
observables $x$ and $D$ (with components $x_j,D_j\ ,\ j=1,2,3,$) known as \emph{location} and \emph{momentum}.
In classical theory knowledge of location and momentum will completely determine the state of the point-system
we consider here. In Quantum Mechanics, we find that the --- so-called --- \emph{dynamical} observables all are
built from combinations of $D$ and (functions of) $x$ : they are differential operators.

The Fourier transform $F$, defined as
$$
F\psi(\xi)=\psi^\wdg(\xi)=\kpd\int dx e^{-ix\xi}\psi(x)\ ,\ 
\lo{(2.14)}
$$
will define a unitary operator of $\mclh$ with the property that 
$$
FDF^*=\mbox{multiplication by }x \ ,\ FxF^*=-D\ .
\lo{(2.15)}
$$

We observe that our quantum theory might just as well be performed by using the Fourier transformed states
$\psi^\wdg$ and observables $A^\wdg=FAF^*$ instead of $\psi$ and $A$. We then  might speak of the momentum
representation, since then the momentum observables $D$ will be `diagonal' (i.e., will be multiplication operators).

For a differential operator observable $A$ the operator $A_t=U^*(t)AU(t)$ in general will not be a differential operator.
But we find it a rewarding problem to look at observables with the property that $A_t$ is a 
\emph{pseudodifferential operator}.

\section{Some Global Pseudodifferential Operator Algebras on $\mbb{R}^3$}

We will discuss here the \emph{calculus of $\psi do$-s} of 3 special algebras of 
pseudodifferential operators (abbrev. $\psi do$-s).

Note, the location observables (of multiplication by) $x_j$ and momentum observables
$D_l$ generate an algebra of differential operators (containing all linear combinations
of finite products of these operators). Clearly $D_j$ and $x_j$ do not commute --- we
get $[D_j,x_j]=\inv{i}$. These differential operators may be written in the form
$$
L=\sum a_\th(x)D^\th\ ,\ \mbox{  or also as },\  L=\sum D^\th \tilde{a}_\th(x)\ ,\ 
\lo{(3.1)}
$$
using multi-index notation, where $a_\th(x)$ and $\tilde{a}_\th(x)$ usually are different
functions.

Calculations among differential operators then are governed by the so-called
\emph{Leibniz formulas}. 

Generally we decide to use the first form of (3.1) when
writing a differential operator, keeping
 multiplications to the left of differentiations. For a polynomial
$a(x,\xi)=\sum_\th a_\th(x)\xi^\th$ in $\xi$ we write 
$$
a(x,D)= \sum_\th a_\th(x)D^\th\ ,\ 
\lo{(3.2)}
$$
then calling $a(x,\xi)$ the \emph{symbol} of the differential operator $a(x,D)$.

\begin{lemma} (Leibniz formulas)
Let $A=a(x,D)\ ,\ B=b(x,D)$
then
$AB=C=c(x,D)\ ,\ A^*=\breve{a}(x,D)$
with symbols given by the formulas
$$
c(x,\xi)=\sum_{j=0}^\i\sum_{|\th|=j}
\frac{(-i)^{|\th|}}{\th!} \p_\xi^\th a(x,\xi)\p_x^\th b(x,\xi)\ ,\ 
\breve{a}(x,\xi)=\sum_{j=0}^\i\sum_{|\th|=j}
\frac{(-i)^{|\th|}}{\th!} \p_\xi^\th\p_x^\th a^*(x,\xi)\ .
\lo{(3.3)}
$$

\end{lemma}

The sums in (3.3) are finite, since the derivatives $\p_\xi^\th$
of a polynomial in $\xi$ vanish as soon as $|\th|$ is larger than
its order. The formulas are easily verified for
$a(x,\xi),b(x,\xi)$ polynomials of order 0 or 1. Then an induction
proof can be given.

With the Leibniz formulas we then can control sums, products and adjoints of differential
operators.

It was the merit of H\"{o}rmander [Hoe2] to design a technique for extending this calculus
of differential operators to a larger class of \emph{symbols}, no longer being polynomials
in $\xi$ , then getting a class of pseudodifferential operators, and providing a meaning
to the Leibniz formulas. We are using this technique here, in a slightly different form,
for construction of some (global) algebras of $\psi do$-s.

First of all we use the Fourier transform (2.13) and (2.14) to write  the action of (3.2) as
$$
a(x,D)u(x)=\inv{(2\pi)^3}\int d\xi\int dy e^{i\xi(x-y)}a(x,\xi)u(y)\ .\ 
\lo{(3.4)}
$$
Clearly we also may write this as
$$
a(x,D)u(x)=\kpd\int d\xi e^{ix\xi}a(x,\xi)u^\wdg(\xi)\ .\ 
\lo{(3.5)}
$$
Both these formulas are easily verified for smooth compactly supported $u(x)$, assuming
$a(x,\xi)$ as a polynomial in $\xi$. But, note, they may be meaningful also
for functions $a(x,\xi)$ which are not polynomials in $\xi$.

Coming to pseudodifferential operators, we then must specify some classes of symbols $a(x,\xi)$
with formulas (3.4)-(3.5) being meaningful, and also find a new meaning of the Leibniz formulas.

\begin{defn}  A smooth function $f(x)$ will be called `of polynomial growth' --- with order $m$ ---
if we have $|f^{(\th)}(x)|=|\p_x^\th f(x)|\leq c_\th(1+|x|)^{m-|\th|}$ as $x\in\mbb{R}^3$, for all multi-indices
$\th$,  with constants $c_\th$  depending on $\th$, but not on $x$.

\end{defn}

Here the order $m$ is allowed to be any real --- positive or negative. For negative $m$ one might rather speak
of a decay, instead of growth.  We also allow order $-\i$, then assuming that $f(x)$ allows \emph{all orders}.
The class of functions of order $-\i$ is usually denoted by $\mcl{S}$. It will serve as source for our functions $u(x)$
in formulas (3.4),(3.5), then guaranteeing existence of all integrals.

We will use 3 spaces of symbols $a(x,\xi)$, in the following, called $\psi c\ ,\ \psi q\ ,\ \psi p$. In essence,
the class $\psi c$ will consist of all $a(x,\xi)$,  defined and smooth for all $x,\xi\in\mbb{R}^3$ which are 
of polynomial growth  --- independently --- in the variables $x$ (with order $m_2$) and $\xi$ (with order $m_1$).
There are two orders then combined into a (double-)order $m=(m_1,m_2)$.

On the other hand, the (larger)  class $\psi q$ will contain all $a(x,\xi)$ such that all $x$-derivatives $\p_x^\io a(x,\xi)$
are of polynomial growth --- order $m$ --- in the variables $\xi$  with constants $c_\th$ of def. 3.2 independent of $x$,
for some real $m$ independent of $\io,\th$.

Finally, the class $\psi p$ consists of all $a(x,\xi)$ in $\psi q$ which are independent of $x_2,x_3$ and periodic (with
period $2\pi/\o$) in $x_1$, with a given fixed (circular) frequency $\o=2\pi\nu$.  

To be precise, let us restate this as follows.

\begin{defn}

(i) The class $\psi c$ of  symbols (we call `strictly classical') consists of all functions $a(x,\xi)$
defined and smooth for all 6 variables $x,\xi$ and such that
$$
|\p_x^\th\p_\xi^\io a(x,\xi)|\leq
c_{\th\io}(1+|\xi|)^{m_1-|\io|}(1+|x|)^{m_2-|\th|}
\lo{(3.6)}
$$
for all multi-indices $\th,\io$, and all $x,\xi\in\mbb{R}^3$ with constants
$c_{\th,\io}$ independent of $x,\xi$.

The class of all such functions $a(x,\xi)$, for a given \emph{order}
$m=(m_1,m_2)$ will be denoted by $\psi c_m$.
We also define
$\psi c=\psi c_\i=\cup_{m}\psi c_m\ ,\ \psi c_{-\i}=\cap_{m}\psi c_m\ .\ $

\medskip

(ii) The class $\psi q$ consists of all smooth functions $a(x,\xi)$,
defined for $(x,\xi)\in \mbb{R}^6$ such that
$$
|\p_{x}^\io\p_\xi^\th a(x,\xi)|\leq c_{\th,\io}(1+|\xi|)^{m-|\th|}
\mbox{\ \ \ for some } m\in\mbb{R}
\mbox{ and all  } \io , \th\ ,\ x,\xi\ .
\lo{(3.7)}
$$
We again use $\psi q_m$ for the class of symbols of order $m$, and define 
$\psi q=\psi q_\i=\cup_{m}\psi q_m\ ,\ \psi q_{-\i}=\cap_{m}\psi q_m\ .\ $

\medskip

(iii) The class $\psi p_m$ consists of all $a(x_1,\xi)\in \psi q_m$ , independent of
$x_2,x_3$ and $2\pi/\o$-periodic in $x_1$, where $\o>0$ is some given fixed positive number;
we again set $\psi p=\psi p_\i=\cup_{m}\psi p_m\ ,\ \psi p_{-\i}=\cap_{m}\psi p_m\ .\ $

\end{defn}

We refer to [Co5], ch.1 for a proof of the fact that the integrals  at right of (3.4)-(3.5) exist, in the
order stated, whenever $u\in\mcls$ and $a(x,\xi)\in \psi q$, defining a continuous operator
$A=a(x,D)$ on the space $\mcls$ --- and then also on the space $\mcls'$ of temperate distributions.
The classes of such operators then will be called $Op\psi c\ ,\ Op\psi p\ ,\ Op\psi q$ , etc.

\bigskip

We again must refer to ch.1 of [Co5] to see that there are \emph{Leibniz formulas with integral remainder}
valid, in the sense that, for product and adjoint among operators $a(x,D),b(x,D)\in Op \psi q$, the 
differences $c(x,\xi)-\sum_0^N\cdots\ ,\ \breve{a}-\sum_0^N\cdots$ in (3.3) may be expressed as
certain integrals, involving very singular integrals (called `finite parts') of derivatives of the symbols
involved. Using these we then get the following result\footnotemark .

\begin{thm}
$Op\psi c=\cup Op\psi c_m$ and $Op\psi q =\cup Op\psi q_m$ are adjoint invariant graded algebras.
The Leibniz formulas (3.3) for product and adjoint hold in the sense of asymptotic convergence
(mod $Op\psi c_{-\i}$) and (mod $Op\psi q_{-\i}$), resp., of the infinite series $\sum_{j=0}^\i$
occurring . The classes $Op\psi c_{-\i}$ and $Op\psi q_{-\i}$ are
two-sided $*$-ideals of $Op\psi c$ and $Op\psi q$, respectively.

\end{thm}

\footnotetext{The assumptions made in ch.1 of [Co5] match ours here for $Op\psi c$, but not for $Op\psi q$.
However, we checked in detail, that the arguments used there may be literally extended to the case
of $Op\psi q$, as shall be lined out explicitly in [Co17].}

In thm. 3.4 we used the following concepts.

\begin{defn} 

(i) A sequence $\{a_n(x,\xi)\in\psi c\}$ is said to converge asymptotically (mod $\psi c_{-\i}$) to
$a(x,\xi)$ if  the order $m=(m_1,m_2)$ of the difference $a(x,\xi)-a_n(x,\xi)$ tends to $(-\i,-\i)$ 
as $n\rta \i$ . Then also we
shall say that $A_n=a_n(x,D)$  tends to $A=a(x,D)$ asymptotically (mod $Op\psi c$).

(ii) A sequence $\{a_n(x,\xi)\in\psi_q\}$ is said to converge asymptotically (mod $\psi q_{-\i}$) to
$a(x,\xi)$ if  the order of the difference $a(x,\xi)-a_n(x,\xi)$ tends to $-\i$ as $n\rta \i$ . Then also we
shall say that $A_n=a_n(x,D)$  tends to $A=a(x,D)$ asymptotically (mod $Op\psi q$).

\end{defn}

The essence of the proof of thm.3.4 then will be that the `integral remainders' representing the 
differences $c(x,\xi)-\sum_0^N\cdots\ ,\ \breve{a}-\sum_0^N\cdots$ in (3.3), must be shown to
be symbols of orders tending to $-\i$, as $N\rta\i$.

We also need

\begin{prop} Let $r=c$ or $r=q$.
For any   sequence of symbols $\{a_j(x,\xi):j=0,1,2,\cdots\}$ with $a_j\in \psi r_m^j$ 
where $m^j\rta -\i$ resp. $m^j_l\rta -\i\ ,\ l=1,2,$
there exists a symbol  $a(x,\xi)\in \psi r_m^0$ such that 
$$
a(x,\xi)=\sum_{j=0}^\i a_j(x,\xi) (\mbox{ (mod }\psi r_{-\i}))
\lo{(3.8)}
$$
\end{prop}

A proof (a la Hoermander) may be found in [Co5],(ch.1,lemma 6.4, p.75. (Or else, cf.[Co16] footnote 18 on p.18) 
(for $r=c$ only, but it may be adapted for $r=q$).

\begin{prop} 

(i) The class $Op\psi c_{-\i}$ consists of all integral operators $Ku(x)=\int_{\mbb{R}^3}k(x,y)u(y)dy$
with kernel $k(x,y)$ in $\mcls(\mbb{R}^6)$.

(ii) The class $Op\psi q_{-\i}$ consists of all $\psi do$-s $C=c(x,D)$ with symbol $c$ having all $x$-derivatives
belonging to $\mcls$ in the $\xi$-variable, uniformly  for all $x\in\mbb{R}^3$.

\end{prop}

For the proof of (i) cf. [Co5], ch.3, prop.3.4 on p.111. (ii) is just a reformulation of the definition of $\psi q_{-\i}$.

\bigskip

Finally, among results about $\psi do$-s, we also need to look at a representation of $\psi do$-s involving
both representations (3.1) --- i.e., allowing multiplications left and right from differentiations. This means
generalizing (3.4) by writing
$$
a(M_l,M_r,D)u(x)=\inv{(2\pi)^3}\int d\xi\int dy e^{i\xi(x-y)}a(x,y,\xi)u(y)\ ,\ 
\lo{(3.9)}
$$
where the symbol $a(x,y,\xi)$  now depends on 9 variables $x,y,\xi\in\mbb{R}^3$, and satisfies the estimates
$$
|\p_{x}^\io\p_y^\l\p_\xi^\th a(x,y,\xi)|\leq c_{\th,\l,\io}(1+|\xi|)^{m-|\th|}
\mbox{\ \ \ for some } m\in\mbb{R}\ ,\ \mbox{  all  }x,y,\xi\ ,\
\mbox{  all  }\io,\l,\th\ .
\lo{(3.10)}
$$
The class of all smooth $a(x,y,\xi)$ defined over $\mbb{R}^9$ satisfying (3.10) will be denoted by $\psi qlr_m$, with
$\psi qlr=\cup_m\psi qlr_m$. The notation $a(M_l,M_r,D)$ seeks to remind of the fact that we have
$a(M_l,M_r,D)=p(x)r(D)q(x)$ for $a(x,y,\xi)=p(x)q(y)r(\xi)$.

Such operators $a(M_l,M_r,D)$ belong to $Op \psi q_m$, if the symbol $a(x,y,\xi)$ satisfies (3.10), and there exists
a Leibniz formula (asymptotic (mod  $Op\psi q_{-\i}$)) defining a symbol $b\in\psi q_m$  such that 
$a(M_l,M_r,D)=b(x,D)$. Again, this is a matter of slightly adapting things around f'la. (5.5) on p.70 of [Co5]. 

We shall have to deal intensively with operators of this kind in sections 11 and following. It then even will be necessary to 
discuss some facts regarding Fourier integral operators with symbol and phase functions in $\psi qlr$.
For more detail we refer to sec.12.

\section{Time-Independent Potentials vanishing at $\i$}

We return to the Dirac equation and will assume here that the potentials $\mbf{A}_j,\mbf{V}$ 
of $H$ in (2.1) do not depend on $t$, and will
have the limit zero, as $|x|\rta\i$. Moreover, we shall assume that 
$ \mbf{V}(x) $ and $\mbf{A}_j(x)\ ,\ j=1,2,3$ are of polynomial growth, order $-1$. We then get $H\in Op\psi c_{(1,0)}$, and
$$
h(x,\xi)=\sum_{j-1}^3 \a_j(\xi_j-\mbf{A}_j(x))+\b+\mbf{V}(x)\ .\ 
\lo{(4.1)}
$$

The propagator $U(t)$ then may be written as $U(t)=e^{-iHt}$; it commutes with $H$ for every $t$. However, it does not
belong to $Op\psi c$.  In [Co3],[Co16] (and in
numerous other articles) we then asked the question for observables $A$ with the property that
the Heisenberg transform $A_t=e^{iHt}Ae^{-iHt}$ belongs to $Op\psi c$, for all $t$. In essence this implies that 
$A=a(x,D)$ has its symbol $a(x,\xi)$ commuting with the symbol $h(x,\xi)$ of $H$, for very large 
$|x|+|\xi|$. Recall, the matrix $h(x,\xi)$ has the two eigenvalues $\l_\pm(x,\xi)=\mbf{V}(x)\pm\<\xi-\mbf{A}(x)\>$,
of multiplicity 2 each, and their spectral projections $p_\pm(x,\xi)$ of (2.4) separate the states belonging to
electron and positron, respectively. The fact that $a(x,\xi)$ must commute with $h(x,\xi)$ implies that $a(x,\xi)$
takes the spaces of electron and positron  states into themselves --- in some weakened sense. Clearly, this
should be a desirable property, in view of the various contradictions or paradoxes in older literature, stemming from
violation of this property.

In earlier publications we were using the name \emph{precisely predictable} for observables $A$ with $A_t\in Op\psi c$,
and we proposed that the rule (2.12) of predicting the statistical expectation-value should be applicable only to
precisely predictable observables. While total energy and (often also) total angular momentum trivially are precisely
predictable, other observables -- like $x_j$ and $D_l$ do not have this property, but they are \emph{approximately
predictable} --- with a preset error --- in the sense that there are precisely predictable observables in their close
neighbourhood.

Here we will attempt to describe the essentials of the theory, omitting a discussion of a large amount of technical
proofs, already discussed in close detail in [Co16].

Suppose $A_t=e^{iHt}Ae^{-iHt}$ belongs to $Op\psi c_m$, for some fixed $m=(m_1,m_2)$, and all $t$. So, we have
$A_t=e^{iHt}Ae^{-iHt}=a_t(x,D)$. Assume also that the time-derivative $\dot{a}_t(x,\xi)$ exists and belongs to 
$\psi c_{m-e^2}$ where $e^2=(0,1)$. Differentiating for $t$ we get 
$$
\dot{A}_t=iHe^{iHt}Ae^{-iHt}-ie^{iHt}Ae^{-iHt}H = i[H,A_t]\ .
\lo{(4.2)}
$$
Since $H$ and $A_t$ are $\psi do$-s , by assumption, we may use the Leibniz formula of lemma 3.1 to obtain a symbol
for the commutator $[H,A_t]=HA_t-A_tH$. We get
$$
\mbox{ symbol }([H,A_t])= [h,a_t]-i\{h,a_t\}-\inv{2!}\{h,a_t\}_2+
\frac{i}{3!}\{h,a_t\}_3+\cdots\ ,\
\lo{(4.3)}
$$
where we use the (generalized) Poisson-brackets
$$
\{h,a_t\}=\{h,a_t\}_1=h_{|\xi}a_{t|x}-a_{t|\xi}h_{|x}\ ,\
\{h,a_t\}_2=h_{|\xi\xi} a_{t|xx}-a_{t|\xi\xi}h_{|xx}\ ,\  \mbox{ etc. }
\lo{(4.4)}
$$
In (4.3) the terms at right have orders $m+e^1,m+e^1-e,m+e^1-2e,\cdots$, with $e^1=(1,0)$,
so, the asymptotic sum mod $\psi c_{-\i}$ exists, by prop.3.6. With (4.3) we may express (4.2) symbol-wise in the form
$$
\dot{a}_t=i[h,a_t]+\{h,a_t\}-\frac{i}{2!}\{h,a_t\}_2-
\inv{3!}\{h,a_t\}_3\pm\cdots\ .\
\lo{(4.5)}
$$

\begin{prop}
If we have $A_t=e^{iHt}Ae^{-iHt}=a_t(x,D)$, where $a_t(x,\xi)\in\psi c_m\ ,\ \dot{a}_t(x,\xi)\in\psi c_{m-e^2}\ ,\ $
then the commutator $[h(x,\xi),a_t(x,\xi)]$ --- naturally being of order $m+e^1$, since $h\in\psi c_{e^1}$ --- 
must have the  (lower) order $m-e^2$.

\end{prop}

Indeed, all terms in (4.5) , except the term involving $[h,a_t]$ , have order $m-e^2$ (or lower), hence $[h,a_t]$ also
must be of order $m-e^2$.

So, indeed, we get $[(h(x,\xi)/\<\xi\>),(a_t(x,\xi)/(\<x\>^{m_1}\<\xi\>^{m_2})]=O((\<x\>\<\xi\>)^{-1})$\ ,\ 
i.e., this commutator vanishes as $|x|+|\xi|\rta\i$.

\bigskip

Vice versa, (4.5) suggests, that we might attempt construction of a precisely predictable $A=a(x,D)\in Op\psi c$ by 
starting with a (self-adjoint) $q(x,\xi)\in\psi c_m$ with the property that $[h(x,\xi),q(x,\xi)]=0$ 
for all $x,\xi$, and then trying to find a  $z(x,\xi)\in\psi c_{m-e}$ such that  $a=q+z$  satisfies (4.5).
Noting that the terms at right of (4.5) are of order $(m+e^1)\ ,\ (m+e_1)-e\ ,\ (m+e^1)-2e\ ,\ (m+e^1)-3e\cdots$
with $e=(1,1)$ , we might neglect all terms at right of (4.5) but the first two, then getting an equation valid modulo
$\psi c_{m-e^2-e}$ only:
$$
\dot{a}_t=i[h,a_t]+\{h,a_t\} \mbox{ (mod } \psi c_{m-e^2-e})\ .\ 
\lo{(4.6)}
$$
 Let us assume that we also have $a_t(x,\xi)=q_t(x,\xi)+z_t(x,\xi)$ with 
$[h(x,\xi),q_t(x,\xi)]=0\ \forall x,\xi$, where
$q_t(x,\xi)\in \psi c_m\ ,\  z_t\in\psi c_{m-e}\ ,\
\dot{q_t}\in\psi c_{m-e^2}\ ,\ $
$\dot{z}_t\in\psi c_{m-e^2-e}\ .$ Then we may omit further terms, vanishing or being of order $m-e^2-e$:
$$
\dot{q}_t=i[h,z_t]+\{h,q_t\} \mbox{ (mod } \psi c_{m-e^2-e})\ .\ 
\lo{(4.6)}
$$
We start an iteration by assuming (4.6) as a sharp equation --- not only modulo $\psi c_{m-e^2-e}$. Assuming $q_t$ known we
obtain an equation for $z_t$:
$$
[h,z_t]=i(\{h,q_t\}-\dot{q_t})\ .\ 
\lo{(4.7)}
$$
Attempting to solve this matrix-commutator equation for $z$ we observe the following:

\begin{prop}
Equation (4.7) has no solution, unless the right hand side $Z_t=i(\{h,q_t\}-\dot{q_t})$ satisfies
$$
p_+(\{h,q_t\}-\dot{q_t})p_+=0\ ,\ p_-((\{h,q_t\}-\dot{q_t})p_-=0\ ,\ \mbox{ for all } x,\xi\ .\
\lo{(4.8)}
$$
If (4.8) holds, then an infinity of solutions is given by
$$
z_t=\inv{\l_+-\l_-}(p_+Z_tp_--p_-Z_tp_+) +c_t=\halb\inv{\<\xi-\mbf{A}(x)\>}(p_+Z_tp_--p_-Z_tp_+) +c_t\ ,\ 
\lo{(4.9)}
$$
with the eigenvalues $\l_\pm$ of $h(x,\xi)$, where $c_t(x,\xi)$ may be any symbol commuting with $h(x,\xi)$ --- i.e.,
we must have $c_t=p_+c_tp_+ +p_-c_tp_+$.

\end{prop}

The proposition is easily verified, using facts on spectral projections:\ \  $p_++p_-=1\ ,\ p_+^2=p_+\ ,\ p_-^2=p_-\ ,\ 
p_+p_-=p_-p_+=0\ ,\  h=\l_+p_++\l_-p_-$\ .

The interesting fact now is that --- while we know $q_t$ only for $t=0$ (where we should have $q_0=q$), the solvability
conditions (4.8) will resolve into a set of partial differential equations determining $q_t$ for all $t$,  from its 
initial-value $q_0$, so that we then indeed may use (4.9) to obtain the desired $z_t$ (including $z=z_0$).
Moreover, this set of differential equations relates to the classical equations determining the propagation of the particle,
as we shall see.

Of course, this will only supply a solution to equation (4.7), not the real thing (4.5). However, then, we shall set up an
iteration, getting us a solution of (4.5) modulo $\psi c_{-\i}$, using prop.3.6. In combination with prop.3.7 this indeed will be
enough to construct a precisely predictable observable $a(x,D)=q(x,D)+z(x,D)$  in $Op\psi c_m$, with lower order $z$, 
starting from an arbitrarily given  symbol $q\in\psi c_m$, commuting with $h$.

There is a mountain of technicalities in our way, all discussed in detail in [Co16]. Here we shall focus on the above first step,
solving eq. (4.7). 

Let us try to evaluate the conditions (4.8).
The assumption $[h,q]=0$ implies that   $q=q^++q^-$, where $q^+=p_+qp_+\ ,\ q^-=p_-qp_-$. We first 
work with a simplifying assumption that $q^+$
and $q^-$ are scalar multiples of $p_+$ and $p_-$, resp., a condition trivially satisfied by symbols being scalar multiples of
the $4\x 4$-unit matrix. In that case we shall be successful if we assume the same for $q_t^+=p_+q_tp_+\ ,\ 
q_t^-=p_-q_tp_-$ . So, we first look at the special case where
$$
q_t=q_t^+p_++q_t^-p_- \mbox{    with scalar (complex-valued) symbols   } q^+,q^-\ .\ 
\lo{(4.10)}
$$

\begin{prop}

With above assumptions on $q_t$ we get
$$
p_+\{h,q_t\}p_+=\{\l_+,q^+_t\}p_+\ \ ,\ p_-\{h,q_t\}p_-=\{\l_-,q^-_t\}p_-\ .
\lo{(4.11)}
$$

\end{prop}

The proof is a calculation (cf. [Co16], p.93). Applying this to (4.8), using (4.10),
these equations assume the form
$$
\dot{q}_t^+=\{\l_+,q_t^+\}\ ,\ \dot{q}_t^-=\{\l_-.q_t^-\}\ ,\ 
\lo{(4.12)}
$$
with the eigenvalues $\l_\pm(x,\xi)=\mbf{V}(x)\pm\<\xi-\mbf{A}(x)\>$ of $h(x,\xi)$, noted in sec.1.

Two things are interesting here: First of all,  the two equations (4.8) have split into separate equations
for $q_t^+$ and $q_t^-$ --- the first involves only $q_t^+$, the second only $q_t^-$.
Secondly, both these equations now are first order partial differential equations for a scalar 
dependent variable: 
$$
\dot{q}_t^+=\l_{+|\xi} q_{t|x}^+ - \l_{+|x} q_{t|\xi}^+\ ,\ 
\dot{q}_t^+=\l_{-|\xi} q_{t|x}^- - \l_{-|x} q_{t|\xi}^- \ .\     
\lo{(4.13)}
$$

Solving the initial-value problem for equations (4.13) is a simple matter, just involving ordinary differential
equations:
For the first equation (4.13) look at the first order system of 6 ODE-s
$$
\dot{x}=\l_{+|\xi}\ ,\ \dot{\xi}=-\l_{+|x}\ \ ,\ \ \ \ \ \l_+=\mbf{V}(x)+\<\xi-\mbf{A}(x)\>\ ,\ 
\lo{(4.14+)}
$$
in the 6 unknown functions  $x(t),\xi(t)$, of the single variable $t$.
Given any initial real 6-vector $(x^0,\xi^0)$ there is a unique curve $x(t),\xi(t)$ in $\mbb{R}^6$
solving (4.14+), passing through $(x^0,\xi^0)$ at $t=0$. In fact, the entire `phase space' $\mbb{R}^6$
is filled with such `orbits' with no two of them intersecting.

We then may look at the first (4.13) along such a curve $x(t),\xi(t)$. Substituting (4.14+) we get
$$
\p_t q_t^+(x(t),\xi(t))=\p_xq_t^+(x(t),\xi(t))\dot{x}(t)+\p_\xi q_t^+(x(t),\xi(t)) \dot{\xi}(t)\ ,
\lo{(4.15)}
$$
amounting to
$\frac{d}{dt} q_{-t}^+(x(t),\xi(t))=0\ .\ $ Or else, $q_{-t}^+(x(t),\xi(t))$ must be a constant --- independent
of $t$ --- along any such curve.

Here we consider the \emph{flow} defined by the system (4.14+): For any fixed $t$ introduce the diffeomorphism
$\nu_t^+:\mbb{R}^6\rta\mbb{R}^6$ defined by letting $(x,\xi)$ move along the solution curve of (4.14+) through it
for a distance $t$ (positive or negative, according to the sign of $t$). 
Then consider the expression $q_\t^+(\nu_{t-\t}(x,\xi))=q_\t^+(x_{t-\t}(x,\xi),\xi_{t-\t}(x,\xi))$, as a function of $\t$.

This function is constant --- independent of $\t$,  as a consequence of the above. Thus , setting $\t=t$ and $\t=0$, 
and using that $\nu_0(x,\xi)=(x,\xi)$, we get
$$
q_t(x,\xi)=q_t(\nu_0(x,\xi))=q_0(\nu_t(x,\xi))=q(\nu_t(x,\xi))\ .
\lo{(4.16)}
$$
So --- since $q_0=q$ is given, we indeed have obtained a well defined $q_t(x,\xi)=q(\nu_t^+(x,\xi))$ 
as the only possible candidate for solving (4.7).

\begin{obs}

It should be noted here that the differential equations are the classical 
equations of motion\footnotemark for a (spinless) electron
moving in the electromagnetic field defined by $\mbf{V}$ and $\mbf{A}_j$.

A similar discussion --- of course --- will hold for the second condition (4.8), resulting 
in another (Hamiltonian) system 
$$
\dot{x}=\l_{-|\xi}\ ,\ \dot{\xi}=-\l_{-|x}\ ,\ \ \ \ \ \l_-=\mbf{V}(x)-\<\xi-\mbf{A}(x)\>\ ,\    
\lo{(4.14-)}
$$
a corresponding flow $\nu_t^-(x,\xi)$ and a $q_t^-(x,\xi)=q(\nu_t^-(x,\xi))$. In each case we also get
a $z_t^+\ ,\ z_t^-$ from (4.9), and a $q_t+z_t$ solving (4.6), thus completing the first step of our
iteration. The flow $\nu_t^-$ will describe the classical motion of a spinless positron.

\end{obs}

A discussion of the elements of the proof of thm.4.5, below,
is given in [Co16], chapters 4 and 5. We also might refer to
[Co3] and [Co5] where the same facts are discussed.

The more general case, where $q^+\ ,\ q^-$ are not necessarily multiples of the identity, is more complicated --- 
and, perhaps, more interesting, since magnetic spin-problems will appear. It will be discussed in the next following section.

\footnotetext{Explicitly, the system (4.14+)  for
$\l_+=\<\xi-\mbf{A}\>+\mbf{V}$ looks like this:
$$
\dot{x}=\inv{\<\xi-\mbf{A}(x)\>}(\xi-\mbf{A}(x))\ ,\
\dot{\xi}=\inv{\<\xi-\mbf{A}(x)\>}\sum_j(\xi_j-\mbf{A}_j(x))\mbf{A}_{j|x}(x)
-\mbf{V}_{|x}(x)\ .
\lo{(4.17)}
$$
The first equation may be solved for $\xi-\mbf{A}$: We get
$$
\xi-\mbf{A}=\frac{\dot{x}}{\sqrt{1-\dot{x}^2}}\ ,\
\<\xi-\mbf{A}\>=\inv{\sqrt{1-\dot{x}^2}}
\lo{(4.18)}
$$
Equating the derivative $\dot{\xi}$ of (4.18) with the second (4.17) gives
$$
(\frac{\dot{x}}{\sqrt{1-\dot{x}^2}})^{^\cdot}+\p_t\mbf{A}(x(t))
=-\mbf{V}_{|x}+\sum_j\dot{x}_j\mbf{A}_{j|x}\ .
\lo{(4.19)}
$$
In (4.19) we get $\p_t\mbf{A}(x(t))=\sum_l \dot{x}_l(t)\mbf{A}_{|x_l}(x(t))$.
Now we use the relation
$$
\dot{x}\x\mbox{curl}\ \mbf{A}=
\sum_l(\dot{x}_l\mbf{A}_{l|x}-\dot{x}_l\mbf{A}_{|x_l})\ .
\lo{(4.20)}
$$
As a consequence (4.19) assumes the form
$$
(\frac{\dot{x}}{\sqrt{1-\dot{x}^2}})^{^\cdot}=
-\mbf{V}_{|x}(x(t))+\dot{x}\x \mbox{curl}\ \mbf{A} (x(t)).
\lo{(4.21)}
$$
But electric and magnetic field $\mcl{E}$ and $\mcl{H}$ as functions of
$\mbf{A}$ and $\mbf{V}$ are given by the formulas
$$
\mcl{E}=-\mbf{A}_{|t}-\mbox{grad}\ \mbf{V}\ ,\
\mcl{B}=\mbox{curl}\ \mbf{A}\ \ ,
\lo{(4.22)}
$$
and the relativistic mass (of the particle with rest mass 1)
will be $\inv{\sqrt{1-\dot{x}^2}}$,
in the physical units we employ here. Accordingly (4.20) reads
$$
(\frac{\dot{x}}{\sqrt{1-\dot{x}^2}})^{^\cdot}=\mcl{E}+\dot{x}\x \mcl{B}\ .
\lo{(4.23)}
$$
Clearly this exactly describes the acceleration of the electron
under the force of the (time-dependent) electromagnetic field acting on it.}

\begin{thm}

Assume a given symbol $q\in\psi c_m$ with $[h(x,\xi),q(x,\xi)]=0\ \forall(x,\xi)$, and such that 
we have\footnotemark \newline $p_+qp_+=q^+(x,\xi)p_+ \ ,\ p_-qp_-=q^-(x,\xi)p_-$.
with (scalar) complex-valued symbols $q^+(x,\xi)\ ,\ q^-(x,\xi)$ . Then there exists a
symbol $a_t(x,\xi)=q_t(x,\xi)+z_t(x,\xi)\in \psi c_m$ \ ,\ for all $t$,  satisfying
$$
a_t(x,D)=e^{iHt}a_0(x,D)e^{-iHt}\ ,\ 
\lo{(4.24)}
$$
and such that $z_t\in\psi c_{m-e}$ while 
$$
q_t(x,\xi)=q^+(\nu^+_t(x,\xi))p_+(x,\xi)+q^-(\nu_t^-(x,\xi))p_-(x,\xi)\ ,\ 
\lo{(4.25)}
$$
with the two flows $\nu_t^+\ ,\ \nu_t^-$ generated by the classical motions of the spinless electron
and positron, respectively.

The symbol $z_t(x,\xi)$ may be chosen such that $a_t(x,\xi)$ is self-adjoint for all $x,\xi$. Then the 
operator $A=a(x,D)=q(x,D)+z_0(x,D)$ is a precisely predictable observable.

\end{thm}

\footnotetext{This condition simply means that $q(x,\xi)$ is a scalar multiple of the identity in 
each of the two eigenspaces $S_\pm=S_\pm(x,\xi)$ of the symbol $h(x,\xi)$.}

\section{A General Commuting $q(x,\xi)$ and a Magnetic  $3$-Vector $\kp_t(x,\xi)$}

In this section we shall discuss the more general case where $q$ is not scalar in the eigenspaces of $h$.
We still look  for a solution of the commutator equation (4.7), i.e.,
$$
[h,z_t]=i(\{h,q_t\}-\dot{q_t})\ ,\ 
\lo{(5.1)}
$$
keeping in mind prop. 4.2 with solvability condition (4.8). But we must replace prop.4.3 :

\begin{prop} We get
$$
p_+\{h,q_t\}p_+=p_+\{\l_+,q_t^+\}p_+ +2\<\z\>p_+\{p_+,q_t^+\}p_+\ , \ 
\lo{(5.1_+)}
$$
$$
p_- \{h,q_t\}p_-=p_-\{\l_-,q_t^-\}p_- -2\<\z\>p_-\{p_-,q_t^-\}p_-\ ,
\lo{(5.1-)}
$$
with $\z=\xi-\mbf{A}(x)\ ,\ q_t^+=p_+q_tp_+\ ,\ q_t^-=p_-q_tp_-$.

\end{prop}

\noindent
\tbf{Proof}
Clearly we have
$$
p_+\{h,q_t\}p_+=p_+\{\l_+p_+,q_t\}p_++p_+\{\l_-p_-,q_t\}p_+ =Z_1+Z_2\ ,\ 
\lo{(5.2)}
$$
where 
$Z_1=p_+\{\l_+,q_t\}p_++\l_+p_+\{p_+,q_t\}p_+$, while

$Z_2=\l_{-|\xi}p_+p_-q_{t|x}p_+-\l_{-|x}p_+q_{t|\xi}p_-p_+
+\l_-p_+\{p_-,q_t\}p_+
=\l_-p_+\{p_-,q_t\}p_+$, since $p_+p_-=p_-p_+=0$. Also, $p_++p_-=1$ implies
$p_{-|x}=-p_{+|x}$ and $p_{-|\xi}=-p_{+|\xi}$, hence,
$\{p_-,q_t\}=-\{p_+,q_t\}$, so that,
$Z_2=-\l_-p_+\{p_+,q_t\}p_+$. Together we get
$$
p_+\{h,q_t\}p_+=p_+\{\l_+,q_t\}p_+ +(\l_+-\l_-)p_+\{p_+,q_t\}p_+\ .
\lo{(5.3)}
$$
Simplifying (5.3) we first recall that $\l_+-\l_-=2\<\z\>=2\<\xi-\mbf{A}\>$.
Furthermore we get 
 $q_t=q_t^++q_t^-$, where $p_+\{\l_+,q_t^-\}p_+=0$,  since $\l_+$
is a scalar, so that $\{\l_+,q_t^-\}=\l_{|\xi}q^-_{t|x}-\l_{|x}q^-_{t|\xi}$,
and $p_+q^-_{t|x}p_+=p_+q^-_{t|\xi}p_+=0$, implied by
$p_+q_t^-=0\Rightarrow p_+q^-_{t|x}=-p_{+|x}q_t^-$, etc.
So, in the first term at right of (5.3) we may replace $q_t$ by $q_t^+$.

The same follows for the second term, so that (5.1+) follows:
Indeed, we get

\noindent
$p_+\{p_+,q_t^-\}p_+=p_+p_{+|\xi}q^-_{t|x}p_+-p_+q^-_{t|\xi}p_{+|x}p_+
=-p_+p_{+|\xi}q_t^-p_{+|x}p_++p_+p_{+|\xi}q_t^-p_{+|x}p_+$
$=0$,
where we again used that $p_+q_t^-=q_t^-p_+=0$
implies $p_+q^-_{t|\xi}=-p_{+|\xi}q_t^-\ ,\ q^-_{t|x}p_+=q_t^-p_{+|x}$. 
A similar argument yields (5.1-), q.e.d.

After prop.5.1 it is clear that we again have split the two
solvability conditions (4.8)
into separate systems for $q_t^{\pm}$: The first cdn. involves
only $q_t^+$, the second only $q_t^-$.
Using (5.1+) and differentiating along the
solution curves of the Hamiltonean system (4.14+)
for $\l_+$ we may rewrite
the first (4.8) as
$$
p_+ q_t^{+'} p_+ - 2\<\z\>p_+\{p_+,q_t^+\}p_+ = 0 \ ,
\lo{(5.4+)}
$$
where  ``$'$'' denotes the directional derivative $\p_t-\l_{\pm|\xi}\p_x + \l_{\pm|x}\p_\xi$,
used for $\l_+$. Similarly
$$
p_- q_t^{-'} p_- + 2\<\z\>p_-\{p_-,q_t^-\}p_- = 0 \ ,
\lo{(5.4-)}
$$
with ``$'$"  for $\l_-$.

\bigskip

In the case of a $q(x,\xi)$ scalar in the two eigenspaces $S_\pm(x,\xi)$, as discussed in thm.4.5, 
 we only needed the two eigenvalues $\l_\pm(x,\xi)$
to set up our first approximation. In the present more general case we will obtain explicit $2\x 2$- matrices of $q^\pm(x,\xi)$
with respect to a natural orthonormal base of $S_\pm(x,\xi)$ of the symbol $h(x,\xi)$.
Getting restricted to only use the Dirac matrices $\a_j,\b$ of (2.8), so that $\b$ is the diagonal matrix
defined there, we recall the orthogonal matrix  $\Upsilon$  of  (2.5), known to satisfy (2.6), i.e.,
$$
h(x,\xi)\Upsilon(x,\xi)=\Upsilon(x,\xi)(\mbf{V}(x)+\<\xi-\mbf{A}(x)\>\b)\ ,\ 
\Upsilon=\inv{\sqrt{2(1+\u_0)}}\mat{1+\u_0\ \ }{\ \ \ -i\s\u\ \  }{\ -i\s\u\ \ \ \ }{\ 1+\u_0\ }\ ,
\lo{(5.5)}
$$
with $\u_0=\<\z\>^{-1}\ ,\ \u=\z/\<\z\>\ ,\ \z=\xi-\mbf{A}(x)\ .$ We may rewrite this as
$$
h(x,\xi)\U_\pm(x,\xi)=\l_\pm(x,\xi)\U_\pm(x,\xi)\ ,\ \mbox{ with }
\U_+=\vc{1+\u_0}{-i\s\u}\ ,\ 
\U_-=\vc{-i\s\u}{1+\u_0}\ .
\lo{(5.6)}
$$
The columns of the $4\x 2$-matrices $\U_\pm(x,\xi)$  are 
eigenvectors to $\l_\pm(x,\xi)$, of length $\sqrt{2(1+\u_0)}$, and mutually orthogonal.

We then have $q_t^\pm(x,\xi)$ represented by the $2\x 2$-matrices 
$$
\kp_t^\pm=((\kp_{jl}^\pm))_{j,l=1,2}=\inv{2(1+\u_0)}\U_\pm^*q_t^\pm\U_\pm\ .
\lo{(5.7)}
$$
Writing $\U_\pm=(\vi^\pm_1,\vi^\pm_2)$ column-wise, we may introduce the $4\x 4$-matrices
$$
p^\pm_{jl}=\inv{2(1+\u_0)}\vi_j^\pm\>\<\vi_l^\pm\ \ \ \ \ ,\ \ \ \ \  
 \vi_j^\pm\>\<\vi_l^\pm=\vi_j^\pm\vi_l^{\pm*}\ ,\ 
\lo{(5.8)}
$$
and then get
$$
q_t^\pm=\sum_{j,l=1}^2\kp_{tjl}^\pm p_{jl}^\pm\ \ \ \ ,\ \ \ \ p_\pm=p_{11}^\pm+p_{22}^\pm\ .
\lo{(5.9)}
$$

We now use (5.9) to translate (5.4+) into a $2\x 2$-matrix form.
Using that $p^2=p$ for $"\pm"$ implies
$p_\pm p'_\pm p_\pm =0$ for any directional derivative ``$'$'',
(5.4+) and (5.9) yield
$$
\sum_{jl} p_{jl}\kp'_{tjl} +\sum_{jl}\kp_{tjl}(pp'_{jl}p)
-2\<\z\>\sum_{jl}\kp_{tjl}p\{p,p_{jl}\}p=0\ ,
\lo{(5.10)} 
$$
where we restricted to ``+'' and dropped the ``+'' in notation. Evidently,
the first term of (5.10) has the matrix $((\kp'_{tjl}))$. The matrices of
the other two terms may be written as $W_t^+\kp_t$ with a certain linear map
$W_t^+$ taking $2\x 2$-matrices to $2\x 2$-matrices. Thus (5.10) may be written as
$$
(\kp_t^+)'+W_t^+\kp_t^+=0\ 
\lo{(5.11)}
$$
with ``$'$" of (5.4+).
Using the hamiltonian system (4.14+) this again will turn into a system
of 4 ODE-s along the classical electron-particle flow
 for the 4 scalar functions $\kp^+_{-t,jl}(x(t),\xi(t))$.

\begin{prop}

Relation (5.11) may be rewritten as
$$
(\kp_t^+)'+[\Th^+,\kp_t^+]=0\ ,\ 
\lo{(5.12)}
$$
with the directional derivative ``$'$"  of (5.4+) and the $2\x 2$-matrix
commutator $[.,.]$, where the  
$2\x 2$-matrix $\Th^+$ is defined as
$$
\Th^+=\inv{2(1+\u_0)}(\U_+^*\U_+'-2\<\z\>\U^*p_{+|\xi}p_{+|x}\U)\ .
\lo{(5.13)}
$$

\end{prop}

\noindent
\tbf{Proof:}
Indeed, (dropping ``+'', and with ``$'$''= any directional
derivative) we have $pp_{jl}=p_{jl}$, hence $pp_{jl}'+p'p_{jl}=p_{jl}'\
\Rta\ (1-p)p_{jl}'=p'p_{jl}$, also, $pp'=p'(1-p)$ as already used.
Thus $pp_{|\xi}p_{jl|x}p=pp_{|\xi}(1-p)p_{jl|x}p=(pp_{|\xi}p_{|x})p_{jl}$,
and, similarly, $pp_{jl|\xi}p_{|x}p=p_{jl}(pp_{|\xi}p_{|x}p)$.
This will give
$$
\sum \kp_{tjl}p\{p,p_{jl}\}p=[pp_{|\xi}p_{|x}p,\sum\kp_{tjl}p_{jl}]
=[pp_{|\xi}p_{|x}p,q_t^+]\ ,\ 
\lo{(5.14)}
$$
showing that the last term in (5.10) has the desired commutator form 
giving the second term at right of (5.13) 

For the second term of (5.10) note that  
$p_{jl}'=\vi_j'\>\<\chi_l+\vi_j\>\<\chi_l'$, where we wrote $\chi_l=\inv{2(1+\u_0)}\vi_j$,
for a moment.  The $\vi_j$ and $\chi_l$ satisfy $\<\chi_l,\vi_q\>=\d_{lq}$, implying that 
$\<\chi_l',\vi_q\>=-\<\chi_l,\vi_q'\>$.
The coefficients of the $2\x 2$-matrix of $pp_{jl}'p$ then will be
$\<\chi_p,(\vi_j'\>\<\chi_l+\vi_j\>\<\chi_l')\vi_q\>=\<\chi_p,\vi_j'\>\d_{lq}-\d_{pj}\<\chi_l,\vi_q'\>$ . 
Accordingly, the second term of (5.10) will give 

$\sum_j\kp_{tjq}\<\chi_p,\vi_j'\>-\sum_l\kp_{tpl}\<\chi_l,\vi_q'\>\ ,\ $ giving the first term at right of (5.13).
Q.E.D.

Of course there is an analogous consideration for ``$-$'' which will be left to the reader.   

\bigskip

Here let us pass from the $2\x 2$-matrix representation of $q_t^+$ to 
the so-called Garding-Wightman representation of $2\x 2$-matrices:

\begin{lemma}

Every complex $2\x 2$-matrix $a=((a_{jk}))$ may be uniquely written in the
form
$$
a=\kp_0+\vec{\kp}.\s\ ,\ \mbox{ where } \kp_0=\halb \mbox{ trace }(a)\ ,\
\kp_j=\halb\trace(\s_ja)\ ,\ j=1,2,3\ ,\ 
\lo{(5.15)}
$$
with the Pauli-matrices $\s_j$ of (2.7),
where $\kp_0,\vec{\kp}$ are real if and only if $a$ is self-adjoint.

\end{lemma}

The proof of lemma 5.3 is trivial. 

\bigskip

If we substitute $\kp_t^+=\kp_{t0}+\s.\vec{\kp}_t^+\ ,\  
\Th^+=-\frac{i}{2}(\mcl{F}_0+\s.\vec{\mcl{F}})\ $
into (5.12) we get
$$
\kp_{0t}'=0\ \ \ \ ,\ \ \ \ \ (\vec{\kp}_t)'+\vec{\mcl{F}}\x \vec{\kp}_t=0\ .\ 
\lo{(5.16)}
$$
Here we used the well known formula
$$
(\s\xi)(\s\eta)=\xi.\eta+i\s.(\xi\x\eta)\ ,\ \xi,\eta\in\mbb{R}^3\ .
\lo{(5.17)}
$$

\bigskip

The first equation (5.16) states what we already know from sec.3: If $q_t$ is a scalar
in $S_+$  the we have $\kp_t^+$ a multiple of the identity, so that $\kp_t^+=\kp_{t0}$
while $\vec{\kp}_t=0$. So, $\kp_t $ is constant on the flow $\nu^+_t$. Assuming that the
corresponding also holds for $q_t^-$ we then again get  the statement of thm.4.5.

For the second equation (5.16), we again involve the system 
(4.14+) of ODE-s and its flow $\nu^+_t$.  We get
$$
\frac{d}{d t}\vec{\kp}_{-t}(x(t),\xi(t))=-\vec{\mcl{F}}(x(t),\xi(t))\x \vec{\kp}_{-t}(x(t),\xi(t))
\lo{(5.18)}
$$
along any solution curve $x(t),\xi(t)$ of the system (4.14+). With the flow $\nu_t$ we get
$$
\frac{d}{d\t}\vec{\kp}_{t-\t}(\nu_\t(x,\xi))+\vec{\mcl{F}}(\nu_\t(x,\xi))\x \vec{\kp}_{t-\t}(\nu_\t(x,\xi))=0\ ,\ 
\lo{(5.19)}
$$
a system of 3 ODE-s in 3 unknown functions of the variable $\t$. We know the solution
$\vec{\kp}_{t-\t}(\nu_t(x,\xi))$ at $\t=t$ where it becomes $\vec{\kp}_0(\nu_t(x,\xi))$
with the matrix $\kp_0$ of $q_0^+=q^+$. Thus $\vec{\kp}_{t-\t}(\nu_t(x,\xi))$ is
completely determined for all $\t$, and especially for $\t=0$, where we get $\vec{\kp}_{t}((x,\xi))$.
The components of $\vec{\kp}$ remain symbols in $\psi c_m$, as a consequence of our discussion
in [Co16], ch.5. Corresponding statements hold for ``-", and the existence result of thm.4.5 will
be following again.

\bigskip

\begin{obs}

The second (5.15) appears interesting from a different viewpoint: Clearly 
the expression $\frac{d}{d\t}\vec{\kp}_{t-\t}(\nu_\t(x,\xi))|_{\t=0}$ may be interpreted as the \emph{rate of change
(in time)} of the real 3-vector $\vec{\kp}_{t}(x,\xi)$ progressing on its orbit through $(x,\xi)$, while subtracting
the orbital rate of change. According to (5.15), this vector equals a vector product
$-\vec{\mcl{F}}(x,\xi)\x \vec{\kp}_t(x,\xi)$ with a certain 3-vector $-\vec{\mcl{F}}(x,\xi)$.

As will be shown in sec.6, below, the vector $\vec{\mcl{F}}(x,\xi)$ will be a linear combination of magnetic vectors ---
the magnetic induction $\mcl{B}(x,\xi)$ and a vector of the form $\dot{x}\x \mcl{E}$, at $(x,\xi)$, where we used (4.18)
to replace $\z=\xi-\mbf{A}(x)$ by $\dot{x}$ --- the velocity of the particle.

So, we might have reason to regard the vector  $\vec{\kp}^+_t(x,\xi)$ as a magnetic moment vector, traveling with the
particle --- since it reacts to the fields at the location $(x,\xi)$ of the particle. But, as we shall find, the magnetic field, this
vector `sees', is not the relativistic field of the moving particle at the point $(x,\xi)$. So, while we are tempted to 
interpret $\vec{\kp}_t$ as a magnetic spin-vector, traveling with the particle, there will be some paradoxes appearing,
possibly to be eliminated by a better setup?

\end{obs}

\section{Extension of Theorem 4.5}

It now will be a matter of a (lengthy) calculation to verify that
the vector $\vec{\mcl{F}}$ plays the role of a
magnetic field vector.

\begin{prop}

The $3$-vector $\vec{\mcl{F}}$ is explicitly given as
$$
\vec{\mcl{F}}=\inv{\<\z\>(1+\<\z\>)}
(-\z\x\mcl{E}+\inv{\<\z\>}(|\z|^2\mcl{B}-(\z\mcl{B})\z))
-\inv{\<\z\>^2}(\mcl{B}+\inv{1+\<\z\>}(\z\mcl{B})\z)
\lo{(6.1)}
$$
with\footnotemark $\z=\xi-\mbf{A}(x)$ and the field vectors
$$
\mcl{E}=-\grad{\mbf{V}}\ ,\ \mcl{B}=\curl \mbf{A}\ .
\lo{(6.2)}
$$

\end{prop}

\footnotetext{Note, we have $\z=\xi-\mbf{A}=\frac{\dot{x}}{\sqrt{1-\dot{x}^2}}\ ,\
\<\z\>=\<\xi-\mbf{A}\>=\inv{\sqrt{1-\dot{x}^2}}$, by (4.19), if we relate $(x,\xi)$
to $(x,\dot{x})$ using the classical equations of motion of (4.14+).}

\noindent
\tbf{Proof.}
To simplify calculations,
we note that the matrices $\Th$ occur only in
the commutator of equation (5.13). When we evaluate them we may
omit any additive term giving a scalar multiple of the $2\x 2$-
identity matrix, because its contribution to the commutator will vanish.
We shall write `$a=b(mod\ 1)$' if $b-a$ is a scalar multiple of the
$2\x 2$-identity matrix. In other words, the term $\mcl{F}_0$ of the 
decomposition of $\Th$ is irrelevant, hence shall be ignored.

Again we shall focus on `` $+$", and shall omit +-sub-(super-)scripts in notation
wit some exceptions.

Let us write $\O_+=\inv{2(1+\u_0)}\U_+$,  then we get
$$
\U_+=(1+\u_0)\vc{\ \ 1\ \ }{-i\s\g}\ ,\ \O_+=\halb \vc{\ \ 1\ \ }{-i\s\g}\ ,\ \mbox{with}\ \
\g=\frac{\u}{1+\u_0}=\frac{\z}{1+\<\z\>}\ ,
\lo{(6.3)}
$$
where we recall that $\u_0=1/{\<\z\>}\ ,\
\u=(\u_1,\u_2,\u_3)\ ,\ \u_j=\z_j/\<\z\>\ ,\
\z_j=\xi_j-\mbf{A}_j(x)\ ,\ j=1,2,3$.

First we look at (the $2\x 2$-matrix)
$$
\Th^\sim=\inv{2(1+\u_0)}\U_+^*\U_+'=\O_+^*\U_+'=-\O_+^{*'}\U_+\ ,\ 
\lo{(6.4)}
$$
recalling that we have $\O_+^*\U_+=1\ ,\ $ hence $\O_+^*\U_+'=-\O_+^{*'}\U_+$ .
From (6.3) we get $\O_+'=\halb \vc{\ \ 0\ \ }{-i\s\g'}$\ , and,
$$
\Th^\sim = -\halb (1+\u_0)(\s\g')(\s\g)=-\halb(1+\u_0)i\s.(\g'\x\g)\ ,\ 
\ (\mbox{mod}\ 1)\ .
\lo{(6.5)}
$$
using (5.17) again.

Note, $\g=\z/(1+\<\z\>)$ is a scalar multiple of $\z=\xi-\mbf{A}$, hence
$\g\x\g'= \inv{(1+\<\z\>)^2}\z\x\z'$, since $\z\x\z=0$.
Thus we get - all (mod $1$) -
$$
\Th^\sim=
\frac{i}{2}\inv{\<\xi-\mbf{A}\>(1+\<\xi-\mbf{A}\>)}\s.((\xi-\mbf{A})\x
(\xi-\mbf{A})')\ .
\lo{(6.6)}
$$

Next we calculate
$$
(\xi-\mbf{A})'=\{\p_t-\sum_j\l_{|\xi_j}\p_{x_j}+\sum_j\l_{|x_j}\p_{\xi_j}\}(\xi-\mbf{A})\ ,
\lo{(6.7)}
$$
where $\l_{|\xi_j}=(\xi_j-\mbf{A}_j)/\<\xi-\mbf{A}\>$ and
$\l_{|x_j}=\mbf{V}_{|x_j}
-\sum_l \mbf{A}_{l|x_j}(\xi_l-\mbf{A}_l)/\<\xi-\mbf{A}\>$\ .
The result is this:        
$$
(\xi_k-\mbf{A}_k)'=-\mbf{V}_{|x_k}+\sum_j\frac{\xi_j
+\mbf{A}_j}{\<\xi-\mbf{A}\>}(\mbf{A}_{k|x_j}-\mbf{A}_{j|x_k})\ ,\ k=1,2,3.
\lo{(6.8)}
$$
The last term equals
$-\inv{\<\xi-\mbf{A}\>}($curl $\mbf{A}\x(\xi-\mbf{A}))_k$. Thus we have
$$
\z'=(\xi-\mbf{A})'=\mbf{V}_{|x}
-\inv{\<\xi-\mbf{A}\>}\mbox{curl}\ \mbf{A}\x(\xi-\mbf{A})=-\mcl{E}-\inv{\<\z\>}\mcl{B}\x\z\ ,
\lo{(6.9)}
$$
and we get
$$
\z\x\z'=-\z\x\mcl{E}-\inv{\<\z\>}\z\x(\mcl{B}\x\z)= -\z\x\mcl{E}
-\inv{\<\z\>}(|\z|^2\mcl{B}-(\z.\mcl{B})\z)\ .
\lo{(6.10)}
$$
All together we get
$$
\Th^\sim=\frac{i}{2}\inv{\<\z\>(1+\<\z\>)}\s.(\z\x\mcl{E}
+\inv{\<\z\>}(|\z|^2\mcl{B}-(\z.\mcl{B})\z))\ .
\lo{(6.11)}
$$

Next we set out to calculate the other part $-2\<\z\>\Th^1$ of the matrix
$\Th$ of (5.13). Here it might be some help to go back and write
$$
2\<\z\> pp_{|\xi}p_{|x}p=pp_\xi h_{|x}p\ ,
\lo{(6.12)}
$$
noting that $2\<\z\>pp_{|\xi}p_{|x}p=\l_+pp_{|\xi}p_{|x}p
+\l_-pp_{|\xi}p_{-|x}p$ while $pp_{|\xi}p=0$.

We get
$$
2\<\z\>\Th^1=\halb (1+\u_0)(1,i\s\g)p_{|\xi}h_0(\z)_{|x}\vc{\ \ 1\ \ }{-i\s\g}
\ (\mbox{mod}\ 1)
\lo{(6.13)}
$$
with $h_0(\z)= \a\z+\b$,
since the term $\mbf{V}_{|x}(1,i\s\g)\vc{\ \ 1\ \ }{-i\s\g}
=\mbf{V}_{|x}(1+|\g|^2)$
is scalar, using (5.17).

Now we get $p=\halb(1+\frac{h_0(\z)}{\<\z\>})=\halb(1+\u_0h_0(\z))$, hence
$p_{|\xi_k}=\halb (\u_{0|\xi_k}/\u_0)h_0(\z)+\halb \u_0\a_k$ where the first
term at right will generate a scalar multiple of $1$, hence may be
ignored.
Also, $h_0(\z)_{|x_k}=(\sum \a_j(\xi_j-\mbf{A}_j)+\b)_{|x_k}=-(\a.\mbf{A})_{|x_k}$.
Substituting into (6.13) we get
$$
2\<\z\>\Th^1=-\frac{\u_0}{4}(1+\u_0)
(1,i\s\g)(\sum_{jl} \mbf{A}_{j|x_l}\a_j\a_l)\vc{\ \ 1\ \ }{-i\s\g}\ .
\lo{(6.14)}
$$
But we have
$$
\sum_{jl}\mbf{A}_{j|x_l}\a_j\a_l=
\mbox{div}\ \mbf{A} -i\rho.\ \mbox{curl}\ \mbf{A}\ ,
\lo{(6.15)}
$$
with $\rho=\mat{\s}{0}{0}{\s}$, 
where again the first term may be ignored, when we substitute this into
(6.14). We get
$$
2\<\z\>\Th^1=
-\frac{i}{4}\u_0(1+\u_0)(1,i\s\g)\rho .\mcl{B}\vc{\ \ 1\ \ }{-i\s\g}\ .
\lo{(6.16)}
$$
A matrix calculation then gives
$$
(1,i\s\g)\mat{\s\mcl{B}}{\ 0\ }{\ 0\ }{\s\mcl{B}}\vc{\ \ 1\ \ }{-i\s\g}
=(\s\mcl{B})+(\s\g)(\s\mcl{B})(\s\g)
=\s((1-|\g|^2)\mcl{B}+2(\g \mcl{B}\g)\ .
\lo{(6.17)}
$$
We have $1-|\g|^2=\frac{2}{1+\<\z\>}$ so (5.17) equals
$$
\frac{2}{1+\<\z\>}(\mcl{B}+\inv{1+\<\z\>}(\z\mcl{B})\z)\ .
\lo{(6.18)}
$$
All together we then get
$$
2\<\z\>\Th^1=
-\frac{i}{2}\inv{\<\z\>^2}\s.(\mcl{B}+\inv{1+\<\z\>}(\z\mcl{B})\z)\ .
\lo{(6.19)}
$$

Collecting things, up to here: We have

$$
\Th=\Th^\sim-2\<\z\>\Th^1
\lo{(6.20)}
$$
with $\Th^1$ of (5.19) and
$$
\Th^\sim=\frac{i}{2}\inv{\<\z\>(1+\<\z\>)}\s.(\z\x\mcl{E}
+\inv{\<\z\>}(|\z|^2\mcl{B}-(\z.\mcl{B})\z))\ .
\lo{(6.21)}
$$
We then may write $\Th=-\frac{i}{2}\s.\ \vec{\mcl{F}}$
to get (6.1), proving prop.6.1, q.e.d.

\begin{thm}
We consider (time-independent) local potentials $\mbf{V}(x),\mbf{A}(x)$
satisfying (3.6) with $m=(0,-1)$ as described early in sec.4.
Assume we have a symbol $q(x,\xi)\in\psi c_m$ such that $q(x,\xi)$ commutes
with $h(x,\xi)$ for all $x,\xi$. Let $\kp^+(x,\xi)$ and $\kp^-(x,\xi)$ be the
matrices representing $q(x,\xi)$ in its two eigenspaces $S_\pm(x,\xi)$,
with respect to the orthonormal bases given by the columns of the
$4\x4$-matrix $\U(x,\xi)$ of (2.5) with Dirac matrices $\a,\b$ of (2.8),
and let $\kp_0^\pm=\trace (\kp^\pm)$ and $\vec{\kp}^\pm=\trace{\s\kp^\pm}$
be given by the Garding-Wightman decomposition of $\kp^\pm$.

Then there exists a symbol $z(x,\xi)\in\psi c_{m-e}\ ,\ e=(1,1)$
such that $A=a(x,D)=q(x,D)+z(x,D)$ is a precisely predictable observable.
In particular, we have $A_t=e^{iHt}Ae^{-iHt}=a_t(x,D)=q_t(x,D)+z_t(x,D)\in Op\psi c_m$,
where  $z_t(x,\xi)\in\psi c_{m-e}$ while $q_t^+(x,\xi)\ ,\ q_t^-(x,\xi)\in\psi c_m$
are defined by giving $\kp_{t0}^\pm\ ,\ \vec{\kp}_t^\pm$ of the 
Garding-Whigtman decomposition (5.15) of their $2\x 2$-matrices, with respect
to orthonormal bases linked to the diagonalization (2.6) of $h(x,\xi)$,
 as follows:

(i) We have
$$
\trace q_t^+(x,\xi)=\trace q^+(\nu_t^+(x,\xi))\ ,\
\trace q_t^-(x,\xi)=\trace q^-(\nu_t^-(x,\xi))\ ,\
\lo{(6.22)}
$$
where $\nu_t^\pm:\mbb{R}^6\rta\mbb{R}^6$ is the flow, letting each point $(x,\xi)$ wander
along the solution $(x(t),\xi(t))$ of (4.14$\pm$) for a time-length $t$ counted positive or
negative.
Here we should remind of the fact that the system (4.14$\pm$) may be rewritten as
a set of second order equations in $x$ only of the form
$$
(\frac{\dot{x}}{\sqrt{1-\dot{x}^2}})^{^\bullet}=\mcl{E}+\dot{x}\x \mcl{B}\ ,
\lo{(6.23+)}
$$
for $\l=\l_+$\ ,\ and,
$$
(\frac{\dot{x}}{\sqrt{1-\dot{x}^2}})^{^\bullet}=-\mcl{E}-\dot{x}\x \mcl{B}\ .
\lo{(6.23-)}
$$
for $\l=\l_-$, with electrical field strength $\mcl{E}$ and magnetic
induction $\mcl{B}$ induced by $\mbf{V}$ and $\mbf{A}$.

(ii) The two real 3-vectors $\vec{\kp}_{\t-t}^\pm(\nu_\t^\pm(x,\xi))$ will satisfy the
equations 
$$
\inv{\sqrt{1-\dot{x}^2}}\frac{d}{d\t}\vec{\kp}_{t-\t}^{+}(\nu_\t)|_{\t=0}=\vec{\kp}_t^+\x\mcl{B}^\sim
\ ,\ \mbox{where}
\ \ \mcl{B}^\sim=\mcl{B}+\inv{1+\sqrt{1-\dot{x}^2}}\dot{x}\x\mcl{E}\ ,
\lo{(6.24+)}
$$
$$
\inv{\sqrt{1-\dot{x}^2}}\frac{d}{d\t}\vec{\kp}_{t-\t}^{-}(\nu_\t)|_{\t=0}=-\vec{\kp}_t^-\x\mcl{B}^\sim
\ ,\ \mbox{where}
\ \ \mcl{B}^\sim=\mcl{B}+\inv{1+\sqrt{1-\dot{x}^2}}\dot{x}\x\mcl{E}\ ,
\lo{(6.24-)}
$$
with initial-values
$\vec{\kp}_0^+=\vec{\kp}^+\ ,\ \vec{\kp}_0^-=\vec{\kp}^-$.

(iii) Formulas (6.22) and (6.24) are valid only asymptotically,
modulo $\psi c_{m-e}$, assuming that the initial symbol $q(x,\xi)$
belongs to $\psi c_m$. That is, they may be trusted if either $|x|$ is large
or if $\dot{x}\approx 1$ = velocity of light --- or both.

However,  an infinite sequence of improvements can be constructed,
by solving (iteratively) a system of differential equations similar to (5.16),
leading to exact symbols $a=q+z\ ,\ a_t=q_t+z_t$ with (6.22), (6.24) being
true asymptotically, modulo $\psi c_{-\i}$.

\end{thm}

\section{The $\vec{\kp}$-vectors of Total Angular Momentum}

Most of the dynamical observables, generally considered, are scalar 
in $\mbb{C}^4$, so also scalar in the two eigenspaces $\mcl{S}_\pm$, 
implying that the two vectors
$\vec{\kp}_t^\pm$ will vanish identically, for all $x,\xi$.
An exception is the \emph{total angular momentum} defined as
$J=S+L $ , where $L=x\x D$ is the \emph{orbital angular momentum}
while $S=\halb\mat{\s}{0}{0}{\s}$ usually is interpreted as the 
(mechanical) \emph{spin} of the particle. It is known that the self-adjoint operator
$J$ commutes with $H$, assuming that $\mbf{A}=0\ ,\ \mbf{V}=\mbf{V}(|x|)$, 
so that $e^{iHt}Je^{-iHt}=J$. So, $J$ is precisely predictable, if $\mbf{V}(|x|)$ 
satisfies our assumptions.
On the other hand, the spin $S$ , as defined above, certainly is not precisely predictable.
Neither is $L$, although thm.4.5 allows construction of a lower order correction
$L_{corr} $, such that $L+L_{corr}$ \emph{is precisely predictable}.
Note, we have $L\in\ Op\psi c_{(1,1)}$, hence $L_{corr}\in Op\psi c_{(0,0)}$.
We may write $J=(J+L_{corr})+(S-L_{corr})$ and then reinterpreted
the (precisely predictable) observable
$S_{corr}=S-L_{corr}\in Op\psi c_{(0,0)}$ as the spin. Checking this symbol-wise one finds that
(modulo lower order) we get 
$$
\mbox{ symb}(S_{corr})=p_+(x,\xi)\ S\ p_-(x,\xi)+p_-(x,\xi)\ S\ p_-(x,\xi)\ ,
\lo{(7.1)}
$$
where the right hand side makes sense also for general potentials, and then commutes with $h(x,\xi)$
also for general potentials, not necessarily $(0,\mbf{V}(|x|))$. We then proposed to 
generally redefine the spin observable, using the right hand side of (7.1).

Here we are interested only in the two vectors $\vec{\kp}^\pm$ for the
(corrected) spin and the total angular momentum.
Note, the orbital angular momentum $L$ is scalar in
$\mcl{S}_\pm$, hence will not contribute to the $\vec{\kp}^\pm$. So, both
$J$ and $S_{corr}$ have the same $\vec{\kp}_t^\pm$-vectors. In fact, it
suffices to just calculate the $2\x 2$-matrices
$\kp^\pm$ of the (uncorrected) spin-observable
$S=\halb\mat{\s}{0}{0}{\s}$, and then calculate its corresponding
vectors $\vec{\kp}^\pm$.

\begin{prop}

Looking at the $2\x 2$-matrices $\kp^{j\pm}(x,\xi)$ of the matrices
$p_+(x,\xi)S_jp_+(x,\xi)$ and \newline $p_-(x,\xi)S_jp_-(x,\xi)$
for a spin component $S_j$ with respect to
the orthonormal bases of $\mcl{S}_\pm$ used in sec.5 and sec.6, we get
$$
\kp^{j+}=\kp^{j-}=\halb\sqrt{1-\dot{x}^2}
\{\s_j+\inv{\sqrt{1-\dot{x}^2}(1+\sqrt{1-\dot{x}^2})}\dot{x}_j(\dot{x}\s)\}
 \ ,\
\lo{(7.2)}
$$
where we have replaced the $\xi$-variable by $\dot{x}$ with the
relation $\dot{x}=\l_{|\xi} \ \Leftrightarrow \
\xi=\mbf{A}(x)+\frac{\dot{x}}{\sqrt{1-\dot{x}^2}}$, as this was
done in the two earlier sections.

\end{prop}

Using (7.2) we then at once obtain the components of the vectors
$\vec{\kp}^{j\pm}$ by using (5.15):
$$
\vec{\kp}^{j+}_l=\vec{\kp}^{j-}_l=\halb\sqrt{1-\dot{x}^2}
\{\d_{jl}+\inv{\sqrt{1-\dot{x}^2}(1+\sqrt{1-\dot{x}^2})}
\dot{x}_j\dot{x}_l\}
\lo{(7.3)}
$$
To express this alternately:
$$
\vec{\kp}^{j+}_l=\vec{\kp}^{j-}_l=\halb\sqrt{1-\dot{x}^2}
\{\d_{jl}-\frac{\dot{x}_j\dot{x}_l}{\dot{x}^2}\}
+\halb\frac{\dot{x}_j\dot{x}_l}{\dot{x}^2}\ ,\  
\mbox{ as } \dot{x}\neq 0\ .
\lo{(7.4)}
$$

\begin{obs}

At speed $\dot{x}=0$ the three vectors $\vec{\kp}^j$ are just the three
unit vectors $\vec{\kp}^{j+}=\vec{\kp}^{j-}=\halb e^j$
(with $e^j_l=1$ in j-th
row, and zero elsewhere)--- except for a factor $\halb$.
At arbitrarily speeds $\dot{x}$ there will
be a relativistic shortening in the perpendicular directions, and
no shortening in the parallel direction --- with respect to
$\dot{x}$.

\end{obs}

\noindent
\tbf{Proof of prop.7.1.}

We discuss the "+" case only, with "-" going similarly. Using
(6.3)  we get $\kp^+=\U_+^*S\O_+$ with $\g$ as stated there. 
That is, we get
$$
\kp^{j+}=\kp^{j-}=\inv{4\<\z\>}(1+\<\z\>)\{\s_j+(\g\s)\s_j(\g\s)\}
 \ ,\ \g=\frac{\z}{1+\<\z\>}\ ,\ \z=\xi-\mbf{A}(x)\ .
  \lo{(7.5)}
$$
A calculation gives
$$
(\g\s)\s_j(\g\s)=2\g_j(\g\s)-\g^2\s_j\ ,\
\lo{(7.6)}
$$
$$
\s_j+(\g\s)\s_j(\g\s)=(1-\g^2)\s_j+2\g_j(\g\s)
=\frac{2}{1+\<\z\>}\{\s_j+\inv{1+\<\z\>}\z_j(\z\s)\}\ ,\
\mbox{\ \ so,\ \ }
\lo{(7.7)}
$$
$$
\kp^{j+}=\kp^{j-}=\inv{4\<\z\>}\{\s_j+\inv{1+\<\z\>}\z_j(\z\s)\} \ ,\
\lo{(7.8)}
$$
Transforming onto the variable $\dot{x}$ again we get
the desired equation (7.2). Q.E.D.

\section{An Electron under Electro-Magnetic Radiation}

We next consider a time-dependent Dirac operator of the form
$$
H=\a_1D_1+\a_2(D_2-\e_0\sin\o(x_1-t))+\a_3D_3+\b\ ,\ 
\lo{(8.1)}
$$
where we use the Dirac matrices $\a,\b$ of (2.9). Symbolwise we may write 
$H=h(t,x,D)$ with $h(t,x,\xi)=h_0(\xi)-\e_0\a_2sin\o(x_1-t)\ ,\ h_0(\xi)=\a\xi+\b$ .

Clearly we then have the potentials $\mbf{V}=\mbf{A}_1=\mbf{A}_2=0\ ,\ \mbf{A}_2=\e_0\sin\o(x_1-t)$.
The corresponding electro-magnetic field then is defined as
$$
\mcl{E}=-\dot{\mbf{A}}-\grad \mbf{V}=\e_0\o \cos\o(x_1-t)(0,1,0)^T\ ,\ 
\mcl{B}=\curl \mbf{A} = \e_0\o \cos\o(x_1-t)(0,0,1)^T\ ,\ 
\lo{(8.2)}
$$
corresponding to a plane polarized wave of (circular) frequency $\o$ propagating in
the positive $x_1$-direction, with $\mcl{E}$ and $\mcl{B}$ oscillating in the 
$(x_1,x_2)$- and $(x_1,x_3)$-plane, respectively.

This Dirac operator $H$ does not belong to $Op\psi c$. But it will belong to the class 
$\psi p_1$ of def.3.3 (iii). Since $H=H(t)$ now depends on $t$, the propagator $U(t)$ no longer
is an exponential function. However, due to the special form of time-dependency,
we find that $U(t)$ is a product of two exponentials:

\begin{prop}

The propagator $U(t)$ such that $U(t)\psi_0=\psi(t,x)$ solves 
$\dot{\psi}+iH(t)\psi=0$ (with $H(t)$ of (8.1)),
and $\psi(0,x)=\psi_0(x)$, has the form
$$
U(t)=T_{-t}e^{-iKt}\ ,\ \mbox{ with } K=H(0)-D_1\ ,\ \mbox{ and the translation }T_t\psi(x)=\psi(x_1+t,x_2,x_3)\ .
\lo{(8.2)}
$$
Moreover, the propagator $U(\t,t)$ solving the problem with initial-values at $t=\t$ may be written as
$$
U(\t,t)=T_{-t}e^{-iK(t-\t)}T_\t\ .
\lo{(8.3)}
$$
\end{prop}

\noindent
\tbf{Proof.}
We get $(T_{-t}H(0)T_t\psi)(x)=(H_0\psi)(x)-\a_2(T_{-t}\mbf{A}_2(x_1)T_t\psi)(x)=H(t)\ ,\ $
since $H_0$ is translation invariant.  Thus we may write
$\dot{\psi}+iH(t)\psi=0$ as
$$
T_t\dot{\psi}+iH(0)T_t\psi=0\ .
\lo{(8.4)}
$$
Here we set $\chi(t,x)=T_t\psi(t,x)=\psi(t,x+te^1)$, and use that
$$
\dot{\chi}(t,x)=\p_t(\psi(t,x_1+t,x_2,x_3))=T_t\dot{\psi}(t,x)+
\p_{x_1}\chi(t,x)\ .\
\lo{(8.5)}
$$
Equation (8.4) then may be written as
$$
\dot{\chi}+i(H(0)-D_1)\chi=0\ .
\lo{(8.6)}
$$
In other words, the substitution $\chi(t,x)=T_t\psi=\psi(t,x+te^1)$
converts the Dirac equation into equ. (8.6), where now the operator
$ H(t)$ of (8.1) is replaced by the (time-independent) operator
$$
K=H(0)-D_1=H_0-\a_2\mbf{A}_2(x_1)-D_1\ .
\lo{(8.7)}
$$

It is evident then that (8.6) will be solved by
$$
\chi(t,x)=e^{-itK}\chi(0,x)\ .
\lo{(8.8)}
$$
Or else, we may write this as
$$
\psi=T_{-t}e^{-iKt}\psi_0\ ,\
\lo{(8.9)}
$$
proving (8.2), while (8.3) then follows trivially. Q.E.D.

\bigskip

Note, for this Dirac operator, the total energy $H(t)$ is not constant ---
it fluctuates periodically, with period $2\pi/\o$. For $t=0$ the spectral
decomposition of $K$, not of $H(0)$ will provide the split between
electron and positron. The spectral theory of $K$ can be worked out
explicitly. We shall find that $K$ has continuous spectrum
along all of $\mbb{R}$. But there is a strong singularity at $t=0$.
We shall set
$$
\mcl{H}=\mclh_e\ \ \oplus\ \ \mclh_p\ ,\ 
\lo{(8.10)}
$$
with the spectral spaces $\mclh_e\ ,\ \mclh_p$ of $K$ belonging to the 
intervals  $(0,\i)$ and $(-\i,0)$ respectively. Then $\mclh_e$ and $\mclh_p$
are defined as the spaces of electron states and positron states, resp., at $t=0$.

It may be seen that these spaces converge towards the well known electron and
positron spaces for $H_0=\a D+\b$ as the amplitude $\e_0$ tends to $0$, so that
$H(t)\rta H_0$.

As time $t$ progresses, the spaces $\mclh_e\ ,\ \mclh_p$  will change; at time $t$ we
will set 
$$
\mclh_e(t)=T_{-t}\mclh_e\ ,\ \mclh_p(t)=T_{-t}\mclh_p\ .
\lo{(8.11)}
$$

However, when looking at propagation of  states, while solving
Dirac's equation, we shall find that still electrons remain electrons and 
positrons remain positrons, as time progresses.

Indeed, a state $\psi_0\in \mclh_e$ will propagate to 
$\psi(t,x)=U(t)\psi_0=T_{-t}e^{-iKt}\psi_0$,
where $e^{-iKt}\psi_0\in\mclh_e$, since 
$e^{-iKt}$ leaves all spectral spaces of $K$ invariant.
So, it follows that $\psi(t,.)\in \mclh_e(t)$ --- indeed,
an eletron state remains an electron state. Similar with
positron states.

Regarding prediction of expectation values, things remain as discussed
earlier: For a state $\psi_0\in\mclh$ and an observable $A$ we get the
expectation value $\<\psi_0,A\psi_0\>$ For a future time then, if 
$\psi_t=U(t)\psi_0$ or also $A_t=U^*(t)AU(t)$ the predicted expectation value
then will be $\<\psi_t,A\psi_t\>=\<\psi_0,A_t\psi_0\>$, marking 
Schr\"{o}dinger or Heisenberg representation.

\begin{lemma} 

We have
$$
U^{-1}(t)H(t)U(t)=H(0)+e^{iKt}D_1e^{-iKt}-D_1
                 =K+U^{-1}(t)D_1U(t)\ .
\lo{(8.12)}
$$
That is, the changes of expectation values of total energy and
of momentum component $D_1$ at time $t$ are related: 
Defining $A_t=U^*(t)AU(t)$ for an arbitrary observable $A$, we get
$$
(H(t))_t-H(0)=(D_1)_t-D_1\ .
\lo{(8.13)}
$$

\end{lemma}

\noindent
\tbf{Proof.}
We get
$$
e^{iKt}T_tH(t)T_{-t}e^{-iKt}=e^{iKt}H(0)e^{-iKt}
=e^{iKt}Ke^{-iKt}+e^{iKt}D_1e^{-iKt}=K+(D_t)_t\ .\ 
\lo{(8.14)}
$$
Q.E.D.

\bigskip

We shall need details of the spectral theory of the operator $K$ but
will discuss this in a later section. Right now let us focus on an
attempt to repeat the procedures of earlier sections, regarding 
potentials vanishing at $|x|=\i$, for the present Dirac operator
$H(t)$ of (2.1). As already observed, we no longer have
$H(t)\in Op\psi c$, but rather have $H(t)\in Op\psi p_1\subset Op\psi q_1$,
with the larger symbol classes of sec.3.
With some exceptions we then shall focus entirely on time-propagation of 
symbols of the
form $q(\xi)$ ---  independent of $x$, with $q\in\psi c_{(m,0)}$, and with
$q(\xi)$ commuting with $h_0(\xi)=\a\xi+\b$, for all $\xi$.
Of special interest will be the case of $q(\xi)=\xi_1$ (and also
$q(\xi)=\xi_j\ ,\ j=2,3$), --- that is, of the momentum observables.

For such a symbol $q(x)$ the operator $q(D)$ is translation invariant: Especially we get
$T_tq(D)T_{-t}=q(D)$, implying that
$$
(q(D))_t=U^*(t)q(D)U(t)=e^{iKt}q(D)e^{-iKt}\ .
\lo{(8.15)}
$$
Therefore our attempt to repeat earlier arguments for the case of a $q\in\psi c$
will focus on the assignment $a(x,\xi)\rta a_t(x,D)=e^{iKt}a(x,D)e^{-iKt}$ equivalent to the ODE-initial-value
problem
$$
\dot{a}_t(x,D)=i[K,a_t(x,D)]\ \ \ \mbox{\ \ as\ \ }-\i<t<\i\ ,\ a_0(x,\xi) \mbox{\ \ given\ \ }\ .
\lo{(8.16)}
$$

The theorem, below, will address the initial-value problem (8.16) modulo $\psi q_{-\i}$. We shall
require another lengthy argument involving calculus of Fourier integral operators (to be discussed
in sec's 11 f.) to also cover the corresponding Heisenberg transform $U^*(t)AU(t)$.
However, the results of sec.10, below, addressing only the case of a simple photon-collision ,
will not be affected by these more complicated things.

\begin{thm}

Given any self-adjoint ($4\x 4$-matrix-valued) symbol
$q(\xi)\in\psi c_{(m,0)}$,  independent
of the location variable $x$\ ,\ 
depending on the momentum variable $\xi$ only,
and such that the commutator
$[h_0(\xi),q(\xi)]=h_0(\xi)q(\xi)-q(\xi)h_0(\xi)$
vanishes, for all $\xi$.

I) There exists a (lower order) `correction symbol'
$z(x_1,\xi)\in\psi p_{m-1}$ with \newline $[h_0(\xi),z(x_1,\xi)]_+
=h_0(\xi)z(x_1,\xi)+z(x_1,\xi)h_0(\xi)=0$
for all $x_1,\xi$, such that the initial-value problem (8.16)
with $a_0(x,\xi)=q(\xi)+z(x_1,\xi)$ admits a solution $a_t(x,\xi)$ 
modulo $\psi q_{-\i}$ of the form
$$
a_t(x,D)=q_t(x_1,D)+z_t(x_1,D)\ \ \mbox{  (mod }Op\psi q_{-\i})\ ,\ 
\lo{(8.17)}
$$
where $q_t(x_1,\xi)\in\psi p_m\ ,\ [h_0(\xi),q_t(x_1,\xi)]=0,\forall x_1,\xi\ ,\ 
z_t(x_1,\xi)\in\psi p_{m-1}\ ,\ [h_0(\xi),z_t(x_1,\xi)]_+=0\ ,\ 
\forall x_1,\xi ,\newline \dot{q}_t(x_1,\xi)\ ,\
\dot{z}_t(x_1,\xi)\in\psi p_{m-1}$, and, $q_0(x_1,\xi)=q(\xi)\ ,\
z_0(x_1,\xi)=z(x_1,\xi)$.

II) The symbols $q_t(x_1,\xi)\ ,\ z_t(x_1,\xi)$
have $x_1$-Fourier-series-expansions
$$
q_t(x_1,\xi)=\sum q_{t,n}(\xi)e^{in\o x_1}\ ,\
z_t(x_1,\xi)=\sum z_{t,n}(\xi)e^{in\o x_1}\ ,\
q_{t,n},z_{t,n}\in\psi p_m\ ,\ 
\lo{(8.18)}
$$
where the sums over $n$ are \emph{finite} if looked at modulo $\psi p_{m-j}$,
for every $j=1,2,\ldots$. That is, for every $j=1,2,\ldots$ only a finite
number of the coefficients $q_{t,n},z_{t,n}$ are not in $\psi p_{m-j}$.

Accordingly, the corresponding $\psi do$-s are of the form
$$
q_t(x_1,D)=\sum e^{in\o x_1}q_{t,n}(D)\ ,\
z_t(x_1,D)=\sum e^{in\o x_1}z_{t,n}(D)\ .
\lo{(8.19)}
$$

III) In \emph{momentum space} --- looking at the
Fourier transformed operators
$q_t(x_1,D)^\wdg=Fq_t(x_1,D)F^*$
\ ,\  $z_t(x_1,D)^\wdg=Fz_t(x_1,D)F^*$ ---
f'las (8.19) assume the form
$$
q_t(x_1,D)^\wdg=\sum  T_{-n\o}q_{t,n}(\xi)\ ,\
z_t(x_1,D)^\wdg=\sum T_{-n\o}z_{t,n}(\xi)\ ,\ 
\lo{(8.20)}
$$
with the translation operator $T_\kp u(x)=u(x_1+\kp,x_2,x_3)$.

IV) In general the "corrected operator" $A(t)=q(D)+z(x_1-t,D)$ of (8.17) 
may not be self-adjoint, so, it may not count as an observable.
However, we may take the self-adjoint operator
$$
\breve{A}(t)=\halb\{A(t)+A^*(t)\}\ ,\ 
\lo{(8.21)}
$$

noting that
$$
q_t(x_1,D)^*=\sum e^{in\o x_1}q^*_{t,-n}(D+n\o e^1)\ ,\
z_t(x_1,D)^*=\sum e^{in\o x_1}z^*_{t,-n}(D+n\o e^1)\ ,\ 
\lo{(8.22)}
$$
so that
$$
\breve{A}_t=U^*(t)\breve{A}(t)U(t)=\breve{q}_t(x_1,D)+\breve{z}_t(x_1,D)
\lo{(8.23)}
$$
with
$$
\breve{q}_t(x_1,\xi)=\sum \breve{q}_{t,n}(\xi)e^{in\o x_1}\ ,\
\breve{z}_t(x_1,\xi)=\sum \breve{z}_{t,n}(\xi)e^{in\o x_1}\ .
\mbox{ where }
\lo{(8.24)}
$$
$$
\breve{q}_{t,n}(\xi)=\halb\{q_{t,n}(\xi)+q^*_{t,-n}(\xi+n\o e^1)\}\ ,\
\breve{z}_{t,n}(\xi)=\halb\{z_{t,n}(\xi)+z^*_{t,-n}(\xi+n\o e^1)\}\ .\
$$
In particular note that
$$
A(t)=q(D)+\breve{z}_0(x_1-t,D)    \ ,\ 
\lo{(8.25)}
$$
with $\breve{z}_t$ of (8.24) for $t=0$, now is self-adjoint, hence counts as
an observable.

On the other hand, it is important to emphasize that we no longer have
$[h_0(\xi),\breve{q}_t(x_1,\xi)]\newline
=[h_0(\xi),\breve{z}_t(x_1.\xi)]_+=0$,
although both still are symbols of one order lower than required.

V) Going into momentum space again, we find that
$$
\breve{q}_t(x_1,D)^\wdg=\halb\sum  T_{-n\o}\{q_{t,n}(D)+
q^*_{t,-n}(D+n\o e^1)\}\ ,\
\lo{(8.26)}
$$
$$
\breve{z}_t(x_1,D)^\wdg=\halb\sum T_{-n\o}\{z_{t,n}(D)+
z^*_{t,-n}(D+n\o e^1)\}\ ,\
$$

\end{thm}

In contrast to our procedure of previous sections --- where we were simplifying
previously published things, we shall attempt to discuss a full proof of thm.8.3
in sections below.

\section{The Photon Hypothesis}

Note, in thm.8.3 we were including the Fourier transformed operators, defined as 
$A^\wdg=FAF^{-1}$ for an important reason: This will transform us to the
\emph{momentum representation}, where the momentum observables $D_j$
appear as multiplication operators $\psi^\wdg \rta \xi_j\psi^\wdg(\xi)$.
Formally, a $\psi do \ \ \  a(x,D)$ will have $(a(x,D))^\wdg=a(-M_r,D)$, with
notation as in (3.9). Especially, we get 
$$
(e^{-in\o x_1}a(D))^\wdg = T_{n\o}a(\xi)\ .
\lo{(9.1)}
$$

This latter formula we find interesting: Looking at (8.20) it appears that , for a
$q(D)$  as in thm.8.3 the Heisenberg transformed $(a_t(x,D))^\wdg$ splits up into
a (discrete) sum of terms consisting of products $T_{n\o}f(\xi)$. So, these terms
have their momentum variable translated by an integer multiple of $n\o$ in the
$x_1$-direction ---  the direction of our radiation. Recalling our constants
$\hbar=c=m_e=|e|=1$, we get  dimensions right when we claim this $n \o$
as an integer multiple of $\hbar\o/c=h\nu/c$. With that, there arises the
suspicion that this points to a collision of the electron (positron) with 
a discrete number of particles, all having momentum $h\nu/c$ --- so, with
\emph{Photons} ?

We will work on such assumption, when we now sketch a proof of thm.8.3,
focusing on the special case of $q(\xi)=\xi_j\ ,\ j=1,2,3.$
At the same time this will prepare us for the proof of the general case.

Recalling the operator $K=H(0)-D_1$ of (8.7), we consider the expression
$A_t=e^{iKt}Ae^{-iKt}$ and assume that $A_t=a_t(x,D)$ is a $\psi do$, for
all $t$, and then write
$$
\dot{a}_t(x,D)=i[K,a_t(x,D)]\ ,\ 
\lo{(9.2)}
$$
then seeking to write this symbolwise, assuming that we work with symbols
$a(x_1,\xi)\in\psi p$, as defined in def.2.3(iii), independent of $x_2,x_3$.

\begin{prop}

For a $\psi do$ $C=c(x_1,D)\in\psi p_m$ we have
$$
symbol([K,C])=[h_0(\xi),c(x_1,\xi)]-\e_0\sin\o x_1 [\a_2,c(x_1,\xi)]-
i(\a_1-1)c_{|x_1} -\frac{i\e_0}{2}\a_2Xc(x_1,\xi)\ ,\ \mbox{ where }
$$
$$
h_0(\xi)=\a\xi+\b\ ,\ 
Xc(x_1,\xi)=\{(c(x_1,\xi+\o e^1)-c(x_1,\xi))e^{i\o x_1}
                     +(c(x_1,\xi)-c(x_1,\xi-\o e^1))e^{-i\o x_1}\}\ .
\lo{(9.3)}
$$

\end{prop}

\noindent
\tbf{Proof.}
For $H_0=h_0(D)$ we get 
$$
\mbox{ symbol }([H_0,C])=[h_0(\xi),c(x_1,\xi)] -i\sum_j \a_j c_{|x_j}(x_1,\xi)\ ,\
\mbox{ symbol }([D_1,C])=-ic_{|x_1}(x_1,\xi)\ .
  \lo{(9.4)}
$$
by using the Leibniz formula (3.3) (with the infinite series there breaking off). For
the term $\e_0\a_2\sin\o x_1$ we proceed directly.
For $[\sin\o x_1\ ,c(x_1,D)]$ get
$$
c(x_1,D)(u(x)\sin\o x_1)=\kpd\int d\xi
e^{ix\xi}(u\sin\o x_1)^\wdg(\xi)c(x_1,\xi)\ ,\
\lo{(9.5)}
$$
where $(ue^{\pm\o ix_1})^\wdg(\xi)=\kpd\int dx u(x_1)e^{-ix(\xi\mp\o e^1)}
=u^\wdg(\xi\mp\o e^1)$, hence

$(u\sin\o x_1)^\wdg=\frac{i}{2}(u^\wdg(\xi+\o e^1)-u^\wdg(\xi-\o e^1))$\ ,\
so that
$$
c(x_1,D)(u(x)\sin\o x_1)=\inv{2i}\kpd\int d\xi e^{ix\xi}
\{e^{i\o x_1}c(x_1,\xi+\o e^1)-e^{-i\o x_1}c(x_1,\xi-\o e^1)\}u^\wdg(\xi)\ .\
\lo{(9.6)}
$$
Accordingly $[\sin\o x_1,c(x_1,D)]$ has the symbol
$$
\frac{i}{2}\{e^{i\o x_1}(c(x_1,\xi+\o e^1)-c(x_1,\xi))+
e^{-i\o x_1}(c(x_1,\xi)-c(x_1,\xi-\o e^1))\}\ .
\lo{(9.7)}
$$
So, we get (9.3), q.e.d.

With prop.9.1 and (9.2)  we then conclude that the symbol $a_t$ of
$A_t=e^{iKt}a(x_1,D)e^{-iKt}$ must satisfy the equation
$$
\dot{a}_t(x_1,\xi)=i[h_0(\xi),a_t(x_1,\xi)]
+(\a_1-1)a_{t|x_1}(x_1,\xi) + (Za_t)(x_1,\xi)\ ,\ 
\lo{(9.8)}
$$
$$
\mbox{ with      } (Zc)(x_1,\xi)=-i\e_0\sin\o x_1 [\a_2,c(x_1,\xi)]
+\frac{\e_0}{2}\a_2(Xc)(x_1,\xi)\ ,\
$$
assuming that $A_t$ and $\dot{A}_t$ belong to $ Op\psi p$.

We note that (9.8) is a differential equation in the variables $t,x_1$,
but also is governed by the commutator $[h_0,a_t]$ representing a term
of order $m+1$, assuming $a_t\in\psi p_m$. Decomposing again
$$
a_t=a_t^+ + a_t^- + a_t^\pm + a_t^\mp\ ,\ \mbox{ where } a_t^+=p_+a_tp_+\ ,\
a_t^-=p_-a_tp_-\ ,\ a_t^\pm=p_+a_tp_-\ ,\ a_t^\mp=p_-a_tp_+\ ,
\lo{(9.9)}
$$
we get
$$
([h_0,a_t])^\pm=2\<\xi\>a_t^\pm\ ,\ ([h_0,a_t])^\mp=-2\<\xi\>a_t^\mp\ ,\
([h_0,a_t])^+=([h_0,a_t])^-=0\ .
\lo{(9.10)}
$$
With $q_t=a_t^+ + a_t^-\ ,\ z_t=a_t^\pm+a_t^\mp$ we get
$a_t=q_t+z_t$ where $[h_0,q_t]=0\ ,\ [h_0,z_t]_+=0.$

Since all terms in (9.8) but the commutator-term are of order $m$ or less
we conclude that
$$
z_t=\inv{2\<\xi\>}\{([h_0,a_t])^\pm-([h_0,a_t])^\mp\}\in\psi p_{m-1}\ .
\lo{(9.11)}
$$
So, we have proven this:

\begin{prop}

If an operator $A=a(x_1,D)\in \psi p_{m}$ has the above property that
$A_t=e^{iKt}Ae^{-iKt}=a_t(x_1,D)$ (mod $\psi q_{-\i}$),
where $a_t$ and $\dot{a}_t$ belong to
$\psi p_m$ (mod $\psi q_{-\i}$) then
(9.9),(9.10),(9.11) lead to a decomposition
$a_t(x_1,\xi)=q_t(x_1,\xi)+z_t(x_1,\xi)$ where
$q_t\in\psi p_m\ ,\ z_t\in\psi p_{m-1}$
all (mod $\psi q_{-\i}$) while
$[h_0,q_t]=0\ ,\ [h_0,z_t]_+=0.$

In particular this decomposition applies to the case $t=0$, so that
also (mod $\psi q_{-\i}$) $a(x_1,\xi)=q(x_1,\xi)+z(x_1,\xi)$ where
$q\in\psi p_m\ ,\ z\in\psi p_{m-1}$ while
$[h_0,q]=0\ ,\ [h_0,z]_+=0.$

\end{prop}

\bigskip

Vice versa, focusing on construction of $\psi do$-s of the form $a(D)$ with
$e^{iKt}a(D)e^{-iKt}\in\psi p$, it is clear then that we
might start with $[h_0,a]=0$, and then have to
add
a "lower order correction" $z(x_1,\xi)\in
\psi p_{m-1}$ (and with $[h_0,z]_+=0$) to make above equ. (9.8) possible.

For this task we will use 
an iteration, starting with a given initial self-adjoint $q(\xi)$
commuting with $h_0(\xi)$, the construction seeking for a $z_t$ of lower
order and a commuting $q_t$ with $q_0=q$ such that $a_t=q_t+z_t$ will
solve (9.8) with higher and higher accuracy, as $|\xi|\rta\i$.

\bigskip

Remembering that (9.8) is an equation for a $4\x 4$ matrix-function $a_t$
we distinguish three steps, to be iterated infinitely:

\begin{quotation}

Step I We omit some lower order terms in (9.8), then trying to solve that as a
sharp equation.

Step II: We multiply the (simplified)  (9.8) left and right by $p_+$ (and left and right by
$p_-$) obtaining two differential equations to be solved. That will
get us an approximate $q_t$.

Step III: We multiply (9.8) left and right by $p_+$ and $p_-$, respectively
(or by $p_-$ and $p_+$, resp.). That will give us equations to obtain an
approximate $z_t$.

\end{quotation}.

These steps, applied alternately, in iteration, will result
in an infinite sequence of
improvements satisfying eq. (9.8) modulo $\psi p_{m-j}$ only, for
$j=1,2,\ldots$. Then an asymptotic limit (mod $\psi q_{-\i}$ (in the sense of prop.3.7)
must be taken to obtain an $a_t^\i=q_t^\i+z_t^\i$
solving (5.6) modulo $\psi p_{-\i}$.

With such $a_t^\i(x_1,\xi)\in\psi p_m$ we then define the operator
$A_t^\i=a_t^\i(x_1,D)$, and then define
$$
B_t=e^{-iKt}A_t^\i e^{iKt}-A_0^\i\ .
\lo{(9.12)}
$$
Clearly we get $B_0=0$, while
$$
\dot{B}_t=e^{-iKt}C_te^{iKt}\ ,\ C_t=\dot{A}_t^\i-i[K,A_t^\i]\ .
\lo{(9.13)}
$$
Here the expression $C_t$ belongs to $Op\psi p_{-\i}$,
since its symbol satisfies (9.8) modulo $\psi p_{-\i}$.
It follows that
$$
e^{-iKt}A_t^\i e^{iKt}-A_0^\i=B_t=\int_0^t d\t e^{-iK\t}C_\t e^{iK\t}\ ,\
\lo{(9.14)}
$$
hence
$$
e^{iKt}A_0^\i e^{-iKt}=A_t^\i -\int_0^t e^{i(t-\t)K}C_\t e^{-i(t-\t)K}\ .
\lo{(9.15)}
$$

Here we are facing a slight difficulty: 

\begin{obs}

Note, the above $C_t$ is the error occurring in our procedure
of solving the ODE-initial-value problem (8.16). That error belongs to $Op\psi q_{-\i}$ ---
its differentiation order is $-\i$. Since it is a $\psi do$,  its momentum representation [i.e., its Fourier transform] 
only provides a negligible contribution if applied to functions with support for very large $\xi$.

On the other hand, the error $A_t^\i-e^{iKt}A_0^\i e^{-iKt}=\G_t$ is given by
$$
\G_t=\int_0^t e^{i\t K}C_{t-\t} e^{-i\t K}\ .
\lo{(9.16)}
$$
We shall show in sec. 13, below, that this kind operator belongs to 
$Op\psi q_{-\i}$ if we assume that  $P_+C_\t P_-=P_-C_\t P_+=0$ for all $\t\in [0,t]$, where
$P_+\ ,\ P_-$ denote the orthogonal projections onto the spaces $\mclh_e$ and $\mclh_p$
of  electron (positron) states, resp.

The projections $P_+\ ,\ P_-$, as spectral projections of $K$, commute with $K$ and with $e^{iKt}$.
Thus, if we introduce a `commuting part' $\kp_c(R)=P_+RP_++P_-RP_-$ , for general operators $R$,
then we get
$$
e^{iKt}\kp_c(A_0^\i)e^{-iKt}=\kp_c(A_t^\i)+ \G_t^\i\ ,\ 
\lo{(9.17)}
$$
where then $\G_t^\i=\int_0^t e^{i\t K}\kp_c(C_{t-\t}) e^{-i\t K}\ \ \in Op\psi q_{-\i}$ 
also is a $\psi do$, so that  the right hand side of (9.17) indeed is a $\psi do$ in  $Op\psi q_m$.
 
\end{obs}

We shall see later that $P_+\ ,\ P_-$ are $\psi do$-s  in $Op\psi q_0$, and that the passage 
$R\rta \kp_c(R)$ to the commuting part may be carried into the infinite series of thm 8.3 with
little or no change. In particular, the discussion in thm. 10.4, involving only the first and second terms
of these infinite series' --- i.e., only a single collision between a Dirac particle and a photon ---
will not be affected at all.

Actually, the projections $p_+(D)\ ,\ p_-(D)$ used in our iteration are close to $P_+$ and $P_-$, resp.,
as shall be seen, so that the commuting terms at each step of the iteration are almost commuting with
respect to $P_+\ ,\ P_-$.

\bigskip

It is easy then to return to our propagator $U(t)=T_{-t}e^{-iKt}$ of the Dirac operator (8.1): Just rewrite
(9.17) as
$$
U^*(t)(T_{-t}\kp_c(A_0^\i)T_t)U(t)=\kp_c(A_t^\i)+ \G_t^\i\ .\ 
\lo{(9.18)}
$$

Setting $\acute{A_t}=\kp_c(A_t^\i)=\acute{a_t}(x_1,D)\in Op\psi q_m$ we shall get
$T_{-t}\kp_c(A_0^\i)T_t=\acute{a}_0(x_1-t,D)\in Op\psi q_m$.

\begin{prop}

We have
$$
U^*(t)\acute{a}_0(x_1-t,D)U(t)=\acute{a}_t(x_1,D)+\G_t^\i\ \ \mbox{\ with\ } \G_t^\i\in Op\psi q_{-\i}\ .
\lo{(9.19)}
$$

\end{prop}

Here the problem remains to relate $\acute{a}_0(x_1,t)$  to the given symbol $q(\xi)$ of thm.8.3. We shall discuss
that in more detail in sec 13, after we control the operators $P_+\ ,\ P_-$.

\section{The Momentum Observables $D_1,D_2,D_3$}

Focusing on the 3
momentum coordinates as observables, we start with the initial self-adjoint
symbol $q(\xi)=\xi_j\in\psi c_{(m,0)}$ with $m=1$ , for fixed $j=1,2,3$.
where $j=1$ will give the momentum coordinate in the direction of our
radiation.  In particular we recall (8.13), i.e.,
$$
(H(t))_t-H(0)=(D_1)_t-D_1\ ,
\lo{(10.1)}
$$
indicating a relation between the development of the observables $H(t)$
and $D_1$, looking at their Heisenberg transforms.

We then want to apply thm.8.3 to the special cases of $q(\xi)=\xi_j\ ,\ j=1,2,3$,
and also discuss the details of the iteration, completing the proof of thm.8.3.

So, in (9.8), we set 
$a_t=q_t+z_t$, where $q_t\in\psi p_1\ ,\ z_t\in\psi p_0$ and
$[h_0(\xi),q_t(x_1,\xi)]=0\ ,\ [h_0(\xi),z_t(x_1,\xi)]_+=0$, for all
$x_1,\xi$. In that substitution we tend to ignore all terms of order $m-1 \ (=0$ for $q=\xi_j$).
In addition, $\dot{z_t}$ also will be regarded as of order $m-1$, and will be
ignored, a fact to be confirmed later on, after solving for $q_t,z_t$
modulo $\psi p_0$ --- assuming that initially, at $t=0$, we have
$q_0(x_1,\xi)=\xi_j\ ,\ j=1,2,3$.

\begin{prop}

The operation $c(x,\xi)\rta (Xc)(x,\xi)$ (with $X$ of (9.3))
lowers the differentiation
order $m$ of $c\in \psi p_m$ by one unit -- to $\psi p_{m-1}$.

Also, if a symbol $M(x,\xi)$ commutes with $h_0(\xi)=\a\xi+\b$ then
we get

$p_+[\a_2,M]p_+(x,\xi)=p_-[\a_2,M]p_-(x,\xi)=0\ .$

\end{prop}

Indeed, looking at (9.3) we observe that
$c(x,\xi+\o e^1)-c(x,\xi)=\int_0^\o d\kp c_{|\xi_1}(x,\xi+\kp e^1)$
has differentiation order $m-1$ if $c(x,\xi)$ has order $m$. Similar
with the second term in (9.3), so that $(Xc)$ has order $m-1$.
For the second statement we observe that
$p_+[\a_2,M]p_+=[p_+\a_2p_+,M]$, since $[h_0,M]=0$ implies $[p_+,M]=0$.
But we know that $p_+\a_2p_+=-p_-\a_2p_-=s_2(\xi)=\xi_2/\<\xi\>$ is a
scalar (cf. lemma 2.1). So $p_+[\a_2,M]p_+=[p_+\a_2p_+,M]
=[s_2,M]=0$. Similar for $p_-$ confirming the statement.

\bigskip

We get
$$
\dot{q}_t=i[h_0,z_t]+(\a_1-1)q_{t|x_1}+Z(q_t)\ \ (\mbox{  mod }\psi p_0 )\ .
\lo{(10.2)}
$$

Here we apply the multiplication $p_+\{XX\}p_+$ of `step II', noting that
$p_+[h_0,z_t]p_+=0$, and that
$p_+Z(q_t)p_+\in\psi p_0$, due to prop.10.1, so that (10.2) simplifies to
$$
\dot{q}^+_t=(s_1-1)q^+_{t|x_1}\ \ (\mbox{  mod }\psi p_0)\ .
\lo{(10.2')}
$$
The sharp D.E. (10.2') with initial-value $q^+_0(x_1.\xi)=\xi_jp_+(\xi)$
has the unique solution $q_t^+(x_1,\xi)=\xi_jp_+(\xi)$.
Similarly we get $q_t^-(x_1,\xi)=\xi_j p_-(\xi)\ .$

So, we will get just
$$
q_t(x_1,\xi)=q_t^+(\xi)+q_t^-(\xi)=\xi_j(p_+(\xi)+p_-(\xi))=\xi_j
\ ,\ j=1,2,3\ .
\lo{(10.3)}
$$
Next we apply step III - multiplying $p_+\{XX\}p_-$ with
$a_t=q(\xi)+z_t(x_1,\xi)$ in (9.8), using that $q_t$ is independent of
$x$ and $t$, and that
$$
p_+[h_0,c]p_-=2\<\xi\>c^\pm\ ,\ p_-[h_0,c]p_+=-2\<\xi\>c^\mp\ ,
\lo{(10.4)}
$$
we get
$$
\dot{z}_t^\pm=2i\<\xi\>z_t^\pm +((\a_1-1)z_{t|x_1})^\pm
-i\e_0\sin\o x_1\{[\a_2^\pm,q]+2s_2(\xi)z_t^\pm\}
+\frac{\e_0}{2}(\a_2Xa_t)^\pm\ .
\lo{(10.5)}
$$
Assuming that $\dot{z}_t$ also is of order $m-1$
and omitting all terms of order $m-1$ this reads
$$
2i\<\xi\>z_t^\pm =i\e_0\sin\o x_1[\a_2^\pm,q_t]
\mbox{  (modulo }\psi p_{m-1})\ .\ 
\lo{(10.5')}
$$
Since division by $\<\xi\>$ lowers the order by $1$
we thus get (also, repeating the procedure with $p_-\{XX\}p_+$)
$$
z_t^\pm=\frac{\e_0}{2\<\xi\>}[\a_2^\pm(\xi),q(\xi)]\sin\o x_1\in\psi p_{m-1}
\ ,\ 
z_t^\mp=-\frac{\e_0}{2\<\xi\>}[\a_2^\mp(\xi),q(\xi)]\sin\o x_1\in\psi p_{m-1}
\ .
\lo{(10.6)}
$$
Both, $z_t^\pm$ and $z_t^\mp$ are approximations modulo $\psi p_{m-2}$,\ m=1,
to be improved in the next iteration.

\begin{rem}

Note that our $z_t^\pm\ ,\ z_t^\mp$ of (10.6) also are independent
of $t$, just as the $q_t=q$, so that $\dot{z}_t=0$ while also
$z^\pm_{t|x_1}\ ,\ z^\mp_{t|x_1} \in\psi p_{m-1}$, so that (10.5')
indeed is satisfied modulo $\psi p_{m-1}$.

In our special case where $q(\xi)$ is scalar -- so that it commutes with
the matrices $\a^\pm(\xi)$ -- we even get $z_t^\pm=z_t^\mp=0$.

\end{rem}

With $z_t^\pm=z^\pm\ ,\ z_t^\mp=z^\mp$  of (10.6) (independent of $t$)
we then get
$$
z_t=z^\pm+z^\mp+z_t^++z_t^- \ ,\ 
\lo{(10.7)}
$$
where $z_t^+\ ,\ z_t^-\in\psi p_{m-1}$ still remain undetermined --
they will be fixed in the next iteration.

\bigskip

For the next iteration we return to steps I and II: With above $q_t=q$ and $z_t$
of (10.7) we set
$$
a_t=(q+z_t) +v_t \ ,\ \mbox{ where } v_t\in \psi c_{m-2}\ ,
\lo{(10.8)}
$$
recalling that $z_t$ still has the free symbols $z_t^+$ and $z_t^-$
 belonging to $\psi p_{m-1}$, so that we may assume $v_t^+=v_t^-=0$.
Substituting into (9.8) and multiplying $p_+\{XX\}p_+$ we get

$$
\dot{z}_t^+=(s_1(\xi)-1)z_{t|x_1}^+        
+\a_1^\pm(\xi)z^\mp_{t|x_1}+\a_1^\pm(\xi)v^\mp_{t|x_1}
+(Z(q+z_t+v_t))^+(x,\xi)\ ,\
\lo{(10.9)}
$$
where we used that $\dot{q}^+=q_{|x_1}^+=v_t^+=0$. We want to look at
(10.9) modulo $\psi p_{m-2}$, hence will drop all terms of order $m-2$:
$$
\dot{z}_t^+=(s_1(\xi)-1)z_{t|x_1}^+        
+\a_1^\pm(\xi)z^\mp_{t|x_1}
+(Z(q+z_t^\pm+z_t^\mp))^+(x_1,\xi)\ ,\
\lo{(10.10)}
$$
keeping in mind that $z_t$ is independent of $x_2,x_3$, also that - for 
$c_t=z_t^+,z_t^-$ we have $Z(c_t)^+$ of order $m-2$, by prop.10.1.

\bigskip

Relation (10.10) again will be regarded as a differential equation for
$z_t^+$. We may write it as
$$
\p_tz^+_t(x_1-t(s_1(\xi)-1),\xi)=F_t(x_1-t(s_1(\xi)-1),\xi)\ ,\
\lo{(10.11)}
$$
$$    
\mbox{ with   }
F_t(x_1,\xi)=\a_1^\pm(\xi)z^\mp_{t|x_1}(x,\xi)+(Z(q+z_t^\pm+z_t^\mp))^+(x_1,\xi)\ .\
$$
This (with initial value $z^+_0(x_1,\xi)$) is solved by integration; we get
$$
z_t(x_1-t(s_1(\xi)-1),\xi)=z_0^+(x_1,\xi)
+\int_0^t d\t F_\t(x_1-\t(s_1(\xi)-1),\xi)\ .\
\lo{(10.12)}
$$
Substituting $x_1-t(s_1(\xi)-1)$ by $x_1$ will give us
$$
z^+_t(x,\xi)=z^+_0(x_1+t(s_1(\xi)-1),\xi)
+\int_0^t d\t F_\t(x_1+(t-\t)(s_1(\xi)-1),\xi)\ .
\lo{(10.13)}
$$
We assume $z^+_0=0$ as to leave the original commutative
part $q=q_0$ untouched. Then we get
$$
z^+_t(x_1,\xi)=\int_0^t d\t F_{t-\t}(x_1+\t(s_1(\xi)-1),\xi)\ .
\lo{(10.14)}
$$
We still simplify our $F_t$ of (10.11), omitting more terms of order $m-2$:
Write
$$
p_+Z(q+z_t^\pm+z_t^\mp)p_+=\frac{\e_0}{2}p_+(\a_2X(q+z_t^\pm+z_t^\mp))p_+
-i\e_0\sin\o x_1 p_+[a_2,q+z_t^\pm+z_t^\mp]p_+\ .
\lo{(10.15)}
$$
Applying prop.10.1 we may omit $z_t^\pm$ and $z_t^\mp$ in the first term,
at right and $q$ in the second term, so that
$$
F_t(x_1,\xi)=\a_1^\pm z^\mp_{t|x_1}
+\frac{\e_0}{2}p_+(\a_2X(q))p_+
-i\e_0\sin\o x_1 p_+[a_2,z_t^\pm+z_t^\mp]p_+\ .
\lo{(10.16)}
$$
The last term still simplifies\ :\  $p_+[a_2,z_t^\pm+z_t^\mp]p_+
=\a_2^\pm z_t^\mp -z_t^\pm\a_2^\pm$, so, we get
$$
F_t(x_1,\xi)=\a_1^\pm z^\mp_{t|x_1}
+\frac{\e_0}{2}p_+(\a_2X(q))p_+
-i\e_0\sin\o x_1 (\a_2^\pm z_t3^\mp -z_t^\pm\a_2^\pm)\ .
\lo{(10.16')}
$$

Due to (10.6) this $F_t$ is independent of $t$. It belongs to $\psi p_{m-1}$,
and it is a finite sum  $\sum_{j=-2}^{+2}f_j^+(\xi)e^{j\o x_1}$ with
certain $f_j(\xi)\in\psi c_{(m-1,0)}$. We may
write the integrand of (10.14) as
$\sum_{j=-2}^{+2}e^{ij\o(x_1+(s_1(\xi)-1)\t)}f_j^+(\xi)$. So, (6.16) then
assumes the form
$$                                            
z^+_t(x_1,\xi)=\sum_{j=-2}^{+2}e^{ij\o x_1}f_j^+(\xi)
\int_0^t d\t e^{ij\o\t(s_1(\xi)-1)}\ ,\ f_j^+\in\psi p_{m-1}\ .
\lo{(10.17+)}
$$
The integrals $\int_0^t d\t e^{ij\o\t(s_1(\xi)-1)}$ in (10.17+) belong to
$\psi c_0$ -- they may be evaluated explicitly, of course. So, $z_t^+$
of (10.17+) indeed belongs to $\psi p_{m-1}$ .

A similar procedure, using the multiplication $p_-\{XX\}p_-$ will lead to
construction of a $z_t^-$ of the form
$$
z_t^-=\sum_{j=-2}^{+2}e^{ij\o x_1}f_j^-(\xi)
\int_0^t d\t e^{-ij\o\t(s_1(\xi)+1)}\ ,\ f_j^-\in\psi p_{m-1}\ .
\lo{(10.17-)}
$$

Four our iteration it is important to note that, while
the $x_1$-Fourier series expansion of $z^\pm,z^\mp$ extended only from
$-1$ to $+1$ , it now will go from $-2$ to
$+2$. One will see that all future such correction 
symbols have finite sums, but with range increasing while the order decreases
to $-\i$. As a consequence, even the asymptotic infinite sum to be defined
eventually will have  only a finite number of terms not of order $\mu$, for
any $\mu\in\mbb{R}$.

\bigskip

We now have $q_t=q$ and $z_t=z^\pm+z^\mp+z_t^++z_t^-$ completely determined,
up to an error in $\psi p_{m-1}$ and $\psi p_{m-2}$, respectively.
Applying step III again then will result in corrections (mod $\psi p_{m-2}$)
called $v_t^\pm$ and $v_t^\mp$ for $z^\pm$ and $z^\mp$:
we use the multiplication $p_+\{XX\}p_-$,
omitting terms of order $m-1$, getting $v_t^\pm$ as a quotient
$(\psi p_{m-1})/\<\xi\>$, where we must use that $\dot{v_t}\in\psi c_{m-1}$,
and confirm this later on the calculated $v_t$, recalling that division
by $\<\xi\>$ preserves $\psi p$ and lowers the $\psi p$-order by 1.
Similar for $v_t^\mp$ using $p_-\{XX\}p_+$.

After obtaining the corrections $v_t^\pm$ and $v_t^\mp$ we still may introduce
correction symbols $v_t^+\ ,\ v_t^- \in \psi p_{m-2}$ (so far held zero)
together with new corrections $w_t^\pm\ ,\ w_t^\mp\in \psi p_{m-3}$
and start over with step I and step II on $a_t=q+z_t+v_t+w_t$.

We have discussed the above for general $q(\xi)$ to fill in the iteration, used for
the proof of thm.8.3.  It should be clear now, how this will go, and we regard that proof complete.

However, we must remind of the fact that this $a_t(x,\xi)$ of (8.17) only solves the initial value problem
(8.16) modulo $\psi q_{-\i}$; it  will not yet lead to the Heisenberg transform of $a_0(x,D)$ as a $\psi do$
$a_t(x,D)$ modulo $\psi q_{-\i}$. We have indicated the steps necessary in sec.9 (cf. Obs.9.3).
Still, we will continue to also apply thm.8.3 to $q(\xi)=\xi_j$, noting that an argument of sec.13, below
will get us to the same expansion (mod $\psi q_{-1}$) for our Heisenberg transform.

\bigskip

For the special $q(\xi)=\xi_j$ , we have in the present section, we get $z_t^\pm=z_t^\mp=0$.
For $\xi_2,\xi_3$ we just get
$$
a_t(x_1,\xi)=\xi_j\ \ (\mbox{ mod } \psi p_{-1})
\ ,\ \mbox{ for all } t\ ,\ \mbox{ as } j=2,3\ .
\lo{(10.18)}
$$
So, the observables $D_2,D_3$ will not change in time, modulo $\psi q_{-1}$ .

For $q=\xi_1$ (10.14) assumes the form
$$
z^+_t(x_1,\xi)=\e_0\o s_2(\xi)p_+(\xi)\int_0^t d\t\cos\o(x_1+\t(s_1(\xi)-1))
\lo{(10.19+)}
$$
$$
=\frac{\e_0}{2}\o s_2(\xi)p_+(\xi)\{\g_t(\xi)e^{i\o x_1}+
\bar{\g}_t(\xi)e^{-i\o x_1}\}
$$
with
$$
\g_t(\xi)=\int_0^t d\t e^{i\o\t(s_1(\xi)-1)}\ .
\lo{(10.20)}
$$      
Similarly,
$$
z^-_t(x_1,\xi)=-\e_0\o s_2(\xi)p_-(\xi)\int_0^t d\t\cos\o(x_1-\t(s_1(\xi)+1))
\lo{(10.19-)}
$$
$$
=-\frac{\e_0}{2}\o s_2(\xi)p_-(\xi)\{\g_t(-\xi)e^{i\o x_1}+
\bar{\g}_t(-\xi)e^{-i\o x_1}\}\ ,\ 
$$
with $\g_t(\xi)$ of (10.20).

In this way we have calculated our symbol $a_t=q_t+z_t^++z_t^-$ modulo
$\psi q_{-1}$, for the observable $D_1$. Of course there will be terms modulo $\psi p_{-2}\cdots$
with stronger and stronger decay as $|\xi|\rta\i$, but the above
lists all terms of order 0.\, for the operator $D_1$. The $a_t$ thus obtained
will not give a self-adjoint $(D_1)_t$, but we have pointed out how to
remedy this.

We summarize

\begin{thm}

Regarding the symbols $a_t=q_t+z_t$ and $\breve{a}_t=\breve{q}_t+\breve{z}_t$
for the 3 observables $D_1,D_2,D_3$ modulo $\psi q_{-1}$, we get
$$
a_t(x_1,\xi)=\xi_j\ \ (\mbox{ mod } \psi q_{-1})
\ ,\ \mbox{ for all } t\ ,\ \mbox{ as } j=2,3\ ,
\lo{(10.21)}
$$
that is, for j=2,3, we have
$$
a_t(x_1,\xi)=\xi_j\ ,\ z_t=z_t^+=z_t^-=0\ .
\lo{(10.22)}
$$
For $j=1$ we get (as formulas modulo $\psi q_{-1}$) 
$$
a_t(x_1,\xi)=\xi_1+
\frac{\e_0}{2}\o s_2(\xi)\{(\g_t(\xi)e^{i\o x_1}+
\bar{\g}_t(\xi)e^{-i\o x_1})p_+(\xi)
-(\g_t(-\xi)e^{i\o x_1}+
\bar{\g}_t(-\xi)e^{-i\o x_1})p_-(\xi)\}\ ,\ 
\lo{(10.23)}
$$
In particular, calculating mod $\psi q_{-1}$, the correction
term for self-adjointness of $a_t(x,D)$ also vanishes, so that 
$a_t(x_1,D)$ already is self-adjoint modulo $\psi q_{-1}$.

\end{thm}

We the come to the following: 

\begin{thm}

Set $\th(\xi)=\halb(1-s_1(\xi))$, evaluate (above) 
$\g_t(\xi)=te^{-i\o\th(\xi)t}\vi(\o\th(\xi)t)\ ,\ $
 with  $\vi(\kp)=\frac{\sin\kp}{\kp}\ .$

Then we have
$$
(H(t))_t-H(0)=(D_1)_t-D_1=
$$
$$
\e_0\o t\cos(\o(x_1-t\th(D))s_2(D)\vi(\o\th(D)t)p_+(D)
\lo{(10.24)}
$$
$$
 -\e_0\o t\cos(\o(x_1-t\th(-D))s_2(D)\vi(\o\th(-D))p_-(D)\}\ ,
$$
a relation valid modulo $Op\psi q_{-1}$ (also, with $D_1$ in $(D_1)_t$ entered only mod $Op\psi q_{-1}$
--- cf.thm.13.3).

\end{thm}

The proof is a calculation, mainly focusing on self-adjointness
(mod $Op\psi q_{-1}$) of the corresponding operator terms.

\begin{rem}

Recall again: A special argument, as sketched at end of sec.9, accessible only
through the spectral theory of the operator $K$, is needed to derive thm's 10.3
and 10.4, after clearing thm.8.3. This is to be discussed in sec.13, below.

\end{rem}

\begin{obs}

It is clear that the first term at right of (10.23) addresses the electron
part of the state, while the second term addresses positrons.
The symbol of the electron part may be rewritten as
$$
\e_0\o t\cos(\o(x_1-t\th(\xi))s_2(\xi)\vi(\o\th(\xi)t)p_+(\xi)
=\frac{\e_0}{2\th(\xi)}s_2(\xi)\{\sin(\o(x_1-2\th(\xi)t)-\sin\o x_1\}\ .
\lo{(10.25)}
$$
Note the right hand side is a difference of a time-independent term
and a term propagating like a wave with speed $2\th(\xi)$. For
large $|\xi|$ --- as dominant here --- we have 
$s_1(\xi)\approx \xi_1/|\xi|=\cos \l$ , with the angle $\l$ between
the vector $\xi$ and the radiation direction $\xi_1$. It follows that
$2\th(\xi)\approx (1-\cos\l)=2\cos(\l/2)$. In other words, this
propagation speed will display the same dependence on the 
direction as Compton's wave-length dependence (cf. Sommerfeld [So1], p.50).

Clearly this term, marking a single collision with a photon, is
of one order lower than the original observable. The further terms,
(we shall not calculate), will be of lower and lower order, hence
of lesser and lesser probability since we deal with large $|\xi|$.

Notice also: the term (10.25) vanishes for $t=0$, marking the fact,
that we do not need a correction $z(x_1,t)$ for our present $q(\xi)=\xi_j$,
when working only mod $\psi q_{-1}$.

\end{obs}

\section{Spectral Theory of the Operator $K=H(0)-D_1$}

So far, regarding the proof of thm.9.4,  we have solved the differential equation 
$\dot{a}_t =i.\mbox{ symbol }([K,A_t])$ modulo $\psi q_{-\i}$. But, in order to
get back to our desired $A_t=e^{iKt}Ae^{-iKT}=a_t(x,D) \mbox{  ( mod  } \psi q_{-\i}\ )$\ ,
we now will have to involve Fourier integral operators. Actually, we shall get a representation
of $e^{-iKt}$ as a sum of two Fourier integral operators, if we just invoke the spectral theorem
for the self-adjoint operator $K$. In fact, this even brings about the additional advantage that
the two FIO-s obtained are mutually orthogonal in our Hilbert space: their products vanish.

Considering the spectral theory of the operator $K$,
we may separate off the variables $x_2,x_3$, since the coefficients of $K$ are 
only dependent on $x_1$. In other words, we may take the Fourier transform with respect to
$\tilde{x}=(x_2,x_3)$. This leads us to a new operator
$$
K=(\a_1-1)D_1+(\xi_2-\mbf{A}_2(x_1))\a_2+\xi_3\a_3+\b\ .
\lo{(11.1)}
$$
Recall, we are using the matrices $\a,\b$ of (2.9). Thus we may write (11.1) block-matrix-wise as
$$
K=\mat{\ 2i\p\ }{\ ip\ }{-iq\ \ }{\ \ 0\ \ }\ ,\
p=\s_3(\xi_2-\mbf{A}_2(x_1))+\s_2\xi_3-i\ ,\ q=\s_3(\xi_2-\mbf{A}_2(x_1))+\s_2\xi_3+i\ ,
\lo{(11.2)}
$$
with $\p=\p_{x_1}$, this being the $\tilde{x}$-Fourier-transformed operator
$K$ of (8.7).

Writing $\p_{x_1}f=f'$, and $\psi=\vc{u}{v}$,
the equation $K\psi=\l\psi$ dissolves into this:
$$
-2u'-i\l u=pv\ ,\ qu=i\l v\ .\
\lo{(11.3)}
$$      
As earlier, let $P(\t)=\s_3(\xi_2-\mbf{A}_2(\t))+\s_2\xi_3$ . We observe that
$$
pq=1+(\xi_2-\mbf{A}_2(x_1))^2+\xi_3^2=1+P(x_1)^2=\<P(x_1)\>^2\ ,\
\lo{(11.4)}
$$
is a scalar. So, in particular,
$$
p^{-1}=\inv{1+P^2(x_1)}q\ ,\ q^{-1}=\inv{1+P^2(x_1)}p\ .
\lo{(11.5)}
$$
The two equations (11.3) combine into one (scalar) first order
differential equation
$$
u'=-\frac{i}{2}(\l-\inv{\l}\<P\>^2)u
\lo{(11.6)}
$$
for the variable $u$ only. Equation (11.6) is solved by
$$
u(x_1,\tilde{\xi})=e^{-i\frac{\l}{2}x_1+\frac{i}{2\l}
\int_0^{x_1}\<P\>^2(\t)d\t}c\ ,\ c\in\mbb{C}^2\ .
\lo{(11.7)}
$$
Once we have $u$ explicitly we may use the second (11.3) to also get $v$.
All together we get
$$
\psi(x_1,\tilde{\xi},\l)=\vc{u}{v}(x_1,\tilde{\xi})=
\vc{i\l c}{qc}e^{-i\frac{\l}{2}x_1+\frac{i}{2\l}\int_0^{x_1}
\<P\>^2(\t)d\t}\ ,\ c=c(\l,\tilde{\xi})\in \mbb{C}^2\ , 
\lo{(11.8)}
$$
where $c$ is independent of $x_1$.

Looking at (11.8) we observe that $\psi$, as a function of $x_1$, never will
be $L^2(\mbb{R})$, except for vanishing $c$. Thus there will not be any
point-eigenvalues of the operator of $x_1$. On the other hand, there should
be continuous spectrum on all of $\mbb{R}$ since (for $c$ constant in $\l)$
an integral $\int d\l \psi\ ,\ $ will be $L^2(\mbb{R})$
defining a wave-packet.

One might see that there is some `separation at $\l=0$'
in this continuous spectrum, insofar as the function
$\psi(x_1,\tilde{\xi},\l)$ becomes very discontinuous there. Indeed, the
point $\l=0$ here separates the line $-\i<\l<\i$ into the half-lines
$(-\i,0)$ and $(0,\i)$. The corresponding partition of unity
$$
1=P_{(-\i,0)} +  P_{(0,\i)}
\lo{(11.9)}
$$
with spectral projections $P_\D$ of $K$ will generate the split  into
electron states and positron states: We may write (with
$\mcl{H}=L^2(\mbb{R})$)
$$
\mcl{H}_e=\{u\in\mcl{H}\ :\ P_+u=u\}\ ,\
\mcl{H}_p=\{u\in\mcl{H}\ :\ P_-u=u\}\ ,\
\lo{(11.10)}
$$
where
$$
P_+=\tilde{F}^{-1}P_{(0,\i)}\tilde{F}\ ,\ 
P_-=\tilde{F}^{-1}P_{(-\i,0)}\tilde{F}\ ,\ 
\lo{(11.11)}
$$
with the $\tilde{x}$-Fourier transform $\tilde{F}$.

\bigskip
   
We now want to get the explicit spectral projections of $K$ of (11.2).
 A practical way to
achieve this is a technique of complex analysis developed by
Titchmarsh [Ti1].

Recalling the resolvent representation of spectral projections:

For a self-adjoint $N\x N$-matrix $X$ ,
we may obtain the spectral projection
$P_\D$ for any closed interval $\D$ of the real axis by the formula
$$
P_\D=-\inv{2\pi i}\int_\G (X-\l)^{-1}d\l\ ,\
\lo{(11.12)}
$$
where $\G$ denotes any simple closed (positively oriented)
curve in the complex plane encircling all eigen-values
on $\D$ but none of the others. Indeed, this is true,
because, if $\vi_1,\ldots , \vi_N$ denotes an orthonormal base of
eigenvectors to eigenvalue
$\l_1,\ldots,\l_N$ then we may write
$$
(X-\l)^{-1}=\sum_1^N \inv{\l_j-\l} \vi_j\>\<\vi_j\ .
\lo{(11.13)}
$$
Then the residue theorem will imply (11.12).

In case where the two endpoints of the interval $\D=[\l_1,\l_2]$
are not eigenvalues, we may
build such a curve $\G$ from the two complex segments $\D\pm i\e$ \ ,\
with $\e>0$ small and short vertical
connecting segments from $\l_j-i\e$ to
$\l_j+i\e$. It then is evident that we must have
$$
P_\D=-\inv{\pi}\lim_{\e\rta 0,\e>0}\Im\{\int_\D d\l (X-(\l-i\e))^{-1}\}\ ,\
\lo{(11.14)}
$$
setting $\Im{A}=\inv{2i}(A-A^*)$ for any matrix $A$.

Formula (11.14) also holds for unbounded self-adjoint linear operators
like our $K$ above -- for a more detailed discussion note 
the book [Ti1] of Titchmarsh. 

To implement (11.14) for $K$ of (11.2) we set up the resolvent ODE
 $K\psi-\l\psi=\chi\ ,\ \psi=\vc{u}{v}\ ,\ \chi=\vc{f}{g}$, so that
$\chi=(K-\l)^{-1}\psi$. That is, we must solve the system
$$
2iu'+ipv-\l u=f\ ,\ -iqu-\l v=g\ ,
\lo{(11.15)}
$$
simplifying to
$$
2iu'-(\l-\inv{\l}\<P\>^2)u=f+\frac{ip}{\l}g\ ,\
v=-\inv{\l}(g+iqu)\ .
\lo{(11.16)}
$$
We must pick the unique solution in
$L^2(\mbb{R})$\ ,\  assuming that $if-\inv{\l}pg\in L^2(\mbb{R})$:
Here we assume $\l=\mu-i\e\ ,\ \e>0$; then the homogeneous equation 
$-2u'-i(\l-\inv{\l}\<P\>^2)u=0$ is solved by
$$
u=ce^{-\frac{i}{2}(\l x_1-\inv{\l}\rho(x_1))}=
ce^{-\frac{i}{2}x_1(\l -\inv{\l}\io)}
=ce^{-\frac{i}{2}x_1\mu(1-\io/|\l|^2)}
e^{-\inv{2}\e x_1(1+\io/|\l|^2)}
\lo{(11.17)}
$$
with $\rho(x_1)=\int_0^{x_1} \<P(\t)\>^2d\t\ ,$ and $\io(\t)=\rho(\t)/\t$.
Here $u$ of (11.17) and its inverse vanish exponentially as $x_1\rta\i$, and as
$x_1\rta -\i$, respectively. Hence
the solution of (11.16) in $L^2$ will be
$$
u=-\halb e^{-\frac{i}{2}x_1(\l -\inv{\l}\io(x_1))}
\int_{-\i}^{x_1} d\t e^{\frac{i}{2}\t(\l -\inv{\l}\io(\t))}
(if(\t)-\inv{\l}pg(\t))\ ,\ 
v=-\inv{\l}(g+iqu)\ .\
\lo{(11.18)}
$$        

We also need (11.18) for the adjoint $(K-(\mu-i\e))^{-1*}=(K-(\mu+i\e))^{-1}$,
So, we also must set  $\l=\mu+i\e\ ,\ \e>0$.
 Then the $L^2$-solution of the ODE will change to this:
$$
u=\halb e^{-\frac{i}{2}x_1(\l -\inv{\l}\io(x_1))}
\int_{x_1}^{\i} d\t e^{\frac{i}{2}\t(\l -\inv{\l}\io(\t))}
(if(\t)-\inv{\l}pg(\t))\ ,\ 
v=-\inv{\l}(g+iqu)\ .\
\lo{(11.19)}
$$        
We now must take the difference of the two
operators in (11.18) and (11.19),
setting $\l=\mu-i\e$ in (11.18) and $\l=\mu+i\e$ in (11.19), with same
$\mu,\e\ ,\ \e>0$ small; then that
difference should be integrated $d\mu$ over an interval
$\D=[\mu_1,\mu_2] \subset \mbb{R}$, not containing $0$. Then we should let $\e>0\ ,\ \e \rta 0\ ,\ $ to,
finally, get a constant multiple of the spectral projection $P_\D$ for $K$.

We shall set $\l=\mu-i\e$ in (11.18) and
work with $\bar{\l}=\mu+i\e$ in
(11.19). Then we introduce the `Greens- function-type expressions'
$$
H^1(\mu,x_1,\t)=e^{-\frac{i}{2}(\l(x_1-\t)-\inv{\l}(\rho(x_1)-\rho(\t)))}
\mbox{ as } \t<x_1\ ,\
\lo{(11.20)}
$$
$$
H^1(\mu,x_1,\t)=e^{-\frac{i}{2}(\bar{\l}(x_1-\t)-
\inv{\bar{\l}}(\rho(x_1)-\rho(\t)))}
\mbox{ as } \t>x_1\ ,\
$$
$$
H^2(\mu,x_1,\t)=\inv{\l}e^{-\frac{i}{2}(\l(x_1-\t)-\inv{\l}(\rho(x_1)-\rho(\t)))}
\mbox{ as } \t<x_1\ ,\
\lo{(11.21)}
$$
$$
H^2(\mu,x_1,\t)=\inv{\bar{\l}}e^{-\frac{i}{2}(\bar{\l}(x_1-\t)-
\inv{\bar{\l}}(\rho(x_1)-\rho(\t)))}
\mbox{ as } \t>x_1\ .\
$$
and
$$
H^3(\mu,x_1,\t)=\inv{\l^2}e^{-\frac{i}{2}(\l(x_1-\t)-\inv{\l}(\rho(x_1)-\rho(\t)))}
\mbox{ as } \t<x_1\ ,\
\lo{(11.22)}
$$
$$
H^3(\mu,x_1,\t)=\inv{\bar{\l}^2}e^{-\frac{i}{2}(\bar{\l}(x_1-\t)-
\inv{\bar{\l}}(\rho(x_1)-\rho(\t)))}
\mbox{ as } \t>x_1\ .\
$$

With these three functions, let
$$
\o=\vc{w}{z}=-\inv{2\pi i}\{(K-\l)^{-1}-(K-\bar{\l})^{-1}\}\chi\ ,\ \chi=\vc{u}{v}\ .
\lo{(11.23)}
$$
We then get
$$
w=\inv{4\pi}\int_{-\i}^{+\i}H^1(\mu,x_1,\t)f(\t)d\t
+\frac{i}{4\pi}\int_{-\i}^{+\i}H^2(\mu,x_1,\t)p(\t)g(\t)d\t\ ,\
\lo{(11.24)}
$$
$$
z=\frac{\e}{\pi|\l|^2}g(x_1)
-\frac{i}{4\pi}q(x_1)\int_{-\i}^{+\i}H^2(\mu,x_1,\t)f(\t)d\t
+\inv{4\pi}q(x_1)\int_{-\i}^{+\i}H^3(\mu,x_1,\t)p(\t)g(\t)d\t\ .
$$
Here it will be a matter of integrating $d\mu$ over some interval $\D$,
 and then taking limit $\e\rta 0$. No question, the first term of the second
 line will give zero-contribution, while we may take the integral $d\mu$
 inside the integral $d\t$.

Actually, if we assume $f,g\in C_0^\i(\mbb{R})$, instead of in $L^2$ then
we may be quite careless in the order of integration, etc.
Just set $\e=0$ in (11.20)-(11.21)-(11.22). 

For $\e=0$ we get
$$
H^1=e^{-\frac{i}{2}(x_1-\t)(\l -\io^2\inv{\l})}\ ,\ \mbox{ for all }x_1,\t\ ,\
H^{j+1}=\inv{\l^j}H^1\ ,\ j=1,2\ .
\lo{(11.25)}
$$
where $\io^2(x_1,\t)=(\rho(x_1)-\rho(\t))/(x_1-\t)$.

Writing $P_\D=((P_\D^{jl}))_{j,l=1,2}$ as a $2\x 2$-block matrix, acting on
$\xi=\vc{f}{g}$, we get
$$
P_\D^{11}f=\inv{4\pi}\int_\D d\l \int d\t e^{-\frac{i}{2}(x_1-\t)(\l -\io^2\inv{\l})} f(\t)\ ,\ 
$$
$$
P_\D^{12}g=\frac{i}{4\pi}\int_\D\frac{d\l}{\l}\int d\t e^{-\frac{i}{2}(x_1-\t)(\l -\io^2\inv{\l})}p(\t)g(\t) \ ,\
\lo{(11.26)}
$$
$$
P_\D^{21}f=-\frac{i}{4\pi}q(x_1)\int_\D \frac{d\l}{\l}\int d\t e^{-\frac{i}{2}(x_1-\t)(\l -\io^2\inv{\l})}f(\t)\ ,\ 
$$
$$
 P_\D^{22}g=\inv{4\pi}q(x_1)\int_\D \frac{d\l}{\l^2}\int d\t e^{-\frac{i}{2}(x_1-\t)(\l -\io^2\inv{\l})}p(\t)g(\t)\ .
$$

\bigskip

Being in control of the spectral  projections of the operator $K$, we may apply the spectral theorem, for
a representation  $G(K)=\int G(\l) dP_\l$, where $G(\l)$ denotes any function of the real variable $\l$.
Accounting for the singularity at $\l=0$ we write
$$
G(K)\psi=\int_{-\i}^{+\i} G(\l)dP_\l\psi
=\int_{-\i}^{0} G(\l)dP_\l\psi
+\int_{0}^{+\i} G(\l)dP_\l\psi
=(G(K))_-\psi+(G(K))_+\psi\ ,
\lo{(11.27)}
$$
Clearly then we may use (11.26) to express the differential $dP_\l$
by $d\l$. For $(G(K))_+=G=((G_{jl}))_{j,l=1,2}$ and $\psi=\vc{f}{g}$ we then get
$$
(G(K))_+^{11}f=\inv{4\pi}\int_0^\i d\l G(\l)\int_{-\i}^\i d\t e^{-\frac{i}{2}(x_1-\t)(\l -\io^2\inv{\l})} f(\t)\ ,\ 
$$
$$
(G(K))_+^{12}g=\frac{i}{4\pi}\int_0^\i\frac{d\l}{\l} G(\l)\int_{-\i}^\i d\t e^{-\frac{i}{2}(x_1-\t)(\l -\io^2\inv{\l})}p(\t)g(\t) \ ,\
\lo{(11.28)}
$$
$$
(G(K))_+^{21}f=-\frac{i}{4\pi}q(x_1)\int_0^\i \frac{d\l}{\l}G(\l)\int_{-\i}^\i d\t e^{-\frac{i}{2}(x_1-\t)(\l -\io^2\inv{\l})}f(\t)\ ,\ 
$$
$$
(G(K))_+^{22}g=\inv{4\pi}q(x_1)\int_0^\i \frac{d\l}{\l^2}G(\l)\int _{-\i}^\i d\t e^{-\frac{i}{2}(x_1-\t)(\l -\io^2\inv{\l})}p(\t)g(\t)\ ,
$$
and corresponding formulas for $G_-(K)$, where $\int_0^\i d\t$ has been replaced by $\int_{-\i}^0$.

In (11.28) we interchange integrals and write $G_{jl}=(G(K))_+^{jl}\  ,\  G_{jl}^-=(G(K))_-^{jl}$:
$$
G_{11}f(x_1)=\inv{4\pi}\int_{-\i}^\i d\t f(\t)
\int_{0}^\i d\l e^{-\frac{i}{2}(x_1-\t)(\l -\io^2\inv{\l})}G(\l)
\ ,\
$$
$$
G_{12}g(x_1)=\frac{i}{4\pi}\int_{-\i}^\i d\t g(\t)p(\t)
\int_0^\i\frac{d\l}{\l}
e^{-\frac{i}{2}(x_1-\t)(\l -\io^2\inv{\l})}G(\l)
\ ,\
\lo{(11.29)}
$$
$$
G_{21}f(x_1)=-\frac{i}{4\pi}q(x_1)\int_{-\i}^\i d\t f(\t) 
\int_0^\i\frac{d\l}{\l}
e^{-\frac{i}{2}(x_1-\t)(\l -\io^2\inv{\l})}G(\l)\ ,\
$$
$$
G_{22}g(x_1)=\frac{1}{4\pi}q(x_1)\int_{-\i}^\i d\t g(\t)p(\t) 
\int_0^\i\frac{d\l}{\l^2}
e^{-\frac{i}{2}((x_1-\t)(\l -\io^2\inv{\l})}G(\l)\ .
$$
Note, for the term $(G(K))_-$ of (11.27) we get the same
kind of formulas --- the difference being that the inner integral
now extends from $-\i$ to $0$, instead from $0$ to $\i$ :
$$
G^-_{11}f(x_1)=\inv{4\pi}\int_{-\i}^\i d\t f(\t)
\int_{-\i}^{0} d\l e^{-\frac{i}{2}(x_1-\t)(\l -\io^2\inv{\l})}G(\l)
\ ,\
$$
$$
G^-_{12}g(x_1)=\frac{i}{4\pi}\int_{-\i}^\i d\t g(\t)p(\t)
\int_{-\i}^{0}\frac{d\l}{\l}
e^{-\frac{i}{2}(x_1-\t)(\l -\io^2\inv{\l})}G(\l)
\ ,\
\lo{(11.29-)}
$$
$$
G^-_{21}f(x_1)=-\frac{i}{4\pi}q(x_1)\int_{-\i}^\i d\t f(\t) 
\int_{-\i}^{0}\frac{d\l}{\l}
e^{-\frac{i}{2}(x_1-\t)(\l -\io^2\inv{\l})}G(\l)\ ,\
$$
$$
G^-_{22}g(x_1)=\frac{1}{4\pi}q(x_1)\int_{-\i}^\i d\t g(\t)p(\t) 
\int_{-\i}^0\frac{d\l}{\l^2}
e^{-\frac{i}{2}(x_1-\t)(\l -\io^2\inv{\l})}G(\l)\ .
$$

Here we would like to transform the inner integrals. Substitute
$$
\l-\io^2\inv{\l}=2\mu\ ,\ \l=\mu\pm\sqrt{\io^2+\mu^2}\ ,\
d\l=\pm\frac{\l d\mu}{\sqrt{\io^2+\mu^2}}\ ,\
\lo{(11.30)}
$$
to be used with both (11.29) and (11.29-).
With $\l=\mu+\sqrt{\io^2+\mu^2}$ we get an invertible map
$\mu\leftrightarrow\l$ with $\l>0$ and
$$
\l=0\Leftrightarrow \mu=-\i\ ,\
\l=\i \Leftrightarrow \mu=\i\ ,
\lo{(11.31)}
$$
useful for (11.29), while $\l=\mu-\sqrt{\io^2+\mu^2}$ implies
$\l<0$ and gives an invertible map with $\mu\leftrightarrow\l$ and
$$
\l=0\Leftrightarrow \mu=\i\ ,\
\l=-\i \Leftrightarrow \mu=-\i\ .
\lo{(11.31-)}
$$
So, (11.31) is useful for a transformation of (11.29) while
(11.31-) will work for (11.29-).

For the 4 inner integrals $I_{jl}$ we get
$$
I_{11}=\int_{-\i}^\i (1+\frac{\mu}{\sqrt{\io^2+\mu^2}})d\mu
e^{-i\mu(x_1-\t)}G(\mu+\sqrt{\io^2+\mu^2})
\ ,\ 
$$
$$
I_{12}=I_{21}=\int_{-\i}^\i \frac{d\mu}{\sqrt{\io^2+\mu^2}}
e^{-i\mu(x_1-\t)}G(\mu+\sqrt{\io^2+\mu^2})
\ ,\ 
\lo{(11.32)}
$$
$$
I_{22}=\int_{-\i}^\i
\frac{\sqrt{\io^2+\mu^2}-\mu}{\io^2\sqrt{\io^2+\mu^2}}d\mu
e^{-i\mu(x_1-\t)}G(\mu+\sqrt{\io^2+\mu^2})
\ ,\
$$
and
$$
I_{11}^-=\int_{-\i}^\i (1-\frac{\mu}{\sqrt{\io^2+\mu^2}})d\mu
e^{-i\mu(x_1-\t)}G(\mu-\sqrt{\io^2+\mu^2})
\ ,\ 
$$
$$
I_{12}^-=I_{21}^-=-\int_{-\i}^\i \frac{d\mu}{\sqrt{\io^2+\mu^2}}
e^{-i\mu(x_1-\t)}G(\mu-\sqrt{\io^2+\mu^2})
\ ,\ 
\lo{(11.32-)}
$$
$$
I_{22}^-=-\int_{-\i}^\i
\frac{\sqrt{\io^2+\mu^2}+\mu}{\io^2\sqrt{\io^2+\mu^2}}d\mu
e^{-i\mu(x_1-\t)}G(\mu-\sqrt{\io^2+\mu^2})
\ .\ 
$$
We substitute (11.35$\pm$) into (11.32$\pm$) and interchange integrals
again, renaming integration variables $(\t,\mu)\rta (y_1,-\xi_1)$:
$$
G_{11}f(x_1)=\inv{4\pi}\int_{-\i}^\i d\xi_1\int_{-\i}^{+\i}dy_1
e^{i\xi_1(x_1-y_1)}G(-\xi_1+\sqrt{\io^2+\xi_1^2})
(1-\frac{\xi_1}{\sqrt{\io^2+\xi_1^2}})
f(y_1) \ ,\ 
$$
$$
G^t_{12}g(x_1)=\frac{i}{4\pi}\int_{-\i}^\i d\xi_1\int_{-\i}^{+\i}dy_1
e^{i\xi_1(x_1-y_1)}G(-\xi_1+\sqrt{\io^2+\xi_1^2})
\frac{1}{\sqrt{\io^2+\xi_1^2}}p(y_1)g(y_1)\ ,\ 
\lo{(11.33)}
$$
$$
G_{21}f(x_1)=-\frac{i}{4\pi}\int_{-\i}^\i d\xi_1\int_{-\i}^{+\i}dy_1
e^{i\xi_1(x_1-y_1)}G(-\xi_1+\sqrt{\io^2+\xi_1^2})
\frac{1}{\sqrt{\io^2+\xi_1^2}}q(x_1)f(y_1)\ ,\ 
$$
$$
G_{22}g(x_1)=\frac{1}{4\pi}\int_{-\i}^\i d\xi_1\int_{-\i}^{+\i}dy_1
e^{i\xi_1(x_1-y_1)}G(-\xi_1+\sqrt{\io^2+\xi_1^2})
\frac{\sqrt{\io^2+\xi_1^2}+\xi_1}{\io^2\sqrt{\io^2+\xi_1^2}}
q(x_1)p(y_1)g(y_1)\ ,\
$$
and
$$
G^-_{11}f(x_1)=\inv{4\pi}\int_{-\i}^\i d\xi_1\int_{-\i}^{+\i}dy_1
e^{i\xi_1(x_1-y_1)}G(-\xi_1-\sqrt{\io^2+\xi_1^2})
(1+\frac{\xi_1}{\sqrt{\io^2+\xi_1^2}})f(y_1)
\ ,\ 
$$
$$
G^-_{12}g(x_1)=-\frac{i}{4\pi}\int_{-\i}^\i d\xi_1\int_{-\i}^{+\i}dy_1
e^{i\xi_1(x_1-y_1)}G(-\xi_1-\sqrt{\io^2+\xi_1^2})
\frac{1}{\sqrt{\io^2+\xi_1^2}}p(y_1)g(y_1)
\ ,\ 
\lo{(11.33-)}
$$
$$
G^-_{21}f(x_1)=\frac{i}{4\pi}\int_{-\i}^\i d\xi_1\int_{-\i}^{+\i}dy_1
e^{i\xi_1(x_1-y_1)}G(-\xi_1-\sqrt{\io^2+\xi_1^2})
\frac{1}{\sqrt{\io^2+\xi_1^2}}q(x_1)f(y_1)
\ ,\ 
$$
$$
G^-_{22}g(x_1)=\frac{1}{4\pi}\int_{-\i}^\i dy_1\int_{-\i}^{+\i}dy_1
e^{i\xi_1(x_1-y_1)}G(-\xi_1-\sqrt{\io^2+\xi_1^2})
\frac{\sqrt{\io^2+\xi_1^2}-\xi_1}{\io^2\sqrt{\io^2+\xi_1^2}}
q(x_1)p(y_1)g(y_1)\ .\ 
$$

In (11.33) we recall that
$$
\io^2(x_1,y_1)=(\rho(x_1)-\rho(y_1))/(x_1-y_1)
=(\xi_2-c(x_1,y_1)^2+\xi_3^2+a(x_1,y_1)^2\ ,\ 
\lo{(11.34)}
$$
with
$$
c(x_1,y_1)=\inv{x_1-y_1}\int_{x_1}^{y_1}\mbf{A}_2(\t)d\t\ ,\ 
d(x_1,y_1)=\inv{x_1-y_1}\int_{x_1}^{y_1}\mbf{A}_2^2(\t)d\t\ ,\ 
a^2=1+d-c^2\ ,
\lo{(11.35)}
$$
by a calculation.

Notice that (11.33) already gives
(the $\tilde{x}$-F-transform)
of $G(K)$ as a sum of 2 one-dimensional `formal' $\psi do$-s.
To get back to the operator $K$  of (8.7) we must replace 
$\psi(x_1)=\vc{f}{g}(x_1)$ in above formulas by
$$
\inv{2\pi}\int e^{-i\tilde{\xi}\tilde{y}}
\psi(x_1,\tilde{y})d\tilde{y} \ ,\
\lo{(11.36)}
$$
and then apply the inverse
$\tilde{x}$-F-transform to the $G\psi$. 

\begin{thm}

For the operator $K$ of (8.7) and a function $G(\l):\mbb(R)\rta\mbb{C}$  we have $G(K)=(G(K))_++(G(K))_-$ in the sense
of  (11.27) where $(G(K))_+=((G_{jl}))_{j,l=1,2}\ ,\  (G(K))_-=((G_{jl}^-))_{j,l=1,2}$\ ,\ with
$$
G_{11}f(x)=\inv{16\pi^3}\int d\xi\int dy e^{i\xi(x-y)}
G(-\xi_1+\sqrt{\eta^2+a^2})(1-\frac{\xi_1}{\sqrt{\eta^2+a^2}})f(y) \ ,\ 
$$
$$
G_{12}g(x)=\frac{i}{16\pi^3}\int d\xi\int dy e^{i\xi(x-y)}
G(-\xi_1+\sqrt{\eta^2+a^2})\frac{1}{\sqrt{\eta^2+a^2}}p(y_1)g(y)\ ,\ 
\lo{(11.37)}
$$
$$
G_{21}f(x)=-\frac{i}{16\pi^3}\int d\xi\int dy e^{i\xi(x-y)}
G(-\xi_1+\sqrt{\eta^2+a^2})\frac{1}{\sqrt{\eta^2+a^2}}q(x_1)f(y)\ ,\
$$
$$
G_{22}g(x)=\frac{1}{16\pi^3}\int d\xi\int dy e^{i\xi(x-y)}
G(-\xi_1+\sqrt{\eta^2+a^2})
\frac{\sqrt{\eta^2+a^2}+\xi_1}{(\tilde{\eta}^2+a^2)\sqrt{\eta^2+a^2}}
q(x_1)p(y_1)g(y)\ .\
$$
and
$$
G^-_{11}f(x)=\inv{16\pi^3}\int d\xi\int dy e^{i\xi(x-y)}
G(-\xi_1-\sqrt{\eta^2+a^2})
(1-\frac{\xi_1}{\sqrt{\eta^2+a^2}})f(y)\ ,\ 
$$
$$
G^-_{12}g(x)=-\frac{i}{16\pi^3}\int d\xi\int dy e^{i\xi(x-y)}
G(-\xi_1-\sqrt{\eta^2+a^2})
\frac{1}{\sqrt{\eta^2+a^2}}p(y_1)g(y)\ .\ 
\lo{(11.37-)}
$$
$$
G^-_{21}f(x)=\frac{i}{16\pi^3}\int d\xi\int dy e^{i\xi(x-y)}
G(-\xi_1-\sqrt{\eta^2+a^2})\frac{1}{\sqrt{\eta^2+a^2}}q(x_1)f(y)\ ,\ 
$$
$$
G^-_{22}g(x)=\frac{1}{16\pi^3}\int d\xi\int dy e^{i\xi(x-y)}
G(-\xi_1-\sqrt{\eta^2+a^2})
\frac{\sqrt{\eta^2+a^2}-\xi_1}{(\tilde{\eta}^2+a^2)\sqrt{\eta^2+a^2}}
q(x_1)p(y_1)g(y)\ .\ 
$$
again using the vectors $\eta=(\xi_1,\xi_2-c(x_1,y_1),\xi_3)\ ,\ \tilde{\eta}=(\eta_2,\eta_3)$

\end{thm}

Clearly the operators $G(K)_\pm$ are formal $\psi do$-s , with their symbol 
containing the factors $G(- \xi_1\pm\sqrt{\eta^2+a^2})$. But it will depend
on the choice of the function $G(\l)$ whether these will be operators
belonging to one of our classes $Op\psi q_m$ . We shall find that true if
we choose $G(\l)\equiv 1$ but false for $G(\l)=e^{-i\l t}$. In the latter case
the operators assume a form we shall call FIO-$\psi do$s.

\section{A  Class of Global Fourier Integral Operators}

The functions $G(\l)$, most important for us here, are
$G(\l)\equiv 1\ ,\ G(\l)=\l\ ,\ G(\l)=e^{-i\l t}$. For $G(\l)\equiv 1$ the operators
$(G(K))_\pm$ will give the two projections
$P_+\ ,\ P_-$ separating the spaces of electron and positron states, at $t=0$.

Clearly they appear as formal (left-right-multiplying) $\psi do$-s
$$
P_\pm=p_\pm(M_l,M_r,D)=((p^\pm_{jl}(M_l,M_r,D)))
$$
with the $2\x 2$-block-symbols
$$
p^+_{11}(x,y,\xi)=\halb(1-\frac{\xi_1}{\sqrt{\eta^2+a^2}}) \ \ \ ,\ \ \ 
p^+_{12}(x,y,\xi)=\frac{i}{2}\frac{1}{\sqrt{\eta^2+a^2}}p(y_1)\ ,\ 
\lo{(12.1)}
$$
$$
p^+_{21}(x,y,\xi)=-\frac{i}{2}\frac{1}{\sqrt{\eta^2+a^2}}q(x_1)\ \ \ ,\ \ \ 
p^+_{22}(x,y,\xi)=\frac{1}{2}
\frac{\sqrt{\eta^2+a^2}+\xi_1}{(\tilde{\eta}^2+a^2)\sqrt{\eta^2+a^2}}
q(x_1)p(y_1)\ ,\
$$
and
$$
p^-_{11}(x,y,\xi)=\inv{2}(1+\frac{\xi_1}{\sqrt{\eta^2+a^2}})\ \ \ ,\  \ \ 
p^-_{12}(x,y,\xi)=-\frac{i}{2}\frac{1}{\sqrt{\eta^2+a^2}}p(y_1)\ ,\ 
\lo{(12.1-)}
$$
$$
p^-_{21}(x,y,\xi)=\frac{i}{2}\frac{1}{\sqrt{\eta^2+a^2}}q(x_1)\ \ \ ,\ \ \ 
p^-_{22}(x,y,\xi)=\frac{1}{2}
\frac{\sqrt{\eta^2+a^2}-\xi_1}{(\tilde{\eta}^2+a^2)\sqrt{\eta^2+a^2}}
q(x_1)p(y_1)\ .\ 
$$
We should remind of the fact that we have $c=c(x_1,y_1)\ ,\ d=d(x_1,y_1)$ given by 
(11.35) , or, explicitly, for $\mbf{A}_2(\t)=\e_0\sin\o \t$,  by
$$
c(x_1,y_1)=\e_0\sin\o\frac{x_1+y_1}{2}\phi(\o\frac{x_1-y_1}{2})\ ,\
d(x_1,y_1)=\frac{\e_0^2}{2}-\frac{\e_0^2}{2}\cos\o(x_1+y_1)\phi(\o(x_1-y_1))\ ,\
\lo{(12.2)}
$$
with $\phi(\kp)=\sin\kp /\kp$\ ,\ and that $a^2=1+d-c^2$.

With that it is easily confirmed that we have all the symbols (12.1),(12.1-) belonging to 
$\psi qlr_0$, as defined by the estimates (3.10), so that we verified that
$P_+,P_-\in Op\psi q_0$\ . In addition, with (12.1),(12.1-),  we have obtained explicit $\psi qlr_0$-symbols
of the operators $P_\pm$ --- valid for the Dirac matrices (2.9) only.

Of course, we have $P_+ + P_-=1$. Looking at above symbols (12.1),(12.1-) we note that, indeed,
$ p^+_{11}+p^-_{11}=1\ ,\ p^+_{12}+p^-_{12}=p^+_{21}+p^-_{21}=0$ for all $x,y,\xi$, but
$$
p^+_{22}+p^-_{22}=\frac{q(x_1)p(y_1)}{\eta^2+a^2}\ .
\lo{(12.3)}
$$
The latter symbol is not $\equiv 1$, although it turns out to be $\equiv 1$ for $\mbf{A}_2\equiv 0$,  and, also, 
for general $\mbf{A}_2(\t)$, and $x_1=y_1$. This points to the fact that the assignment $a(x,y,\xi) \rta a(M_l,M_r,D)$ is not bi-unique:
an operator A=$ a(M_l,M_r,D)$ may be represented by many different symbols $b(x,y,\xi)$, exactly one of
them independent of $y$, then giving $a(M_l,M_r,D)=b(x,D)\in Op\psi q$. 

There will be an asymptotic Leibniz formula, to get this $b(x,\xi)$ from $a(x,y,\xi)$:
$$
b(x,\xi)=\sum_{j=0}^\i\inv{j!}\{(-i\p_y\p_\xi)^ja(x,y,\xi)\}_{x=y}
\mbox{   (mod } \psi q_{-\i}\ )\ .
\lo{(12.4)}
$$
Applying this to the symbol (12.3) it is found that the term of order $0$ at right of (12.4) is $\equiv 1$, thus, at least,
confirming that $P_++P_-=1 $  (mod $Op\psi q_{-1}$).

Applying now thm.11.1 to the function $G(\l)=e^{-i\l t}$ we again obtain $e^{-iKt}$ as a sum of two `formal' $\psi do$-s, 
given by  (11.37) and (11.37-). Their $\psi dolr$-symbols are given by 
$$
e^{-it(\xi_1+\sqrt{\eta^2+a^2})}p_+(x_1,y_1,\xi)
\mbox{\ \ \ \ and \ \ \ }  e^{-it(\xi_1-\sqrt{\eta^2+a^2})}p_-(x_1,y_1,\xi)
\mbox{\ \  , resp.\ \ ,\  }
\lo{(12.5)}
$$
 with $p_\pm$ of (12.1).

Evidently the symbols (12.5) do not belong to $\psi qlr$ --- any derivative landing  on the exponential factor producing
no decay in the required sense. However, referring to [Co5], p.53, we observe that these $\psi do$-symbols (12.5) still
belong to the space $ST$ defined there. As a consequence, the `finite-part-singular integrals' defined there still exist;
we have the Beals formulas as well as the Leibniz formulas with integral reminder of ch.1, sec's 4 and 5 valid, although no
asymptotically convergent Leibniz formulas can be derived, for a $\psi do$-calculus.

Actually, a different interpretation then is customary: Following H\"{o}rmander [Hoe4] such operators are written in
the form
$$
Au(x)=\inv{(2\pi)^3}\int d\xi\int dy e^{i\vi(x,y,\xi)}a(x,y,\xi)u(\xi)\ ,\ 
\lo{(12.6)}
$$
with a symbol $a(x,y,\xi)\in\psi q$, as before, and with a (real-valued) `phase function' $\vi(x,y,\xi)$ .
In our present case we will have
$$
\vi(x,y,\xi)=\xi(x-y)+\vi_0(x,y,\xi) \mbox{\ \ \ with\ \ \ } \vi_0=t(\xi_1\pm \sqrt{\eta^2+a^2})\ ;
\lo{(12.7)}
$$
note, we have $\vi(x,y,\xi)\ ,\ \vi_0(x,y,\xi)\in\psi qlr_1$.

H\"{o}rmander introduced the name `Fourier integral operator' (abbrev. FIO) for operators of this form.
One may find an extensive theory of `local' FIO-s --- applicable only to functions defined in a bounded 
subdomain of $\mbb{R}^3$, or also on a compact manifold [cf. also Egorov [Eg1], Maslov[Ms1], Buslaev[Bu1]
for development of general ideas ]. When applied to a function $u(x)$, local FIO-s will move singularities
of $u(x)$. A given local $FIO$ can be given by many different symbols and phase functions.  Composition of
two local FIO-s will give a local FIO again, with construction of new phase function and symbol involving an interesting
but complicated theory, not concerning us here.

The kind of `global' FIO-s over $\mbb{R}^3$, we have here, has been studied by Sandro Coriasco [Cr1] [Cr2],
although only for phase functions and symbols in $\psi c$ --- not in $\psi q$, as we require. In [Cr1] we find results
for composition of our kind of FIO-s, but only for special phase functions: They cover the case of $AB\ ,\ BA$ where
$B$ is a $\psi do$ [it has $\vi_0\equiv 0$], and also the case $A^*BA$, again with $B$ a $\psi do$: Then $A^*BA$ also
is a $\psi do$. Essential ingredient of the discussion is the fact that $e^{i\psi(x,y,\xi)}$ is a symbol in $\psi clr_0$
whenever $\psi(x,y,\xi)\in \psi c_0$, while this is not true for $\psi$ of order $>0$.

Although Coriasco discusses only the case where symbols and phase functions belong to $\psi clr$, we find that his
results extend to symbols and phase functions in $\psi qlr$ if the asymptotic convergence modulo $\psi c_{-\i}$
in his results is replaced by asymptotic convergence modulo $\psi q_{-\i}$. We state the result required here in thm.12.1, below,
without discussing the (very technical) proof, strongly leaning on Coriasco's methods.
 That proof uses the finite part integral, and our `Leibniz formulas with 
integral reminder' of [Co5], ch.1--- still valid here, as noted above. More details about that proof may be found in [Co17].

\begin{thm}

Let $C=c(x,D)\in Op\psi q_{m}$, and let $P_\pm$ be the two projections (12.1),(12.1-). Then we have
$$
e^{iKt}P_+CP_+e^{-iKt}\in Op\psi q_{m}\ \ \ ,\ \ \ e^{iKt}P_-CP_-e^{-iKt}\in Op\psi q_{m}\ ,\ 
\lo{(12.8)}
$$
and, likewise,
$$
U^*(t)P^t_+C_tP^t_+U(t)\in Op\psi q_{m}\ \ \ ,\ \ \ U^*(t)P^t_-C_tP^t_-U(t)\in Op\psi q_{m}\ ,\ \ C_t=T_{-t}CT_t\ ,
\lo{(12.9)}
$$
for the propagator $U(t)$ of our Dirac equation, with $H(t)$ of (8.1), and the projections 
$P^t_\pm=T_{-t}P_\pm  T_t$ onto the electron (positron) spaces $\mclh_e(t)\ ,\ \mclh_p(t)$ at time $t$ of (8.11).

\end{thm}

\section{Returning to the Heisenberg Transform}

Finally, after gaining control on the FIO-analysis of the operators $e^{-iKt}$ we now may address
the gap between thm.8.3 and its application to obtain the operator $U^*(t)a_0^\i(x,D)U(t)$ as a $\psi do$
in $Op\psi q$. We had pointed out at the end of sec.9 that we should replace $A_t^\i=a_t^\i(x,D)$ by the operator
$\kp_c(A_t^\i)=P_+A_t^\i P_+ + P_-A_t^\i P_-$, then landing at (9.17), with its remainder $\G_t^\i\in Op\psi q_{-\i}$ .
Then, however, we should have to carry the operation $R\rta \kp_c(R)$ into the asymptotic expansions (mod $\psi q_{-\i}$)
of thm.8.3. In particular we already stated that the initial expansions of thm's 10.3 and 10.4 will not change by passing
from $A_t^\i$ to $\kp_c(A_t^\i)$.

\begin{obs}
Looking at the quantum mechanical application:  We are mainly interested in predicting an observable $R$ 
in a pure electron (or pure positron) state; that is in a state $\psi$ satisfying $P_+\psi=\psi$  (or, $P_-\psi=\psi$).
If $P_+\psi=\psi$ then the expectation value for an observable $R=q(D) \in Op\psi q$, at time $t$, may be written as 
$$
\<U(t)\psi,RU(t)\psi\>=\<\psi,e^{iKt}T_tq(D)T_{-t}e^{-iKt}\psi\>=
\<\psi,e^{iKt}P_+RP_+e^{-iKt}\psi\>=\<\psi,\kp_c(R)e^{-iKt}\psi\>\ ,\ 
\lo{(13.1)}
$$
using that $q(D)$ is translation invariant, i.e., $T_tq(D)T_{-t}=q(D)$, and  that $P_-\psi=P_-P_+\psi=0$\ ,\ giving
$P_+RP_+\psi=(P_+RP_++P_-RP_-)\psi=\kp_c(R)\psi$.

So, the operator $e^{iKt}\kp_c(R)e^{-iKt}$ really is governing prediction of $R=q(D)$ 
in the sense of the Heisenberg transform, for all times. And, according to thm. 12.1, this operator belongs to
$Op\psi q$, at all $t$.

\end{obs}

\begin{prop}

With the symbols $p_\pm(\xi)=\halb(1\pm\inv{\<\xi\>}h_0(\xi))\ , \ h_0=\a\xi+\b\ ,\ $ we have
$$
P_+- p_+(D)\in Op\psi q_{-1}\ ,\  P_-- p_-(D)\in Op\psi q_{-1}\ ,\ 
\lo{(13.2)}
$$

\end{prop}

\noindent
\tbf{Proof.}
Clearly we obtain a block-matrix representations of the symbols $p_\pm(\xi)$ by setting $\mbf{A}_2=0$ in (12.1)
and (12.1-), where then $\eta^2+a^2=\<\xi\>^2$. Also, modulo $\psi q_{-1}$, we may replace the 
terms $p_{22}$  and $p_{22}^-$  by  $\halb(1\mp \xi_1/\sqrt{\eta^2+a^2})$, as already noted in (12.3).
Looking at (12.2) we observe that the functions $a,c,d$ all are bounded with all their $x_1,y_1$-derivatives.

Taking the differences (13.2) we then note that
$$
\inv{\<\xi\>}-\inv{\sqrt{\eta^2+a^2}}=
\frac{d^2-2c\xi_2}{\<\xi\>\sqrt{\eta^2+a^2}\{\<\xi\>+\sqrt{\eta^2+a^2}\}}\ \ \in\psi q_{-2}\ .\ 
\lo{(13.3)}
$$
This, and similar observations will indeed show the statement, q.e.d.

Now let us come back to formulas (10.21),(10.22),(10.23): According to our arguments, so far, this was 
just a rewriting of (8.17), with its following Fourier series expansion, for the special case of 
$q(D)=D_j\ ,\ j=1,2,3,$ listing the terms of order $0$ and $1$ explicitly, while ignoring all terms of order less than $0$. 
But, recall, this only solves the initial-value problem (8.16) modulo $Op\psi q_{-\i}$; it does  not make 
$e^{-Kt}a_0(x,D)e^{-iKt}$ a $\psi do$ in $Op\psi q$.

On the other hand, looking at  (9.17) --- now established, since we proved thm. 12.1, it is clear that we get
$$
e^{iKt}\kp_c(A_0^\i)e^{-iKt}-\kp_c(A_t^\i)\in Op\psi q_{-\i}\ ,\ 
\lo{(13.4)}
$$

In order to get our formula on Heisenberg's transform, modulo $Op\psi q_{-1}$ it then will be a 
matter of showing that the passing from $A_t^\i$ to $\kp_c(A_t^\i)$ will only produce errors in 
$Op\psi q_{-1}$.

Note,
f'la (10.23) may be written as
$$
a_t(x_1,\xi)=\xi_1+f_+(x_1,\xi)p_+(\xi)+f_-(x_1,\xi)p_-(\xi)\ ,\ \mbox{\ \ (mod\ } \psi q_{-1}\ )\ ,
\lo{(13.5)}
$$
with scalar symbols $f_\pm(x_1,\xi)\in \psi q_0$.

Using (13.2) we may write (13.5) as
$$
a_t(x_1,D)=D_1+f_+(x_1,D)P_+ + f_-(x_1,D)P_-   \ ,\ \mbox{\ \ (mod\ } \psi q_{-1}\ )\ .
\lo{(13.6)}
$$

Here we get
$$
\kp_c(D_1)=D_1+[P_+,D_1]P_++[P_-,D_1]P_-=D_1+[(P_+-p_+(D)),D_1]+[(P_--p_-(D)),D_1]\ ,\ 
\lo{(13.7)}
$$
since $p_\pm(D)$ commute with $D_1$.  Clearly the last two terms in (13.7) belong to $Op\psi q_{-1}$,
since the differences $P_\pm-p_\pm(D)$ are $Op\psi q_{-1}$ while $D_1\in \psi q_{1}$ and because the
commutator with the scalar $D_1$ still has order of the sum of orders decreased by 1. Thus we get
$\kp_c(D_1)-D_1\in Op\psi q_{-1}$.

Next, 
$$
\kp_c(f_+(x_1,D)P_+)=f_+(x_1,D)P_++[P_+,f_+(x_1,D)]P_+\ ,
\lo{(13.8)}
$$
and similarly for $f_-(x_1,D)P_-$\ ,\ 
where again the commutators of $P_\pm$ with the scalar operators $f_\pm(x_1,D)$ are of order $-1$.

As a consequence we get
$$
\kp_c(a_t(x,D))=a_t(x,D)\ \  \mbox{(\ mod\ } \psi q_{-1}\ )\ .\ 
\lo{(13.9)}
$$

With the above we repeat the result of thm 10.4:

\begin{thm}

Set $\th(\xi)=\halb(1-s_1(\xi))$, evaluate (above) 
$\g_t(\xi)=te^{-i\o\th(\xi)t}\vi(\o\th(\xi)t)\ ,\ $
 with  $\vi(\kp)=\frac{\sin\kp}{\kp}\ .$ For any observable $R$ write
$R_t=U^*(t)RU(t).$\ ,\  with the propagator $U(t)$ of the Dirac equation
$\dot{\psi}+iH(t)\psi=0$, with the Dirac operator $H(t)$ of  (8.1), marking
a Dirac particle under the influence of a plane polarized electro-magnetic wave
in the $x_1$-direction.

Then  we have $(H(t))_t-H(0)=(D_1)_t-D_1$  where $(D_1+r_t(x,D))_t$, with a suitable $\psi do$ 
$r_t(x,D)\in Op\psi q_{-1}$ is a $\psi do$ in $Op\psi q_1$ satisfying
$$
(D_1+r_t(x,D))_t-D_1=
\e_0\o t\cos(\o(x_1-t\th(D)))s_2(D)\vi(\o\th(D)t)p_+(D)
$$
$$
 -\e_0\o t\cos(\o(x_1-t\th(-D)))s_2(D)\vi(\o\th(-D)t)p_-(D)\}\ ,
\lo{(13.10)}
$$
a relation valid modulo $Op\psi q_{-1}$.

\end{thm}

We might point again to observation 10.6, above: For our conjecture that the two terms 
at right of (13.10) mark the possibility of a collision between the Dirac particle and a `photon' ,
we can offer only two reasons: (i) the fact that --- in the momentum representation ---
these terms mark a shift of energy by $\pm h\nu$ and of momentum by $\pm h\nu/c$\ ,\  
while multiple collisions will shift by discrete integer multiples of that; (ii) that a
directional shift of propagation speed will enter, similar in nature as that observed by Compton
for the shift of wavelength.

Perhaps others might see more details, in these matters.
\bigskip

\noindent
References   

\bigskip

\footnotesize

\noindent
[Be1] R. Becker, \emph{Theorie der Electrizitaet}; Bd.2, B.G. Teubner
Verlag, Leibzig 1949.

\noindent
[BLT] N. N. Bogoliubov, A. A. Logunov and I. T. Todorov,
{\em Introduction to Axiomatic Quantum Field Theory}, Benjamin, 

Reading, Massachusetts, 1975.

\noindent
[Bu1] V.S.Buslaev, \emph{The generating integral and the canonical Maslov operator
in the WKB-method}; Funct. anal. i 

 ego pril., 3:3 (1969), 17-31. English translation: Funct. Anal. Appl., 3 (1969), 181-193.

\noindent
[CZ] A.P. Calderon and A.Zygmund, \emph{Singular integral operators and
differential equations}; Amer. J. Math. 79 (1957) 

 901-921.

\noindent
[Cp1] A.Compton, Phys.Rev. 21 483 (1923).

\noindent
[Cp2] A.Compton, Phil.Mag. 46 897 (1923).

\noindent
[Cr1] S.Coriasco, \emph{Fourier integral operators in SG-classes (I),
composition theorems and Action on SG-Sobolev spaces}; 

Univ. Politech. Torino 57 (1999) 49-302.

\noindent
[Cr2] S.Coriasco, \emph{Fourier integral operators on SG-spaces (II),
application to SG-hyperbolic Cauchy problems}; Oper. Theory 

Adv. Appl. 126 (2001) 81-91.

\noindent
[CR]  S.Coriasco and L.Rodino, \emph{Cauchy problems for SG-hyperbolic
equations with constant multiplicities}; Ricerche Mat. 

 48 (1999) 25-43.

\noindent
[Co1] H.O.Cordes, \emph{On pseudodifferential operators and smoothness of
special Lie group representations}; Manuscripta Math. 

28 (1979) 51-69.

\noindent
[Co2] H. O. Cordes, \emph{A version of Egorov's theorem for systems of
hyperbolic pseudodifferential equations}, J. of Functional 

Analysis 48 (1982), 285-300.

\noindent
[Co3] H.O.Cordes, \emph{A pseudo-algebra of observables for the Dirac equation};
Manuscripta Math. 45 (1983) 77-105.

\noindent
[Co4] H.O.Cordes, \emph{A pseudodifferential Foldy-Wouthuysen transform};
Communications in PDE 8(13) (1983) 1475-1485.

\noindent
[Co5] H.O.Cordes, {\em The technique of pseudodifferential operators};
London Math. Soc. Lecture Notes 202; Cambridge Univ. 

Press 1995, Cambridge.

\noindent
[Co6] H.O.Cordes, \emph{On Dirac observables}; Progress in Nonlinear DE
42 2000 Birkhaeuser Basel/Switzerland 61-77.

\noindent
[Co7] H.O.Cordes, \emph{Dirac algebra and Foldy-Wouthuysen transform};
Evolution equations and their applications; 

editors Lumer-Weis; 2000 Marcel Dekker inc. New York Basel.

\noindent
[Co8] H.O.Cordes, \emph{A precise pseudodifferential Foldy-Wouthuysen transform
for the Dirac equation}; J. evol. equ. 4 (2004) 

125-138. 

\noindent
[Co9] H.O.Cordes, \emph{Symmetry conditions on Dirac observables}; Proc. Inst. Math.
NAS Ukraine 50 (2004) 671-676.

\noindent
[Co10] H.O. Cordes,  \emph{Lorentz transform of the invariant Dirac
algebra};  Integral equ. oper. theory 34 (1999) 9-27.

\noindent
[Co11] H.O.Cordes, {\em Elliptic pseudo-differential operators - an abstract
theory}; Springer Lecture Notes Math. Vol. 756, 

Springer Berlin Heidelberg New York 1979

\noindent
[Co13] H.O.Cordes, \emph{Remarks about observables for the quantum mechanical
harmonic oscillator}; Operator Theory, Adv., 

Appl., 191 305-321 2009.

\noindent
[Co14] H.O.Cordes, {\em Spectral theory of linear differential operators
and comparison algebras}; London Math. Soc. Lecture 

Notes No.76 (1987); Cambridge Univ. Press; Cambridge.

\noindent
[Co15] H.O.Cordes, \emph{The split of the Dirac Hamiltonian into precisely
predictable energy components}; Fdns. of Phys. 34 

(1004) 1117-1153.

\noindent
[Co16] H.O.Cordes, \emph{Precisely predictable Dirac Observables}; Fundamental
Theories of Physics 154 Springer 2007.

\noindent
[Co17] H.O.Cordes, \emph{On Dirac's first order symmetric hyperbolic system}; to appear.

\noindent
[deV] E.deVries, \emph{Foldy-Wouthuysen transformations and related problems};
Fortschr. d. Physik 18 (1970) 149-182.

\noindent
[DV] D.Dieks and P.Vermaas, {\em The modal interpretation of quantum
mechanics}; 1998 Kluver Akad. Pub., Dordrecht Boston 

London.

\noindent
[Di1] P.A.M.Dirac, {\em The Principles of Quantum Mechanics}; 4-th Edition,
Oxford University Press, London 1976.

\noindent
[DEFJKM] P.Deligne, P.Etingof, D.Freed, L.Jeffrey, D.Kazhdan, and
D.Morrison, {\em Quantum fields and Strings for 

Mathematicians}; Princeton Univ. Press; Princeton 1999.

\noindent
[Eg1] Yu.V. Egorov, \emph{The canonical transformations of
pseudodifferential operators};
Uspehi Mat. Nauk 25 (1969) 235-236.

\noindent
[FS] L.D.Faddeev and A.A.Slawnov, {\em Gauge fields}; Introduction to
Quantum Theory; 

Benjamin/Cummings 1980 Reading MA London Amsterdam Sydney Tokyo.

\noindent
[Far1] G.Farmelo, {\em The Strangest Man, The Hidden Life of
Paul Dirac;} Basic books, Perseus Book Group New York 2009.

\noindent
[FW] L. Foldy, S. Wouthuysen, \emph{On the Dirac theory of spin $-\halb$
particles}.  Phys Rev 78:20-36, 1950.

\noindent
[Gg1]   Gegenbauer, Wiener Sitzungsberichte 88 (1884) 990-1003.

\noindent
[GS] I. Gelfand and G.E.Silov, {\em Generalized Functions},
Vol.1; Acad. Press New York 1964.

\noindent
[GL] M.Gell-Mann and F.Low, \emph{Quantum electrodynamics at small distances};
 Phys. Rev. 95 (1954) 1300-1312 .

\noindent
[Go1] I. Gohberg, \emph{On the theory of multidimensional singular integral
operators}; Soviet Math. 1 (1960) 960-963.

\noindent
[GK] I.Gohberg and N.Krupnik, {\em Einfuehrung in die Theorie der
eindimensionalen singulaeren Integraloperatoren};
Birkhaeuser, 

Basel 1979 (Russian ed. 1973).

\noindent
[GNP] D. Grigore, G. Nenciu, R. Purice,  \emph{On the nonrelativistic limit of the
Dirac Hamiltonian}; Ann. Inst. Henri Poincare 

- Phys. Theor. 51 (1989) 231-263.

\noindent
[Hd1] J. Hadamard, {\em Lectures on Cauchy's problem}; Dover, New York 1953
[Originally published by Yale Univ.Press in 

1923].

\noindent
[Hi1] D. Hilbert, {\em Integralgleichungen}; Chelsea NewYork 1953.

\noindent
[HLP] G.H.Hardy, J.E.Littlewood, and G.Polya, {\em Inequalities}; Cambridge
Univ. Press 1934.

\noindent
[Hs1] W. Heisenberg.  {\em Gesammelte Werke}.
Berlin-New York: Springer, 1984.
    
\noindent
[Hoe1] L. Hoermander, {\em Linear partial differential operators};
Springer New York Berlin Heidelberg 1963.

\noindent
[Hoe2] L.Hoermander, \emph{Pseudodifferential operators and hypo-elliptic
equations}; Proceedings Symposia pure appl. Math. 10 

(1966) 138-183.

\noindent
[Hoe3] L.Hoermander, {\em The analysis of linear partial
differential operators} Vol's I--IV; Springer New York Berlin Heidelberg 

1983-1985.

\noindent
[Hoe4] L.Hoermander, {\em Fourier integral operators I}; Acta.math. 127 (1971) 79-183.

\noindent
[IZ] C. Itzykson and J. B. Zuber, {\em Quantum Field Theory},
McGraw Hill, New York, 1980.

\noindent
[Ka1] T.Kato, {\em Perturbation theory for linear operators};
Springer Verlag Berlin Heidelberg New York 1966.

\noindent
[LS] Laurent Schwartz, {\em Theorie des distributions}; Herman Paris 1966.

\noindent
[MO] W.Magnus and F.Oberhettinger, {\em Formeln und Saetze fuer die
speziellen Funktionen der Mathematischen Physik}; 

2.Auflage, Springer Verlag Berlin Goettingen Heidelberg 1948.

\noindent
[MOS] W.Magnus, F.Oberhettinger and R.P.Soni, {\em Formulas and theorems
for the special functions of Mathematical Physics}; 

3rd edition, Springer Verlag New York 1966.

\noindent
[Ms1] V.P.Maslov, \emph{Theory of perturbations and asymptotic methods};
Moskow Gos. Univ. Moskow, 1965.

\noindent
[Ma1] A. Messiah, {\em Quantum Mechanics}, Vol.I,II; John Wiley NewYork 1958.

\noindent
[Mu] C. M\"uller, {\em Grundprobleme der Mathematischen Theorie
elektromagnetischer Schwingungen}; Springer Verlag, Berlin 

G\"ottingen  Heidelberg 1957.

\noindent
[JvN] J.v.Neumann, {\em Die Mathematischen Grundlagen der Quantenmechanik};
 Springer 1932 New York; reprinted Dover. Publ. inc. 1943; English
translation 1955 Princeton Univ. Press.

\noindent
[PJOA] A.Pais, M.Jacob, D.Olive, M.Atiyah.  {\em Paul Dirac}.
Cambridge: Cambridge University Press, 1998.

\noindent
[Sa] A.Salam, {\em Elementary Particle Theory} N.Svartholm (ed)
Stockholm Almquist Forlag AB 1968

\noindent
[Schr\"{o}1] E. Schr\"{o}dinger.\emph{\"{U}ber den Comptoneffekt;}
Annalen der Physik (4) 82 (1927)

\noindent
[Schr\"{o}2] E. Schr\"{o}dinger.\emph{Quantisierung als
Eigenwertproblem;} Annalen der Physik (4) 79 (1926)

\noindent
[Schr\"{o}3] E. Schr\"{o}dinger.\emph{Collected Papers;}
Friedr. Viehweg und Sohn 1084.

\noindent
[SB] D.Shirkov, N.Bogoliubov. {\em Quantum Fields}.
Reading, MA: Benjamin, 1982.

\noindent
[Sie1]   Siemon, Programm Luisenschule, Berlin (1890)
[Jahrbuch ueber die Fortschritte der Math. (1890) 840-842.]

\noindent
[So1] A.Sommerfeld, {\em Atombau und Spektrallinien},
vol.1. 5th ed. Braunschweig, Viehweg and Sons, 1931.

\noindent
[So2] A.Sommerfeld, \emph{Atombau und Spektrallinien}, Vol.2.
Braunschweig Vieweg and Sons, 1931.

\noindent
[Sn1]   Sonine, Math. Ann. 16 (1880) 38f.

\noindent
[St1]   Struve, Mem. de l'Acad.Imp.des Sci. de St Peterburg (7) 30 (1882)
no. 8; Ann. der Physik, (8) 17 (1882) 1008-1016.

\noindent
[Ta1] M. Taylor, {\em Pseudodifferential operators}; Princeton Univ. Press.,
Princeton, NJ 1981.

\noindent
[Ta2] M.Taylor, {\em Partial differential equations}; Vol.I,II,III;
Springer New York Berlin Heidelberg 1991.

\noindent
[Th1] B.Thaller, {\em The Dirac equation}; Springer 1992 Berlin Heidelberg
New York.

\noindent
[Ti1] E.C.Titchmarsh, {\em Eigenfunction expansions associated with
second order differential equations} Part 1, 2-nd ed.

Clarendon Press, Oxford 1962.

\noindent
[Ti2] E.C.Titchmarsh, {\em Eigenfunction expansions associated
with second order differential equations} Part 2 [PDE];
Oxford 

Univ. Press 1958.

\noindent
[Tr1] F. Treves, {\em Introduction to pseudodifferential and Fourier integral
operators}, Vol's I,II; Plenum Press, New York London 

1980.

\noindent
[Un1] A. Unterberger, \emph{A calculus of observables on a
Dirac particle},
Annales Inst. Henri Poincar\'e (Phys. Th\'eor.), 69 (1998) 

189-239.

\noindent
[Un2] A. Unterberger, \emph{Quantization, symmetries and
relativity}; Contemporary Math. {\bf 214}, AMS (1998), 169-187.

\noindent
[Wa1] G.N.Watson, \emph{A Treatise on the Theory of Bessel Functions};
Cambridge Univ. Press, 1922.

\noindent
[Wb1] S.Weinberg, \emph{on weak forces and gauge theory with SU(2)};
Phys. Rev. Lett. 19 (1967) 1264.

\noindent
[We1] A.Weinstein, \emph{A symbol class for some Schr\"odinger equations on
$\mbb{R}^n$}; Amer. J. Math. (1985) 1-21.

\noindent
[Wk1] J.Walker, \emph{The Analyrical Theory of Light}, Cambridge 1904
392-395.

\noindent
[Wi1] E. Wichmann. {\em Quantenphysik}.  Braunschweig: Viehweg und Sohn, 1985.

\noindent
[YM] C.N.Yang and R.L.Mills, \emph{Conservation of isotopic spin and isotopic
gauge invariance}; Phys.Rev. 96 (1954) 191-195.

\normalsize
\vspace {1 cm}
\noindent
Emeritus Professor\\
Department of Mathematics\\
University of California\\
Berkeley, CA 94720, U.S.A.\\
E-mail: cordes@math.berkeley.edu\\
\vspace{.5cm}
\textsl{(Received: 4 May, 2014)}

\end{document}